\documentclass[preprint,superscriptaddress,nofootinbib,amsmath,amssymb,prd]{revtex4-2}

\usepackage{graphicx}
\usepackage[colorlinks,linkcolor=magenta,anchorcolor=cyan,citecolor=blue]{hyperref}
\usepackage{physics}
\usepackage{subcaption}
\DeclareUnicodeCharacter{2212}{-}
\DeclareUnicodeCharacter{2243}{$\simeq$}
\DeclareUnicodeCharacter{202F}{$\ $}

\begin{document}

\title{Can the sound horizon-free measurement of $H_0$ constrain early new physics?}

\author{Jun-Qian Jiang}
\affiliation{School of Fundamental Physics and Mathematical
    Sciences, Hangzhou Institute for Advanced Study, UCAS, Hangzhou
    310024, China}
\affiliation{School of Physical Sciences, University of Chinese Academy of Sciences, Beijing 100049, China}
\email{jiangjunqian21@mails.ucas.ac.cn}
\email{jiangjq2000@gmail.com}

\author{Yun-Song Piao}
\affiliation{School of Physical Sciences, University of
Chinese Academy of Sciences, Beijing 100049, China}
\affiliation{School of Fundamental Physics and Mathematical
    Sciences, Hangzhou Institute for Advanced Study, UCAS, Hangzhou
    310024, China}
\affiliation{International Center for Theoretical Physics
    Asia-Pacific, Beijing/Hangzhou, China}
\affiliation{Institute of Theoretical Physics, Chinese
    Academy of Sciences, P.O. Box 2735, Beijing 100190, China}
\email{yspiao@ucas.ac.cn}

\begin{abstract}

The sound horizon-independent $H_0$ extracted by using galaxy
clustering surveys data through, e.g., EFTofLSS or ShapeFit
analyses, is considered to have the potential to constrain the
early new physics responsible for solving the Hubble tension.
Recent observations, e.g. DESI, have shown that the sound
horizon-independent measurement of $H_0$ is consistent with $\Lambda$CDM.
In this work, we clarify some potential misuses and
misinterpretations in these analyses. On the one hand, imposing
some prior from other cosmological probes is often used to
strengthen the constraints on the results, however, these priors
are usually derived using the assumption of $\Lambda$CDM, it is
not suitable to apply these so-called $\Lambda$CDM priors (e.g.,
the $n_s$ prior from CMB), which would bias the results, to early
new physics because these early new physics are usually
accompanied by shifts of the $\Lambda$CDM parameters. On the other
hand, the constraints on $H_0$ in the sound horizon-independent
EFTofLSS analysis arise from not only the shape of the power
spectrum ($k_\text{eq}$-based $H_0$), but also the overall
amplitude (when combined with CMB lensing observations) and the
relative amplitudes of the BAO wiggles, thus besides $k_\text{eq}$
other information may also play a role in constraining $H_0$.
We also make forecasts for an Euclid-like survey, which suggest that ongoing observations will also have difficulty constraining early new physics.

\end{abstract}

\maketitle
\newpage

\section{Introduction}
The Hubble tension~\cite{Verde:2019ivm,DiValentino:2021izs,Perivolaropoulos:2021jda,Schoneberg:2021qvd,Shah:2021onj,Abdalla:2022yfr,DiValentino:2022fjm,Hu:2023jqc,Verde:2023lmm} is one of the most severe tensions in recent cosmological observations.
The CMB measurements from \textit{Planck} suggest a smaller value for the Hubble constant $H_0$~\cite{Planck:2018vyg}, while on the other hand, measurements from Cepheid-calibrated Type Ia supernovae, such as those from the SH0ES collaboration, indicate a larger $H_0$~\cite{Riess:2021jrx}. There is a $4 \sim 6 \sigma$ tension between the results of these two types of measurements.
As systematic errors are difficult to fully account for this discrepancy, many modifications to the $\Lambda$CDM model have been proposed in order to resolve this issue.
Meanwhile, the constraints from BAO and uncalibrated Type Ia supernovae on the late-time cosmic expansion history suggest that the Hubble tension may require new physics before recombination to reduce the sound horizon $r_s$~\cite{Bernal:2016gxb,Addison:2017fdm,Lemos:2018smw,Aylor:2018drw,Schoneberg:2019wmt,Knox:2019rjx,Arendse:2019hev,Efstathiou:2021ocp,Krishnan:2021dyb,Cai:2021weh,Keeley:2022ojz,Gomez-Valent:2023uof,Jiang:2024xnu,RoyChoudhury:2024wri,RoyChoudhury:2020dmd}.
Many proposals to reduce the sound horizon have been put forward to resolve the Hubble tension, such as early dark energy (EDE) (see e.g. Refs.\cite{Karwal:2016vyq,Poulin:2018cxd,Kaloper:2019lpl,Agrawal:2019lmo,Lin:2019qug,Smith:2019ihp,Niedermann:2019olb,Sakstein:2019fmf,Ye:2020btb,Gogoi:2020qif,Braglia:2020bym,Ye:2020oix,Lin:2020jcb,Odintsov:2020qzd,Seto:2021xua,Nojiri:2021dze,Karwal:2021vpk,Jiang:2021bab,Ye:2021iwa,Wang:2022jpo,Rezazadeh:2022lsf,Poulin:2023lkg,Odintsov:2023cli}), which accelerates the expansion rate $H(z)$ in the early universe to reduce $r_s$.

The sound horizon is the primary “standard ruler” for measuring the Hubble constant through CMB or LSS observations.
However, there is also other information that can be used to measure $H_0$.
Therefore, investigating sound horizon-free $H_0$ can help clarify the Hubble tension, and especially constrain these potential early new physics.
In Ref.~\cite{Baxter:2020qlr}, it was proposed that the $H_0$ inferred from CMB lensing should be insensitive to the sound horizon, as the BAO wiggles are washed out by the projection integral.
Galaxy surveys, which contain more information than CMB lensing due to having more accessible modes, provide many other ways to sound horizon-independently measure $H_0$.
For example, in Refs.~\cite{Philcox:2020xbv}, it was proposed that using a non-informative prior for the physical baryon density $\omega_b = \Omega_b h^2$ when fitting the galaxy power spectrum can avoid dependence on the sound horizon.
Alternatively, by introducing a parameter that rescales the sound horizon scale (BAO wiggles) and marginalizing over it, information from the sound horizon can be removed~\cite{Farren:2021grl}.
The ShapeFit method~\cite{Brieden:2021edu} provides another similar measurement approach by fitting the parameter $m$ to the power spectrum~\cite{Brieden:2022heh}.
The $H_0$ extracted using these methods is typically considered to be obtained by taking the matter-radiation equality scale $k_\text{eq}$ as the “standard ruler" (which, as will be clarified in this paper, is not accurate), and therefore it is sometimes referred to as the “$k_\text{eq}$-based $H_0$".
In addition, by fitting the turnover scale of the matter power spectrum, the matter-radiation scale $k_\text{eq}$ can be measured in a model-independent way, allowing for the determination of sound horizon-free $H_0$~\cite{Bahr-Kalus:2023ebd}.
The relative amplitude of the BAO wiggles in the matter power spectrum, combined with the $\omega_b$ inferred from BBN and the $\Omega_m$ measured by other methods, can also be used to measure $H_0$~\cite{Krolewski:2024jwj}.

The new physics proposed to address the Hubble tension typically focuses on increasing $H_0$ by reducing the sound horizon $r_s$, which means that these new physics usually do not have a fully consistent impact on the sound horizon-independent $H_0$ inferred from other information.
There are two strategies to utilize these two $H_0$: one is to take $\Lambda$CDM and each potential new physics as assumed models to infer these two $H_0$, respectively.
A correct assumed model should yield two consistent $H_0$ and also be compatible with the $H_0$ measured directly in the local universe if the Hubble tension is to be solved.
Another strategy is to focus on the two $H_0$ inferred by assuming $\Lambda$CDM model.
If $\Lambda$CDM is correct, then the two $H_0$ are naturally self-consistent.
Conversely, the two $H_0$ may not be self-consistent if there is some potential early new physics.
For example, it has been proposed in Ref.~\cite{Farren:2021grl} if some axion-like EDE models are correct, Euclid-like surveys may be able to distinguish these two $H_0$ inferred with $\Lambda$CDM model.
In this paper, we discuss the latter strategy.
\footnote{We would like to caution that the sound horizon-independent $H_0$ inferred by the former strategy may be biased due to prior volume effects~\cite{Smith:2022iax}, since the new physics usually increases the parameters of the model.}

Recently, there have been attempts to infer the sound horizon-independent $H_0$ using real data.
Among these, the $H_0$ inferred from DESI Y1 data~\cite{Zaborowski:2024wpo} following the method of \cite{Farren:2021grl}, and the $H_0$ inferred using the ShapeFit method from BOSS DR12~\cite{Brieden:2022heh}, are the most tightly constrained values.
Neither of them found a significant deviation between the sound horizon-free $H_0$ and the sound horizon-based $H_0$.
These results are argued to put pressure on early new physics models such as EDE.

In this work, we demonstrate that the current observations and analyses actually cannot distinguish between the $\Lambda$CDM model and potential new early physics.
And we also clarify factors that could lead to unfair model comparisons in these methods.
In particular, we explicitly demonstrate that using overly strong priors derived from the assumption of $\Lambda$CDM may lead to incorrect horizon-independent $H_0$ through mock data analysis.
For the methods used in Refs.~\cite{Farren:2021grl,Philcox:2022sgj,Zaborowski:2024wpo}, we analyze how to correctly interpret these methods, which shows that these horizon-independent $H_0$ are not equivalent to the “$k_\text{eq}$-based $H_0$" because other information also contributes to the constraints on $H_0$.
Finally, we question whether early new physics necessarily leads to two different $H_0$ values, and conducted mock data analysis to investigate whether surveys like Euclid can effectively constrain these early new physics using sound horizon-free $H_0$.

\section{Inferring $H_0$ from galaxy surveys} \label{sec:method}

\subsection{Sound horizon-based $H_0$ from BAO}

The BAO wiggles on the matter power spectrum originate from acoustic oscillations before recombination, so their scale is naturally related to the sound horizon $r_s$.
BAO measurements are fitted using a template based on a fiducial cosmology.
Our pipeline carefully follows the analysis of DESI~\cite{DESI:2024uvr}.
The galaxy power spectrum is modeled as
\begin{equation}
    P(k, \mu) = \mathcal{B}(k, \mu) P_\text{lin}^\text{nw}(k) + \mathcal{C}(k, \mu) P_\text{lin}^\text{w}(k) + \mathcal{D}(k) \, ,
\end{equation}
where $P_\text{lin}^\text{nw}(k)$ and $P_\text{lin}^\text{w}(k)$ are the no-wiggle and wiggle components decomposed from the linear matter power spectrum using the “peak average" method~\cite{Brieden:2022lsd}.
The function $\mathcal{B}(k, \mu)$ takes into account the linear galaxy bias $b_1$ and the “Fingers of God" effect, which is characterized by a smoothing scale $\Sigma_s$:
\begin{equation}
    \mathcal{B}(k, \mu) = (b_1 + f \mu^2 )^2 (1 + \frac{1}{2} k^2 \mu^2 \Sigma_s^2)^{-2} \, ,
\end{equation}
where $f$ is the growth rate.
Similarly, the function $\mathcal{C}(k, \mu)$ takes into account the linear galaxy bias and the non-linear damping effect on the BAO component:
\begin{equation}
    \mathcal{C}(k, \mu)=(b_1 +f \mu^2)^2 \exp[- \frac{1}{2} k^2 ( \mu^2 \Sigma_\parallel^2 +(1- \mu^2 ) \Sigma_\perp^ {2} ) ] \,,
\end{equation}
where $\Sigma_\parallel$ and $\Sigma_\perp$ are the damping scales for modes along and perpendicular to the line-of-sight, respectively.
For the post-reconstructed spectrum, they may also need to include the smoothing kernel.
The remaining nonlinear effects are captured by $ \mathcal{D}(k) $, which is composed of a set of piecewise cubic spline functions.
Due to the Alcock–Paczyński effect, the galaxy power spectrum is evaluated at $(1 + \alpha_\parallel) k_\text{obs}$ for the direction parallel to the line of sight and $(1 + \alpha_\perp) k_\text{obs}$ for the direction perpendicular to the line of sight instead of the observed $k_\text{obs}$.
The two parameters $\alpha_\parallel$ and $\alpha_\perp$ are the main results of BAO measurements:
\begin{equation}
    \alpha_\parallel(z) = \frac{H(z)^\text{fid}r_s^\text{fid}}{H(z)r_s}, \quad \alpha_\perp(z) = \frac{D_A(z)r_s^\text{fid}}{D_A(z)^\text{fid}r_s},
\end{equation}
where $D_A(z)$ is the angular diameter distance.
They may be reparameterized as:
\begin{equation}
    \alpha_\text{iso} = (\alpha_\parallel \alpha_\perp^2)^{1/3}, \quad \alpha_\text{AP} = \alpha_\parallel / \alpha_\perp .
\end{equation}
Once the late-time cosmology is assumed as $\Lambda$CDM, which means that the expansion history of the late universe can be fully controlled by $H_0$ and $\Omega_m$\footnote{Here, we ignore the effect of neutrinos on the late-time cosmological expansion history, as their mass has already been tightly constrained~\cite{Vagnozzi:2017ovm,Wang:2024hen,Naredo-Tuero:2024sgf,Du:2024pai,Jiang:2024viw,RoyChoudhury:2018gay,RoyChoudhury:2019hls}.}, then these quantities can be decomposed into constraints on $r_sH_0$ and $\Omega_m$.
To obtain $H_0$, the sound horizon $r_s$ derived from the CMB fits can be used for calibration.
Alternatively, we combine the BAO results with $\omega_b$ inferred from the BBN, as this is the only other quantity needed to compute the sound horizon in the $\Lambda$CDM model (with $N_\text{eff}$ fixed).

\subsection{Sound horizon-free $H_0$}

Here, we briefly summarize two methods for inferring horizon-independent $H_0$ from galaxy surveys.
In addition to galaxy surveys, CMB lensing itself can also be used to infer horizon-independent $H_0$.
Therefore, these galaxy survey methods are sometimes combined with CMB lensing analyses to break some parameter degeneracies, thereby strengthening the constraints.

\subsubsection{EFTofLSS}
The Effective Field Theory of Large-Scale Structure (EFTofLSS) allows us to directly fit the full power spectrum up to the quasi-linear scales.
In this work, we follow Ref.~\cite{DESI:2024jis,DESI:2024hhd,Zaborowski:2024wpo} and perform the analysis using the code \texttt{velocileptors}~\cite{Chen:2020fxs,Chen:2020zjt} under the Lagrangian perturbation theory (LPT) up to third order with $b_3$ fixed to 0.
EFTofLSS will also capture the sound horizon information from the scale of the BAO wiggles.
A feasible way to eliminate the information from the sound horizon is to introduce a free parameter $q_\text{BAO}$ to scale the BAO wiggle scale:
\begin{equation}
    P_\text{lin}(k, q_\text{BAO}) = P_\text{lin}^\text{nw}(k) + P_\text{lin}^\text{w}(q_\text{BAO} \cdot k) \, ,
\end{equation}
where $P_\text{lin}^\text{nw}(k)$ and $P_\text{lin}^\text{w}(k)$ are split from the linear matter power spectrum with the same “peak average" method~\cite{Brieden:2022lsd} as in our BAO analysis.
This method is also robust beyond the $\Lambda$CDM models.
Note that this is different from the method used in Refs.~\cite{Farren:2021grl}.
In principle, the information about the sound horizon also comes from the scale of baryonic Jeans suppression~\cite{Lesgourgues:2006nd}.
However, it has been shown that even for Euclid-like observations, its impact on the results is negligible.
Therefore, we will not consider its effect on $H_0$ here.
EFTofLSS has a large number of parameters whose prior would bias the results~\cite{Simon:2022lde,Holm:2023laa}.
To alleviate this problem, following~\cite{Zaborowski:2024wpo}, we use a physically motivated prior for the EFT parameters (nuisance parameters) and impose an approximate Jeffreys prior~\cite{1946RSPSA.186..453J} on the counter terms and the stochastic terms by analytically~\cite{Maus:2024dzi} fixing them to the best-fit value for every set of other parameters~\cite{Hadzhiyska:2023wae}.
Since loop integrals (and the Boltzmann code) are time-consuming, a 4th-order Taylor series emulator with cosmological parameters (including those modifying the power spectrum) as variables is trained and used for each bin.
These calculations are recovered in the importance sampling.

\subsubsection{ShapeFit}
ShapeFit~\cite{Brieden:2021edu} is a model-independent method for compressing information of the power spectrum into a few quantities.
It characterizes the deviation of the linear power spectrum from the fiducial cosmology using five compressed variables:
\begin{equation}
    \alpha_\text{iso}(z), \ \alpha_\text{AP}(z), \ f(z) \sigma_{s8}(z), \ m(z), \ n(z) \ .
\end{equation}
Compared to traditional BAO and RSD analyses, the additional shape parameters $n$ and $m$ describe the scale-independent and scale-dependent dependencies of the linear power spectrum slope:
\begin{equation}
    P_\text{lin}(k) = P_\text{lin}^\text{fid} \exp{ \frac{m}{a} \tanh[a \ln(\frac{k}{k_p})] + n \ln(\frac{k}{k_p}) } \, ,
\end{equation}
where the pivot scale for ShapeFit is fixed at $k_p = \pi / r_s^\text{fid} \simeq 0.03 h^\text{fid}$ Mpc$^{-1}$ and $a$ is chosen to be $0.6$.
In the $\Lambda$CDM model, the scale dependence of the linear power spectrum is manifested as a logarithmic dependence determined by the scale $k_\text{eq}$.
Therefore, $m$ provides a measurement for $k_{\text{eq}} \sim \Omega_m H_0^2$ (in the unit of Mpc$^{-1}$).
$H_0$ can be determined if a constraint on $\Omega_m$ is provided, which can come from other cosmological probes or from the uncalibrated BAO information in ShapeFit itself.

In this work, we follow the “$\mathfrak{D}_V + F_\text{AP} + m$” analysis of Ref.~\cite{Brieden:2022heh}, which uses the BAO and shape information in ShapeFit to provide sound horizon-independent constraints on $H_0$.
In our notation, we use the compressed variable $\alpha_\text{iso}(z), \alpha_\text{AP}(z), m(z)$ obtained from mock observations to fit the linear power spectrum, while introducing a free nuisance parameter to scale the sound horizon similarly to EFTofLSS.

In order to fit the linear power spectrum template provided by ShapeFit to the observations, nonlinear corrections are also required.
Here, we again use \texttt{velocileptors} and recompute for each set of parameters rather than using approximations.

\begin{table}[]
    \centering
    \scriptsize
    \begin{tabular}{|c|c|} \hline \hline
        \multicolumn{2}{|c|}{\textbf{Cosmological Parameters}} \\ \hline
        $H_0$                & [20, 100] \\
        $\Omega_m$           & [0.01, 1] \\
        $\omega_b$           & $\mathcal{N}(0.02218, 0.00055)$ \\
        $\ln(10^{10} A_{s})$ & [1.61, 3.91] \\
        $n_s$                & $\mathcal{N}(0.9649, 0.042)$ or $\mathcal{N}(0.96, 0.02)$ or [0.8, 1.2] \\
        $m_\nu$              & fixed to 0.06 eV\\
        $N_\text{eff}$       & fixed to 3.046\\ \hline
        \multicolumn{2}{|c|}{\textbf{Power spectrum modification parameter} (fixed to 1 if not used)} \\ \hline
        $q_\text{BAO}$       & [0.9, 1.1] \\
        $A_\text{BAO}$       & [0.8, 1.2] \\
        $A_{A_\text{s, FS}}$   & [0.9, 1.1] \\ \hline
        \multicolumn{2}{|c|}{\textbf{BAO Parameters}} \\ \hline
        $\alpha_\text{iso}$  & [0.8, 1.2] \\
        $\alpha_\text{AP}$ (not used in monopole fitting alone) & [0.8, 1.2] \\
        $\Sigma_\parallel$   & $\mathcal{N}(\Sigma_\parallel^\text{fid}, 2.0)$ (see Table.6 of \cite{DESI:2024uvr}) for the values of $\Sigma_\parallel^\text{fid}$) \\
        $\Sigma_\perp$       & $\mathcal{N}(\Sigma_\perp^\text{fid}, 1.0)$ (see Table.6 of \cite{DESI:2024uvr}) for the values of $\Sigma_\perp^\text{fid}$) \\
        $\Sigma_s$           & $\mathcal{N}(2.0, 2.0)$ \\
        $b_1$                & [0.2, 4] \\
        $d\beta$ (not used in monopole fitting alone) &  [0.7, 1.3] \\
        $a_{0,n}$            & $\mathcal{N}(0, 10^4)$ \\
        $a_{2,n}$ (not used in monopole fitting alone) & $\mathcal{N}(0, 10^4)$ \\ \hline
        \multicolumn{2}{|c|}{\textbf{EFTofLSS Parameters}} \\ \hline
        $(1+b_1)\sigma_8(z)$ & [0.0, 3.0] \\
        $b_2\sigma_8(z)^2$, $b_s\sigma_8(z)^2$ & $\mathcal{N}(0, 5)$ \\
        $\alpha_0, \alpha_2, \alpha_4$ & $\mathcal{N}(0, 12.5)$ + Approx. Jeffreys prior \\
        SN$_0$               & $\mathcal{N}(0, 2) \times 1/\bar{n}_g$ + Approx. Jeffreys prior \\
        SN$_2$               & $\mathcal{N}(0, 5) \times f_\text{sat}\sigma^2_\text{1 eff}/\bar{n}_g$ + Approx. Jeffreys prior \\
        SN$_4$               & $\mathcal{N}(0, 5) \times f_\text{sat}\sigma^4_\text{1 eff}/\bar{n}_g$ + Approx. Jeffreys prior \\ \hline
        \multicolumn{2}{|c|}{\textbf{ShapeFit Parameters} (in addition to the BAO parameters $\alpha_\text{iso}$, $\alpha_\text{AP}$ above)} \\ \hline
        $m - m_\text{fid}$    & [-3, 3] \\
        $n - n_\text{fid}$    & [-0.5, 0.5] or fixed to 1 \\
        $f/f_\text{fid}$      & [0., 2.] \\ \hline
        \hline
    \end{tabular}
    \caption{The prior of the parameters used in this work,
    where $\mathcal{N}(\mu, \sigma)$ represents a normal distribution with mean $\mu$ and standard deviation $\sigma$.
    For some parameters, we explore different choices of priors.}
    \label{tab:prior}
\end{table}

\section{The capacity of DESI Y1 measurements to constrain early new physics} \label{sec:DESIY1}

It has been claimed that the DESI Y1 results show that the sound horizon-free $H_0$ is consistent with the sound horizon-based $H_0$ at the current observation level~\cite{Zaborowski:2024wpo}.
\footnote{We do not consider the constraints of Ref.~\cite{Brieden:2022heh} in this work because of the tight $n_s$ prior there.}
Here, we test whether this analysis has already imposed constraints on some early new physics.
It should be noted that we are not testing all the information in the DESI Y1 data, which includes other information such as the amplitude of the matter power spectrum, to constrain these potential new physics.
Instead, we focus solely on whether the relation between the two $H_0$ can provide constraints.

We assume various early new physics scenarios and generate mock data using observational redshift ranges, survey area, and the number of tracers observed, similar to those of DESI Y1.
The covariance takes into account statistical errors but not systematic errors, which, according to Refs.~\cite{DESI:2024uvr,DESI:2024jis}, are subdominant contributions.
The details of the mock data are provided in Appendix \ref{sec:mockspec}.
Then, we follow a pipeline similar to the DESI Y1 analysis~\cite{DESI:2024uvr,DESI:2024jis,Zaborowski:2024wpo} (partially described in \autoref{sec:method}) to obtain constraints on cosmological parameters from pre-reconstructed BAO in galaxy and quasar clusters, and the sound horizon-free constraints using EFTofLSS.
The parameters used in our analysis and their prior are summarized in \autoref{tab:prior}.
A prior for $\omega_b$ from updated BBN results~\cite{Schoneberg:2024ifp} is always imposed.
The range of fitted scales for EFTofLSS is $0.02 \sim 0.20 h$Mpc$^{-1}$, while for BAO it is $0.02 \sim 0.30 h$Mpc$^{-1}$.
For $n_s$, in this section we impose a $10\times$ looser prior ($n_s \sim \mathcal{N}(0.9649, 0.042)$) from the \textit{Planck} result~\cite{Planck:2018vyg}, except for the case with CMB lensing, where a tighter prior ($n_s \sim \mathcal{N}(0.96, 0.02)$) is imposed.
Although, as will be shown below, some choices of the prior may not be appropriate, here we follow the choice of \cite{Zaborowski:2024wpo} and investigate what would be the expected result if the assumed new physics exists.
For BAO parameter inference and EFTofLSS analysis, we use the No-U-Turn sampler (NUTS)~\cite{Hoffman:2011ukg}, while for inferring cosmological parameters from BAO parameters, we use Cobaya~\cite{Torrado:2020dgo}.
All these MCMC chains converge to achieve the Gelman-Rubin statistic to $R-1<0.03$.
Importance sampling is also performed when emulators are used.

For the sound horizon-free $H_0$, we examine two datasets examined in Ref.~\cite{Zaborowski:2024wpo}: galaxy clustering only and its combined analysis with CMB lensing (\textit{Planck} PR4~\cite{Carron:2022eyg} and ACT DR6~\cite{ACT:2023dou,ACT:2023kun}) and the constraint on $\Omega_m$ by Pantheon+~\cite{Scolnic:2021amr}.
While the DESI galaxy clustering data used in our analysis are mock data generated for simulation purposes, the CMB lensing and Pantheon+ datasets are drawn from real observational data.
The other Type Ia supernovae data is not used as they exhibited similar results to Pantheon+ in \cite{Zaborowski:2024wpo}.
The constraints on $\Omega_m$ from the 3D power spectrum of the DESI Ly-$\alpha$ forest via the Alcock-Paczyński effect are not available at the time we prepare this work.

\begin{figure}[htb!]
    \centering
\begin{subfigure}{0.49\textwidth}
\includegraphics[width=\linewidth]{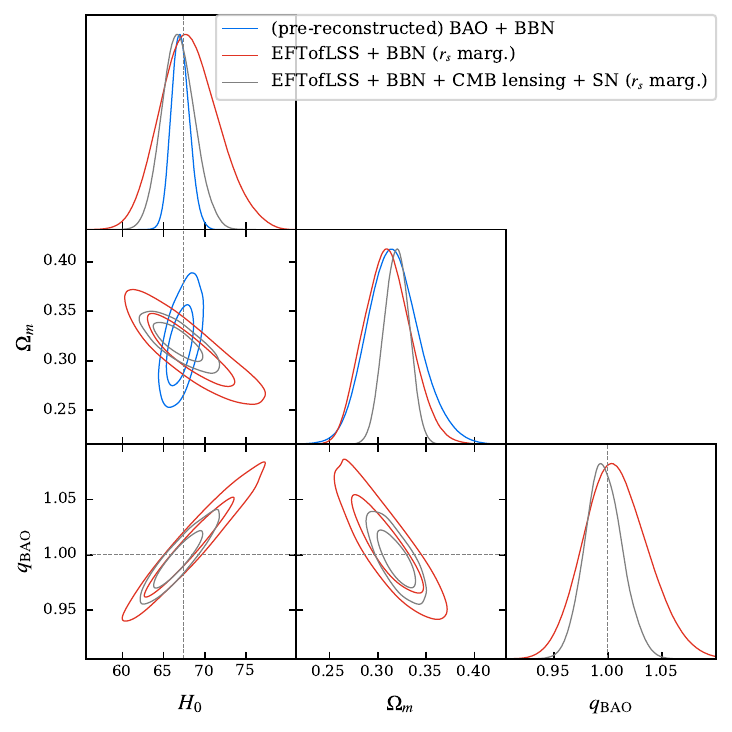}
\caption{Planck2018}
\end{subfigure}
\begin{subfigure}{0.49\textwidth}
\includegraphics[width=\linewidth]{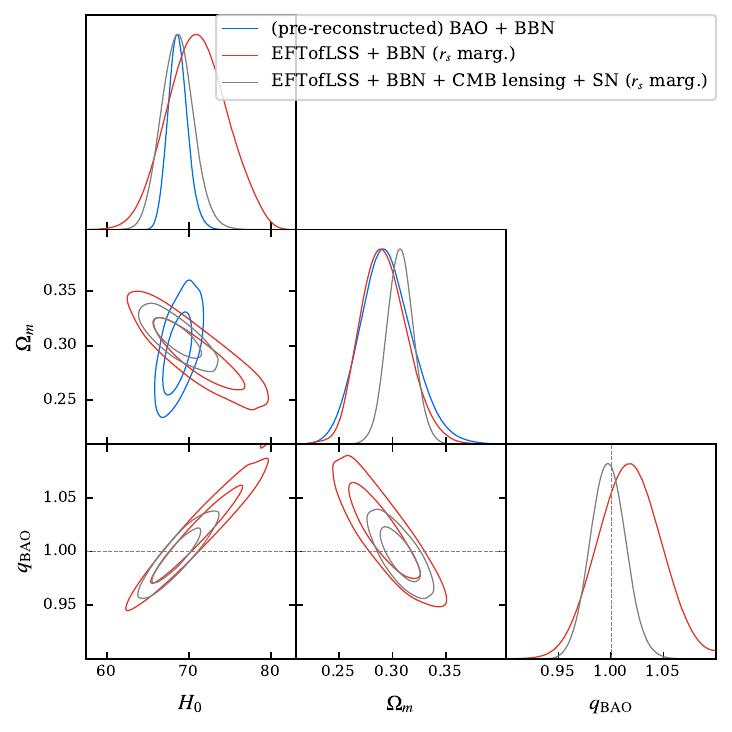}
\caption{Smith2019}
\end{subfigure}
\begin{subfigure}{0.49\textwidth}
\includegraphics[width=\linewidth]{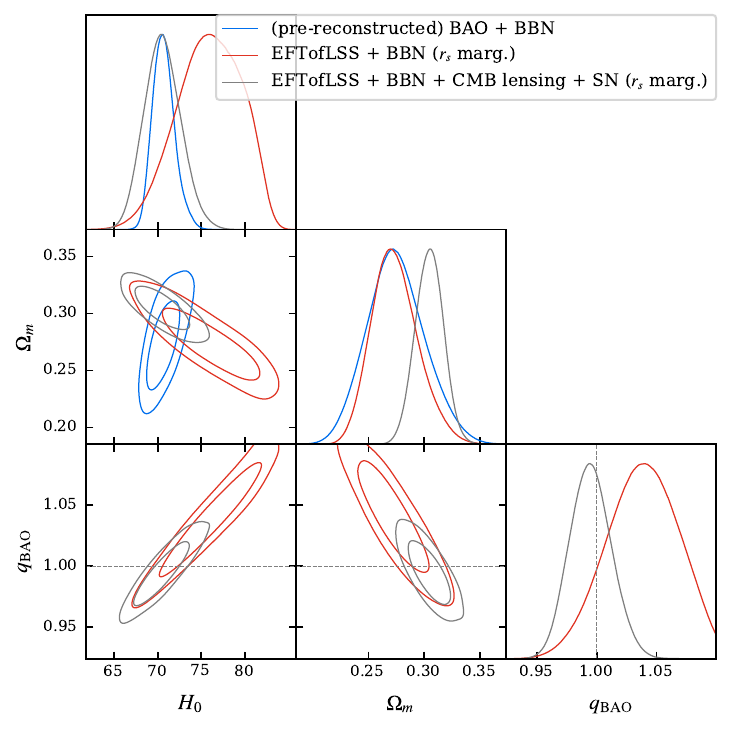}
\caption{Hill2021}
\end{subfigure}
\begin{subfigure}{0.49\textwidth}
\includegraphics[width=\linewidth]{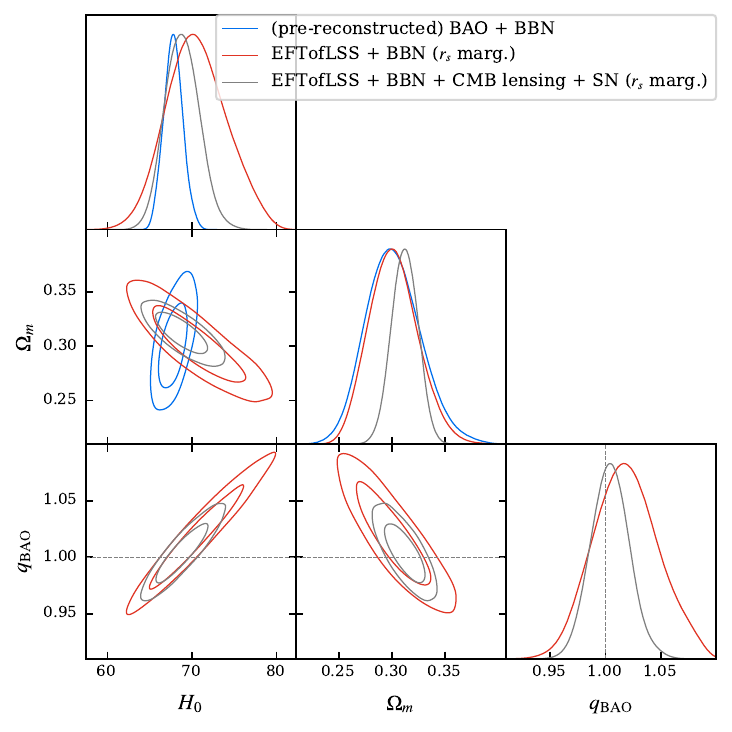}
\caption{Jiang2024}
\end{subfigure}

\caption{The expected constraints ($1 \sigma$ and $2 \sigma$ confidence intervals) of a DESI Y1-like survey on $H_0$, $\Omega_m$ and $q_\text{BAO}$ with mock data generated from a variety of cosmologies.
Real data are used for CMB lensing and Type Ia supernovae (SN), see \autoref{sec:DESIY1} for details.
$q_\text{BAO}=1$ is shown in dashed lines.
In addition, for Planck2018, we indicate its true $H_0=67.36$ km/s/Mpc with a dashed line.}
\label{fig:DESIY1}
\end{figure}

\begin{table}[htb!]
    \centering
    \begin{tabular}{|c|c|c|c|c|} \hline \hline
                              & Planck2018 & Smith2019 & Hill2021 & Jiang2024 \\ \hline
(pre-reconstructed) BAO + BBN & $67.0_{-1.20}^{+0.91}$ & $68.7_{-1.35}^{+0.94}$ & $70.7_{-1.3}^{+1.1}$ & $67.8\pm1.1$ \\
EFTofLSS + BBN ($r_s$ marg.) & $68.2^{+3.4}_{-3.7}$ & $71.0_{-3.7}^{+3.6}$ & $76.0_{-3.7}^{+4.0}$ & $70.6_{-4.3}^{+2.9}$ \\
EFTofLSS + BBN + CMB lensing + SNeIa ($r_s$ marg.) & $66.8\pm2.0$ & $68.6_{-1.9}^{+2.0}$ & $70.6_{-2.1}^{+2.0}$ & $68.8_{-2.2}^{+1.9}$ \\
\hline \hline
\end{tabular}
\caption{The expected constraints (mean values and $1 \sigma$ confidence intervals) of a DESI Y1-like survey on the sound horizon-based $H_0$ (km/s/Mpc, first row) and sound horizon-independent $H_0$ (last two rows) with mock data generated from a variety of cosmologies.
Real data are used for CMB lensing and Type Ia supernovae (SNeIa), see \autoref{sec:DESIY1} for details.}
\label{tab:DESIY1}
\end{table}

We investigate four cosmological models.
The results are shown in \autoref{fig:DESIY1} and \autoref{tab:DESIY1}.
The first is the \textit{Planck} 2018 $\Lambda$CDM model (based on the mean values of the \texttt{base\_plikHM\_TTTEEE\_lowl\_lowE\_lensing} chain, \textbf{Planck2018} hereafter)~\cite{Planck:2018vyg}, which is also the fiducial model for BAO and ShapeFit analysis.
The different degenerate directions of BAO and EFTofLSS in the $H_0$ - $\Omega_m$ plane suggest that they have different sources of parameter constraints.
Both of them unbiasedly recover the $H_0=67.36$ km/s/Mpc we used to generate the mock data within the uncertainties ($< 0.5 \sigma$).
It validates the correctness of our pipeline.
Adding CMB lensing and SNeIa tightens the constraint on $H_0$ but does not significantly change the mean value.
Furthermore, the uncertainties of our EFTofLSS analysis for $H_0$ are comparable to the analysis of the DESI Y1 real data~\cite{Zaborowski:2024wpo}, but slightly tighter for EFTofLSS + BBN which may arise from systematic errors and the more accurate modeling of the footprints and the redshift distributions of galaxies in DESI Y1.

For potential early new physics, we examine three EDE cosmologies.
Two of them are axion-like EDE models~\cite{Poulin:2018cxd}, chosen as the best fit for \textit{Planck} 2015 + SH0ES + BAO + SNeIa ($f_\text{EDE}=0.122$, $H_0=72.19$ km/s/Mpc, \textbf{Smith2019} hereafter)~\cite{Smith:2019ihp} and the best fit for ACT DR4 ($f_\text{EDE}=0.241$, $H_0=77.6$ km/s/Mpc, \textbf{Hill2021} hereafter)~\cite{Hill:2021yec}, respectively.
They have been discussed in Refs.~\cite{Farren:2021grl,Zaborowski:2024wpo}.
In addition to them, we consider a $\phi^4$ AdS-EDE model~\cite{Ye:2020btb} with best-fit values for state-of-the-art CMB + BAO + SNeIa + $H_0$ data ($f_\text{EDE}=0.11$, $H_0=72.22$ km/s/Mpc, \textbf{Jiang2024} hereafter)~\cite{Jiang:2024nha}.
We will discuss the differences between these cosmologies and the implications for $H_0$ inference in \autoref{sec:Euclid}.
Here, we focus on the face values of the two classes of $H_0$ inferred from a DESI Y1-like survey.
It should be noted that the inferred $H_0$ is not required to recover the true $H_0$ of the models since the analysis pipeline is carried out under the $\Lambda$CDM model, instead we need to be aware of the consistency between the two classes of $H_0$.
We find that the discrepancies between $H_0$ inferred from EFTofLSS + BBN and (pre-reconstructed) BAO + BBN are $0.60\sigma$ (Smith2019), $1.37\sigma$ (Hill2021) and $0.63\sigma$ (Jiang2024) respectively.

Since the late universe is $\Lambda$CDM in all the models we consider, both pre-reconstructed BAO and post-reconstructed BAO should be able to unbiasedly infer $r_sH_0$ and $\Omega_m$ if the systematic errors are negligible, and in combination with BBN under the $\Lambda $CDM assumption to infer the same $H_0$ except that the latter has a smaller error.
We show in \autoref{sec:exp_BAO} that our (pre-reconstructed) BAO analysis pipeline indeed recovers the theoretically predicted constraints on the cosmological parameters.
Therefore, for the post-reconstructed BAO from all tracers (including the Ly-$\alpha$ forest), we simply use the central values of the pre-reconstructed BAO results and the uncertainties from the DESI Y1 real data results as the expected constraints.
With this assumption, the discrepancies between $H_0$ inferred from EFTofLSS + BBN and (post-reconstructed) BAO + BBN are $0.61\sigma$ (Smith2019), $1.40\sigma$ (Hill2021) and $0.64\sigma$ (Jiang2024) respectively.
Both Smith2019 and Jiang2024 are very close to the results of the corresponding real data ($0.64 \sigma$)~\cite{Zaborowski:2024wpo}.
This result suggests that the relationship between the two $H_0$ inferred from the DESI Y1 data does not place pressure on these two models.

It has been noticed in Ref.~\cite{Zaborowski:2024wpo} that for real data, $q_\text{BAO}$ will shift from 1 to a smaller value after the CMB lensing and SNeIa data were added.
In our mock data analysis, none of the models (including Planck2018) present a preference for $q_\text{BAO}<1$.
However, for all these EDE models, $q_\text{BAO}$ is shifted towards smaller values when CMB lensing and SNeIa data are added, which is consistent with the trend in \cite{Zaborowski:2024wpo}.
Interestingly, the mean values of $q_\text{BAO}$ are all shifted to around 1 and cause $H_0$ to be very consistent with the sound horizon-based results.
This shift is related to the constraints on $\Omega_m$ by SNeIa.
Pantheon+ (and other recent SNeIa data) favours a relatively high $\Omega_m$, which corresponds to a higher $H_0$ and smaller $q_\text{BAO}$, whereas the EDE model we examine here favours a lower $\Omega_m$ in the EFTofLSS analysis.
It can also be found in DESI Y1 real data~\cite{Zaborowski:2024wpo} that sound horizon-independent EFTofLSS analyses favour a lower $\Omega_m$.
It is not clear whether this is a coincidence or a hint for new physics.

\section{Toward the fair comparison: the role of $n_s$ prior} \label{sec:ns1LCDM}

In order to strengthen the constraints, priors for cosmological parameters obtained from other cosmological probes are sometimes included in analysis~\cite{Zaborowski:2024wpo,Brieden:2022heh}.
However, some priors require the assumption that the early universe is $\Lambda$CDM.
In fact, in addition to $H_0$, early new physics usually requires a simultaneous shift of cosmological parameters to fit the observations.
For example, $n_s$ is significantly raised for the EDE model.
The use of CMB-based $A_s$ and $n_s$ in sound horizon-independent $H_0$ analyses was questioned in Ref.~\cite{Smith:2022iax}.
Here we explicitly show that the $n_s$ prior will induce bias on the results by mock data analysis.

We consider a DESI Y1-like survey and an Euclid-like survey, which is similar to \cite{Chudaykin:2019ock,Farren:2021grl}, see Appendix \ref{sec:mockspec} for details.
The BAO and EFTofLSS analysis pipeline for the Euclid-like survey is similar to \autoref{sec:DESIY1}, but all the analyses use the same range of scales ($k \in [0.01, 1] h$Mpc$^{-1}$).

We use $\Lambda$CDM cosmologies to generate mock data.
The cosmological parameters are identical to Planck2018, except $n_s$, which is chosen to be $1$.
This choice of $n_s$ comes from conjecture based on the results of the early new physics needed to solve the Hubble tension~\cite{Ye:2021nej,Jiang:2022uyg,Cruz:2022oqk,Jiang:2022qlj,Jiang:2023bsz,Peng:2023bik,Wang:2024tjd,Wang:2024dka} and some corresponding inflation models (e.g. \cite{Kallosh:2022ggf,Ye:2022efx}).
A fair $\Lambda$CDM-based analysis should be able to recover the value of $H_0$ without bias since the mock data is generated under the same model.

\begin{figure}[htb!]
    \centering
\begin{subfigure}{0.49\textwidth}
\includegraphics[width=\linewidth]{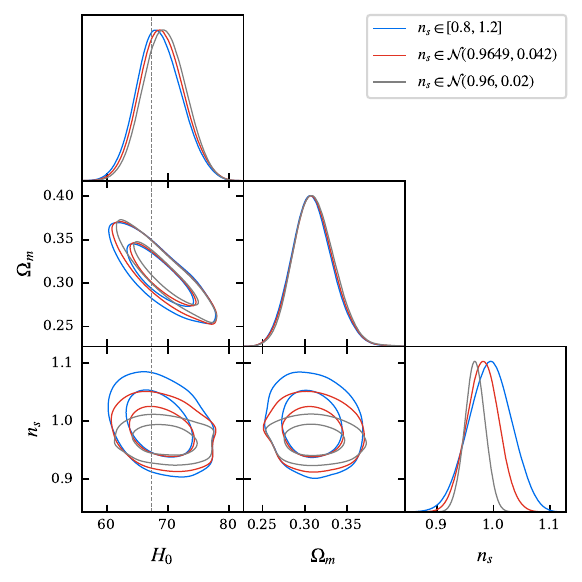}
\caption{DESI Y1-like survey}
\end{subfigure}
\begin{subfigure}{0.49\textwidth}
\includegraphics[width=\linewidth]{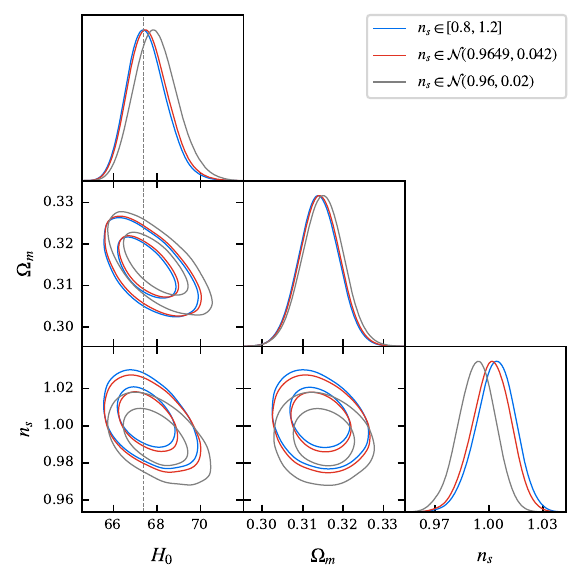}
\caption{Euclid-like survey}
\end{subfigure}

\caption{Sound horizon-free EFTofLSS + BBN constraints ($1 \sigma$ and $2 \sigma$ confidence intervals) of different surveys on $H_0$, $\Omega_m$ and $n_s$ with mock data generated from a $n_s=1$ $\Lambda$CDM model.}
\label{fig:ns1LCDM}
\end{figure}

\begin{table}[htb!]
\centering
\begin{tabular}{|c|c|c|c|} \hline \hline
$n_s$ prior & [0.8, 1.2] & $\mathcal{N}(0.9649, 0.042)$ & $\mathcal{N}(0.96, 0.02)$ \\ \hline
DESI Y1-like survey & $68.6_{-3.7}^{+3.4}$ & $69.0_{-3.6}^{+3.4}$ & $69.5\pm3.5$ \\
Euclid-like survey & $67.54_{-1.01}^{+0.76}$ & $67.63_{-1.01}^{+0.80}$ & $67.96_{-1.08}^{+0.87}$ \\
\hline \hline
\end{tabular}
\caption{Sound horizon-free EFTofLSS + BBN constraints (mean values and $1 \sigma$ confidence intervals) of different surveys on $H_0$ (km/s/Mpc, $r_s$ marginalized) with mock data generated from a $n_s=1$ $\Lambda$CDM model.}
\label{tab:ns1LCDM}
\end{table}

We investigate here sound horizon-independent EFTofLSS analyses with three $n_s$ priors:
an uninformative flat prior in the range [0.8, 1.2],
$10\times$ looser prior ($n_s \sim \mathcal{N}(0.9649, 0.042)$) from the \textit{Planck} result~\cite{Planck:2018vyg},
and a tighter prior ($n_s \sim \mathcal{N}(0.96, 0.02)$).
The latter two come from the priors used in~\cite{Zaborowski:2024wpo}.
The constraints on the cosmological parameters are shown in \autoref{fig:ns1LCDM} and \autoref{tab:ns1LCDM},
where we can find a flat prior can infer $H_0$ with a small bias ($0.18\sigma$ for DESI Y1 and $0.24\sigma$ for Euclid).
The imposition of Gaussian priors leads to different levels of bias.
$n_s \sim \mathcal{N}(0.9649, 0.042)$ leads to a $0.46\sigma$ bias for DESI Y1 and $0.27\sigma$ for Euclid.
$n_s \sim \mathcal{N}(0.96, 0.02)$ leads to a $0.61\sigma$ bias for DESI Y1 and $0.56\sigma$ for Euclid.

\begin{figure}
    \centering
    \includegraphics[width=0.5\linewidth]{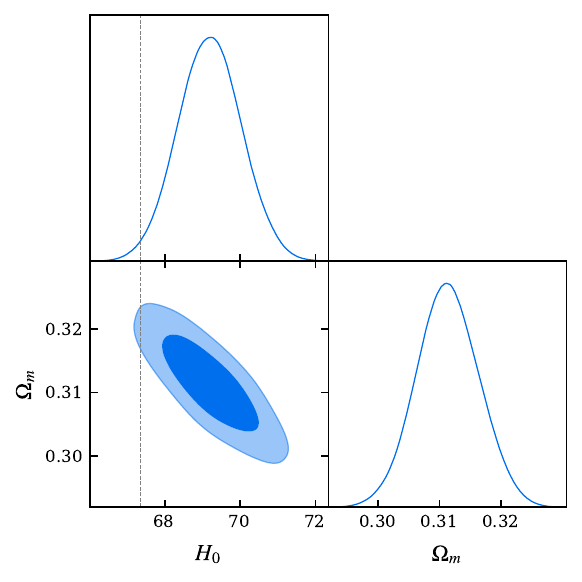}
    \caption{Sound horizon-free ShapeFit + BBN constraints ($1 \sigma$ and $2 \sigma$ confidence intervals) of an Euclid-like survey on $H_0$ and $\Omega_m$ with mock data generated from a $n_s=1$ $\Lambda$CDM model.}
    \label{fig:ns1LCDM_ShapeFit}
\end{figure}

In the ShapeFit analysis, constraints on the compressed variables $m$ and $n$ are usually strongly degenerated.
For this reason, it is common to fix $n$ (which corresponds to fixing $n_s$) to strengthen the constraints on the shape parameter $m$.
Here we investigate the impact of this choice on the sound horizon-independent $H_0$ with an Euclid-like survey.
We perform the ShapeFit analysis with $n$ and $n_s$ fixed to their fiducial value $0.9649$.
The results with the BBN $\omega_b$ prior in \autoref{fig:ns1LCDM_ShapeFit} clearly indicate that $H_0$ does not recover the value $67.36$ km/s/Mpc of the mock data.
We find that $H_0=69.19_{-0.84}^{+0.80}$ km/s/Mpc, which means that fixing $n$ and $n_s$ leads to a $2.2\sigma$ bias.

These analyses show that a strong $n_s$ prior deviating from the true value can significantly bias the results.
In the following analysis, we use an uninformative flat prior for $n_s$.

\section{Sound horizon-independent $H_0$ is not equivalent to $k_\text{eq}$-based $H_0$} \label{sec:neqkeq}

The constraining power on sound horizon-independent $H_0$ by marginalizing the BAO wiggle scale in EFTofLSS analysis is usually considered to come from the shape of $k \gtrsim k_\text{eq}$ part of the matter power spectrum.
Beyond the power-law dependence determined by the primordial perturbation, it has a logarithmic dependence, which arises from the growth during the radiation-dominant period.
Therefore this shape depends on the scale $k_\text{eq}$ in the $\Lambda$CDM model and the inferred $H_0$ is sometimes referred to as $k_\text{eq}$-based $H_0$.
In this section we address the role of other information in such $H_0$ inference in order to show that sound horizon-independent $H_0$ is not equivalent to $k_\text{eq}$-based $H_0$.

\subsection{The role of the overall amplitude}

In addition to the shape of the matter power spectrum, its overall amplitude is also related to $H_0$~\cite{Smith:2022iax}.
When analyzing galaxy clustering alone (or with a BBN prior), it provides effective information only if an informative $A_s$ prior is provided.
However, when analyzed jointly with the CMB lensing data, since the overall amplitude of the CMB lensing is also related to $H_0$ and $A_s$, but with different degenerate directions, in principle the degeneracy can be broken thus constraining $H_0$.

\begin{figure}[htb!]
    \centering
    \includegraphics[width=\linewidth]{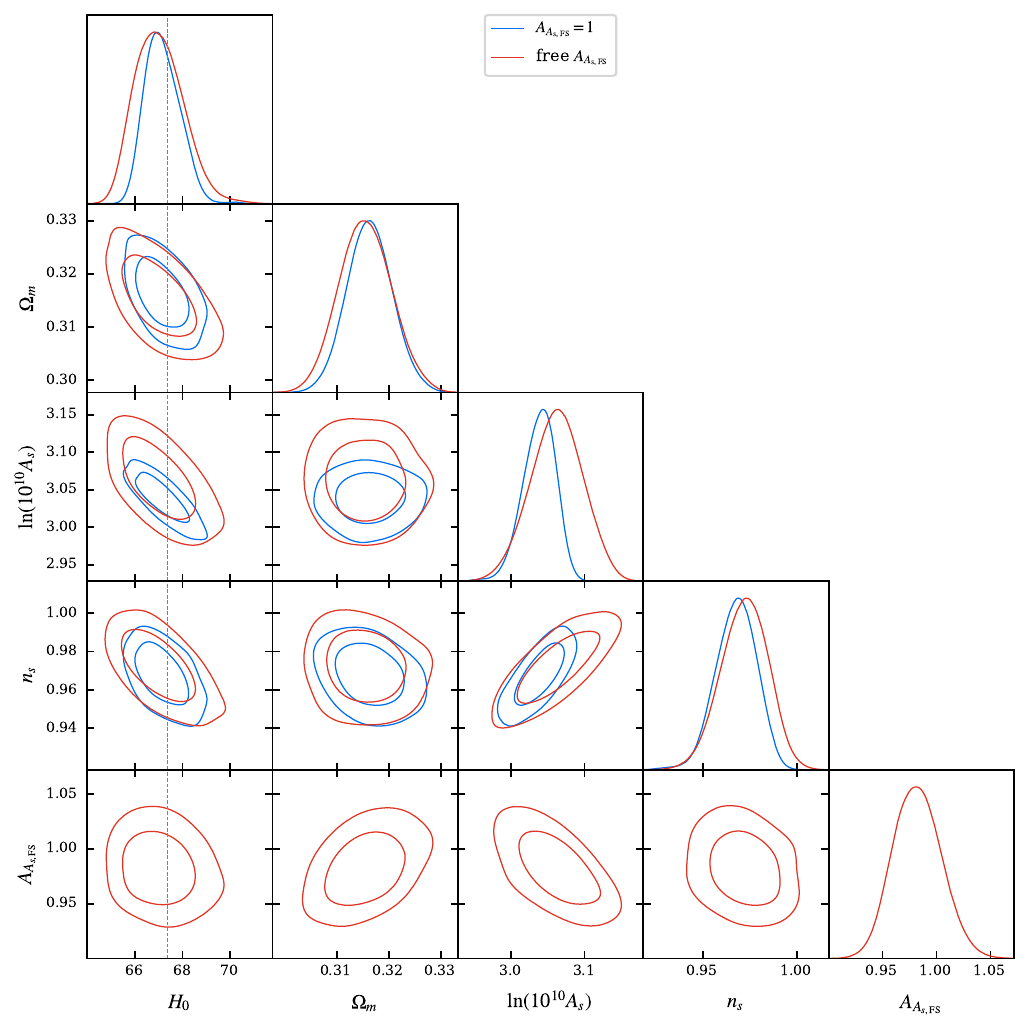}
    \caption{Sound horizon-free EFTofLSS + BBN constraints ($1 \sigma$ and $2 \sigma$ confidence intervals) of an Euclid-like survey with CMB lensing on cosmological parameters with Planck2018 mock data.}
    \label{fig:AAs}
\end{figure}

To test whether this effect provides appreciable contributions in the sound horizon-independent $H_0$, we introduce a parameter $A_{A_\text{s, FS}}$ to individually scale the $A_s$ used to compute the galaxy clustering (full shape) power spectrum:
\begin{equation}
    A_\text{s, FS} = A_{A_\text{s, FS}} A_\text{s} \, ,
\end{equation}
while keeping the $A_s$ used to compute the CMB lensing spectrum unchanged.
By allowing $A_{A_\text{s, FS}}$ to vary freely, the previously mentioned degeneracy breaking cannot occur, thus avoiding information from the overall amplitude.
In \autoref{fig:AAs}, we show constraints from a joint analysis of an Euclid-like survey, BBN and CMB lensing data with fixed $A_{A_\text{s, FS}}=1$ and freely varying $A_{A_\text{s, FS}}$.
The mock galaxy clustering data is generated assuming the Planck2018 model and the CMB lensing data is the real \textit{Planck} PR4 + ACT DR6 data.
We find that allowing a freely varying $A_{A_\text{s, FS}}$ relaxes the standard deviation of $H_0$ from $0.75$ to $1.0$ (km/s/Mpc).
Thus the constraint from the overall amplitude leads to tightening the constraint by $\sim 25\%$ for the dataset we consider here.
Future CMB observations have the potential to strengthen its constraints.

\subsection{The role of the relative amplitude of BAO wiggles}

Another potential source of information is the relative amplitude of BAO wiggles, which is controlled by $\Omega_b/\Omega_m$ in the $\Lambda$CDM model.
It can provide constraints on $H_0$ when it is combined with the $\omega_b = \Omega_bh^2$ prior from BBN and the constraint on $\Omega_m$ from the galaxy clustering itself or other cosmological probes~\cite{Krolewski:2024jwj}.

\begin{figure}[htb!]
    \centering
    \includegraphics[width=\linewidth]{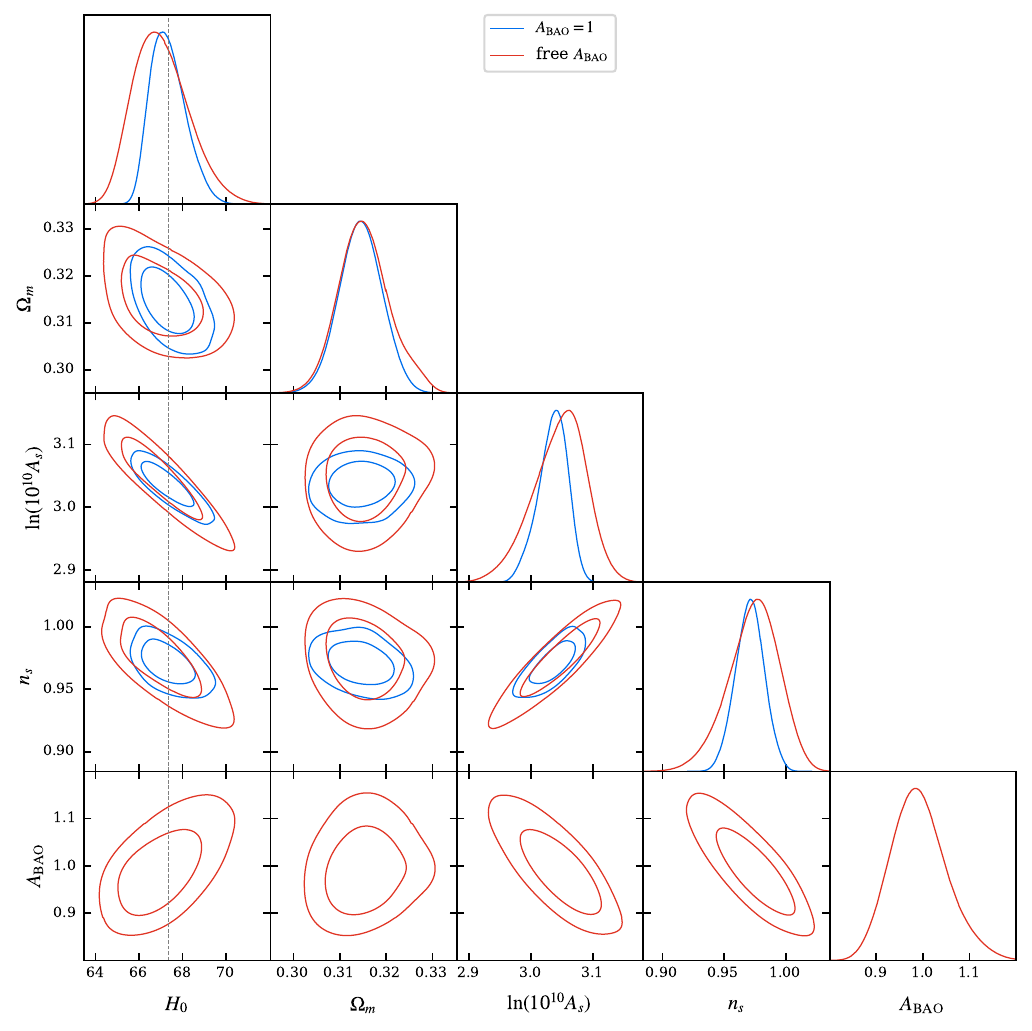}
    \caption{Sound horizon-free EFTofLSS + BBN constraints ($1 \sigma$ and $2 \sigma$ confidence intervals) of an Euclid-like survey on cosmological parameters with Planck2018 mock data.}
    \label{fig:Abao}
\end{figure}

Similar to the overall amplitude, we introduce here a parameter $A_\text{BAO}$ which scales the amplitude of the BAO wiggles:
\begin{equation}
    P_\text{lin}(k, q_\text{BAO}, A_\text{BAO}) = P_\text{lin}^\text{nw}(k) + A_\text{BAO} P_\text{lin}^\text{w}(q_\text{BAO} \cdot k) \, .
\end{equation}
Then we test its effect on the sound horizon-independent $H_0$ by a joint EFTofLSS analysis using BBN and an Eulicd-like survey assuming Planck2018 as the real cosmology.
The resulting constraints on the cosmological parameters (and $A_\text{BAO}$) are shown in \autoref{fig:Abao}.
We find that allowing a freely varying $A_\text{BAO}$ relaxes the standard deviation of $H_0$ from $0.80$ to $1.3$ (km/s/Mpc).
In other words, the information on the relative amplitudes of the BAO wiggles tightens the constraint on $H_0$ by $\sim 38\%$.

In summary, while the shape of the broadband power spectrum (which is determined by $k_\text{eq}$ in $\Lambda$CDM) dominates the constraints on the sound horizon-independent $H_0$, other information also makes non-negligible contributions to the final result.

\section{Implications of sound horizon-independent $H_0$ for early new physics in future surveys} \label{sec:Euclid}

\begin{figure}[htb!]
    \centering
\begin{subfigure}{0.49\textwidth}
\includegraphics[width=0.9\linewidth]{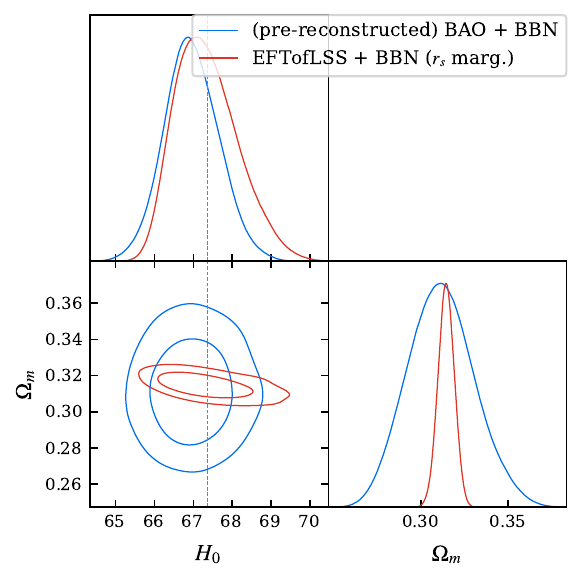}
\caption{Planck2018}
\end{subfigure}
\begin{subfigure}{0.49\textwidth}
\includegraphics[width=0.9\linewidth]{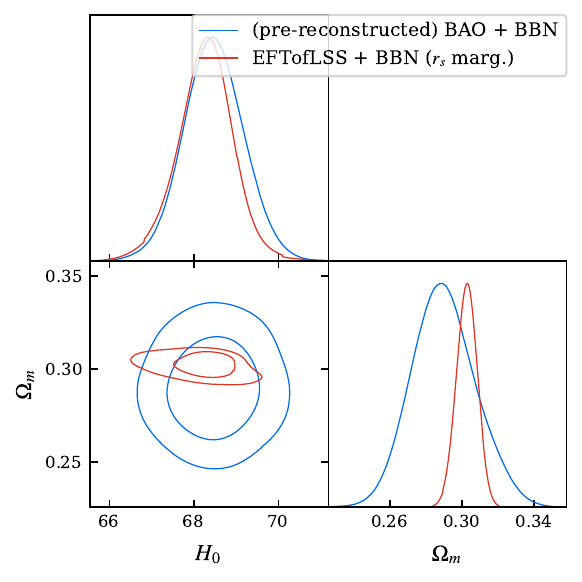}
\caption{Smith2019}
\end{subfigure}
\begin{subfigure}{0.49\textwidth}
\includegraphics[width=0.9\linewidth]{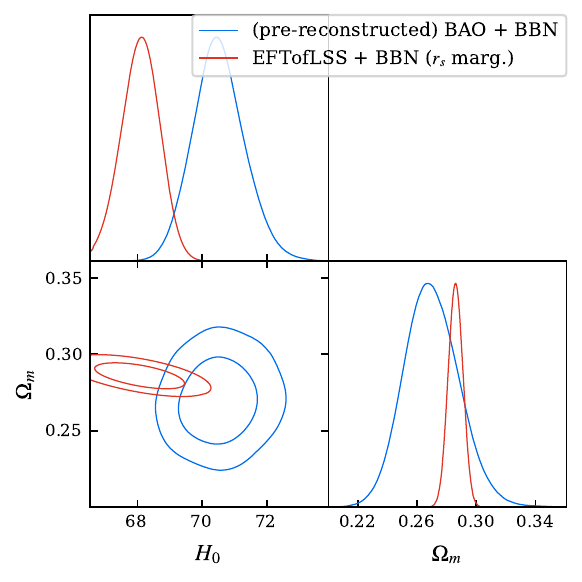}
\caption{Hill2021}
\end{subfigure}
\begin{subfigure}{0.49\textwidth}
\includegraphics[width=0.9\linewidth]{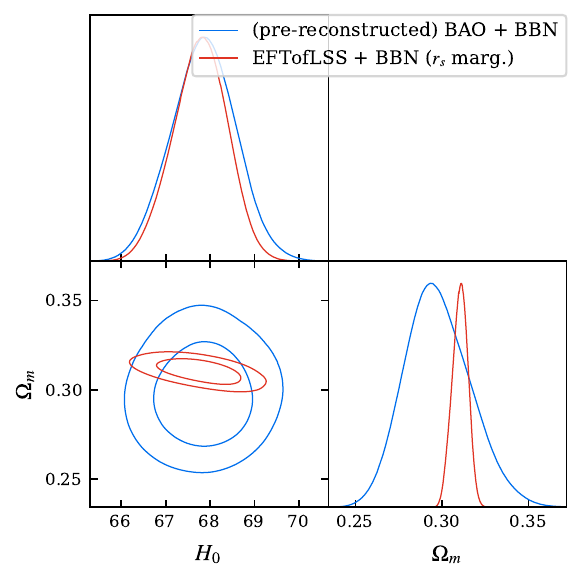}
\caption{Jiang2024}
\end{subfigure}

\caption{Constraints ($1 \sigma$ and $2 \sigma$ confidence intervals) of an Euclid-like survey on $H_0$ and $\Omega_m$ with different real cosmologies.}
\label{fig:Euclid}
\end{figure}

\begin{table}[htb!]
    \centering
    \begin{tabular}{|c|c|c|c|c|} \hline \hline
                              & Planck2018 & Smith2019 & Hill2021 & Jiang2024 \\ \hline
(pre-reconstructed) BAO + BBN & $66.95_{-0.74}^{+0.64}$ & $68.46_{-0.69}^{+0.70}$ & $70.52_{-0.69}^{+0.89}$ & $67.85_{-0.57}^{+0.87}$ \\
EFTofLSS + BBN ($r_s$ marg.) & $67.33_{-0.94}^{+0.61}$ & $68.27_{-0.42}^{+0.31}$ & $68.09_{-0.35}^{+0.31}$ & $67.77_{-0.55}^{+0.62}$ \\
\hline \hline
\end{tabular}
\caption{Constraints (mean values and $1 \sigma$ confidence intervals) of an Euclid-like survey on the sound horizon-based $H_0$ (km/s/Mpc, first row) and sound horizon-independent $H_0$ (second row) with mock data generated from a variety of cosmologies.}
\label{tab:Euclid}
\end{table}

In this section, we investigate whether future galaxy clustering observations can rule out or confirm the presence of early physics through sound horizon-independent $H_0$.
For each of the four cosmologies mentioned in \autoref{sec:DESIY1}, we generate mock galaxy clustering observations for an Euclid-like survey according to the specifications described in Appendix~\ref{sec:mockspec}.
We then perform (pre-reconstructed) BAO and EFTofLSS analyses on them respectively to obtain sound horizon-based and sound horizon-independent $H_0$ and show the constraints for $H_0$ and $\Omega_m$ in \autoref{fig:Euclid}.

Firstly, we notice that, in contrast to the results in \autoref{fig:DESIY1} for DESI Y1, all three EDE models we consider here tend to favor lower sound horizon-independent $H_0$ to different degrees.
\footnote{We verified that even considering the same flat prior on $n_s$, DESI Y1 favors a higher sound horizon-independent $H_0$.}
For the most shifted Hill2021 model, we conduct a test in Appendix~\ref{sec:weak_Euclid} with a weakened Euclid-like observation, which has a higher mean value of the sound horizon-independent $H_0$ with respect to the results in this section.
We suspect that the higher sound horizon-independent $H_0$ preferred by DESI Y1 may be influenced by the prior choice of the EFT parameters, especially given the Jeffreys prior we use here.
And when observations become more constraining, the influence of the prior diminishes.

We show the marginalized $H_0$ posterior distributions in~\autoref{tab:Euclid}.
We find that the two $H_0$ for all EDE models except Hill2021 are difficult to distinguish in an Euclid-like survey.

\begin{figure}[htb!]
    \centering
\begin{subfigure}{0.49\textwidth}
\includegraphics[width=\linewidth]{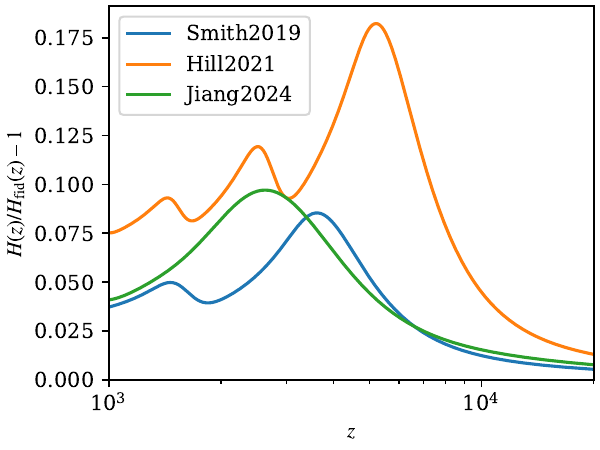}
\end{subfigure}
\begin{subfigure}{0.49\textwidth}
\includegraphics[width=\linewidth]{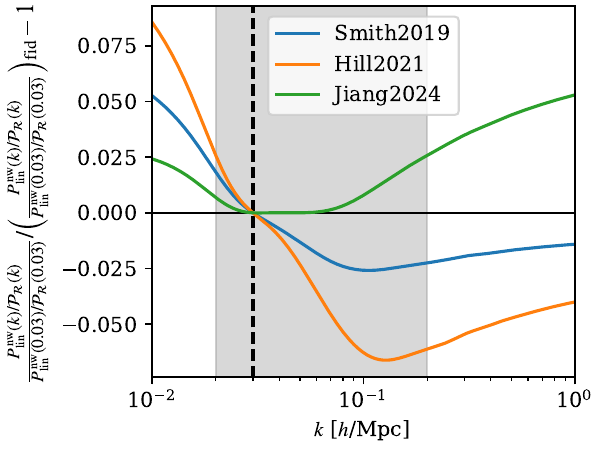}
\end{subfigure}

\caption{\textbf{Left}: The effect of different EDE models on the expansion rate $H(z)$ of the early universe.
\textbf{Right}: The effect of different EDE models on the no-wiggle component $P_\text{lin}^\text{nw}(k)$ of the linear matter power spectrum at $z=0$.
We normalize it with the primordial power spectrum $\mathcal{P}_\mathcal{R}$ and the value at $k=0.03 h$/Mpc.
The fiducial model (Planck2018) is always used as the reference model.
The shaded area indicates the scale range used for the DESI Y1 full-shape analysis.}
\label{fig:cosmos}
\end{figure}

In the case where $n_s$ can vary freely and the nuisance parameter can be constrained precisely,
the full shape analysis of the power spectrum for sound horizon-independent $H_0$ constraints is dominated by constraints from the transfer function, although there are numerous sources of information as shown in \autoref{sec:neqkeq}.
The EDE model brings a faster expansion rate in the early universe.
This causes the modes inside the horizon to experience more Hubble friction and thus be suppressed, yet other $\Lambda$CDM parameter shifts (specifically $\omega_c=\Omega_ch^2$) tend to compensate for it.
\footnote{The details of EDE at the perturbation level (e.g., the sound speed $c_s^2$) also change the matter power spectrum.
However, its contribution is minor, so we ignore it here.}
Hence the transfer function is related by the expansion history in the horizon, which is shown at the left plane of \autoref{fig:cosmos} for the EDE models we consider here.
The EDE actually opens up new parameter spaces ($f_\text{EDE}$ and $z_c$, etc.), which gives it a great degree of freedom to influence the expansion history.
Further, different EDE models can produce different behaviors, e.g. the axion-like EDE model (Smith2019 and Hill2021) produces oscillations whereas the AdS-EDE (Jiang2024) brings only one peak.
In the right plane of \autoref{fig:cosmos}, the effect of these different selections on the transfer function is shown.
We focus on the no-wiggle component of the linear matter power spectrum because it is associated with the sound horizon-independent $H_0$ that we investigate here.
We make comparisons around $k = 0.03h$/Mpc, which is similar to the choices made in the ShapeFit analysis.
Jiang2024 and Smith2019 (especially the former) provide a similar shape of the power spectrum (in the unit of $h$/Mpc) to Planck2018 over the range covered by the observations.
This means that they will have similar $k_\text{eq}^{-1}H_0$ in the context of $\Lambda$CDM.
On the other hand, a viable EDE model should have a similar value of $r_sH_0$ and $\Omega_m$ to the $\Lambda$CDM model since they both have a good fit to the BAO observation (this is not the case for Hill2021 since it was obtained by fitting only to the CMB).
\footnote{At the same time, when $q_\text{BAO}$ is scaling $k$ in the unit of $h$Mpc$^{-1}$, it is actually scaling $r_s^{-1}$ (in the unit of $h$Mpc$^{-1}$) or $(r_sH_0)^{-1}$ (in the unit of Mpc$^{-1}$) so that $q_\text{BAO}$ should also not deviate significantly from 1 when the two $H_0$ are consistent.}
As a result, these EDE models should give similar $r_dH_0$ and $\Omega_m$ in the BAO analysis for $\Lambda$CDM (as shown explicitly in \autoref{sec:exp_BAO}).
Therefore the two $H_0$ they give in $\Lambda$CDM should also be close to each other.

\section{Conclusion}

Galaxy clustering surveys can be used to extract sound horizon-independent $H_0$ through, e.g., EFTofLSS~\cite{Farren:2021grl} or ShapeFit~\cite{Brieden:2022heh} analyses.
It provides insights into Hubble tension that are different from the conventional sound horizon-based $H_0$ and is considered to have the potential to constrain the early new physics introduced for solving Hubble tension.
However, we point out in this work that there are some possible misunderstandings and misuses here.

Imposing some prior from other cosmological probes is often used to strengthen the constraints.
These priors are usually derived using the assumption of $\Lambda$CDM, which is fine for studies confined within $\Lambda$CDM.
However, it is dangerous to apply these conclusions to early new physics because these early new physics are usually accompanied by shifts of $\Lambda$CDM parameters, which means that these priors may not apply.
For example, we show that a strong $n_s$ prior introduces a non-negligible bias in the case where the prior does not match the true value.
Thus, a fair comparison should use uninformative priors or only those that are model-independent (in the context of potential new physical models).

This sound horizon-independent $H_0$ was sometimes called $k_\text{eq}$-based $H_0$ because the starting point for such methods is usually the shape of the power spectrum, which is determined by $k_\text{eq}$ in the $\Lambda$CDM model.
For the EFTofLSS method we investigate here, however, the independence of the sound horizon is obtained by marginalizing the sound horizon with a free parameter.
Thus other information may also play a role in constraining $H_0$.
We show that, at least for an Euclid-like survey, the relative amplitude of the BAO wiggle poses a non-negligible constraint on the sound horizon-independent $H_0$.
And when analyzed jointly with the CMB lensing observations, the overall amplitude of the power spectrum also contributes to the constraints on $H_0$, even if no prior for the amplitude of the primordial scalar perturbation $A_s$ is imposed.
We therefore suggest not to call this $H_0$ as $k_\text{eq}$-based $H_0$ to avoid misinterpretation.

Recently Ref.~\cite{Zaborowski:2024wpo} analyzed DESI Y1 data to obtain the tightest bound to date for the sound horizon-independent $H_0$ with EFTofLSS.
We perform a similar analysis with mock data, which shows that the DESI Y1 results do not constrain some EDE models (e.g.~\cite{Smith:2019ihp,Jiang:2024nha}) obtained by fitting observations including CMB and BAO.
However, it does put pressure on the EDE model obtained by fitting only the ACT DR4 CMB data~\cite{Hill:2021yec}.
Furthermore, we made forecasts for an Euclid-like survey.
We notice that the nuisance prior may constitute some bias in the case of model misspecification, and it will be weak in an Euclid-like survey.
Nonetheless, an Euclid-like survey still has difficulty ruling out the EDE model in~\cite{Smith:2019ihp,Jiang:2024nha}.

Early new physics like EDE typically relieve Hubble tension by opening up new parameter spaces, and there is a wide variety of them.
This makes it difficult to exclude them completely via the sound horizon-independent $H_0$.
Nevertheless, it can be used to narrow the parameter space of the new physics.
In particular, since the main constraining comes from the transfer function, which is closely related to the expansion history of the early universe, these results can also be used to heuristically construct modifications to the expansion history of the early universe.
We leave it for future study.
Finally, we would like to mention that in this paper we assumed the late universe to be $\Lambda$CDM.
However, early new physics may also need to be coupled with some modifications on the late universe to resolve tensions in cosmology (see e.g.~\cite{Vagnozzi:2023nrq} for a review) and there are some hints for deviations form $\Lambda$CDM (e.g.~\cite{Bargiacchi:2021hdp,Bargiacchi:2023rfd,DESI:2024mwx}, see also~\cite{Sakr:2025daj,Sakr:2025fay}).
These modifications may change the constraints on $H_0$ from large-scale structural observations.

\begin{acknowledgments}
YSP is supported by NSFC, No.12075246, National Key
Research and Development Program of China, No. 2021YFC2203004, and
the Fundamental Research Funds for the Central Universities.
We acknowledge the use of high-performance computing services provided by the International Centre for Theoretical Physics Asia-Pacific cluster, the Scientific Computing Center of University of Chinese Academy of Sciences, and the Tianhe-2 supercomputer.
\end{acknowledgments}

\appendix

\section{Specification of mock data} \label{sec:mockspec}

\begin{table}[htb!]
\centering
\begin{tabular}{|c|c|c|c|c|c|} \hline \hline
$z$ range & $z_\text{eff}$ & area [deg$^2$] & tracer & $N_\text{tracer}$ & $b$ \\ \hline
$[0.1, 0.4]$ & 0.295 & 7473 & BGS & 300,017 & 1.5 \\
$[0.4, 0.6]$ & 0.510 & 5740 & LRG & 506,905 & 2.0 \\
$[0.6, 0.8]$ & 0.706 & 5740 & LRG & 771,875 & 2.0 \\
$[0.8, 1.1]$ & 0.930 & 5740 & LRG & 859,824 & 2.0 \\
$[0.8, 1.1]$ & 0.930 & 5924 & ELG &1,016,340& 1.2 \\
$[1.1, 1.6]$ & 1.317 & 5924 & ELG &1,415,687& 1.2 \\
$[0.8, 2.1]$ & 1.491 & 7249 & QSO & 856,652 & 2.1 \\
\hline \hline
\end{tabular}
\caption{Specification for a DESI Y1-like survey used in our work.}
\label{tab:mockspec_DESIY1}
\end{table}

We summarize here the specifications for generating mock survey data in this work.
For the DESI Y1-like survey, it is specified by the parameters in \autoref{tab:mockspec_DESIY1}.
The area of sky occupied by different tracers is taken from Table 2 of~\cite{DESI:2024aax}.
The number of tracers $N_\text{tracer}$ counted in each redshift box was taken from Table 2 of~\cite{DESI:2024uvr} and Table 1 of~\cite{DESI:2024jis}.
The linear bias $b$ for different types of tracer is based on the values in Table 4 of~\cite{DESI:2024uvr}, which is also used by DESI to produce mock data.
We generate mock data in the scale range $k \in [0.02, 0.2] h$Mpc$^{-1}$ for full shape analysis with $\Delta k = 0.005 h$Mpc$^{-1}$.
For BAO analysis $k \in [0.02, 0.3] h$Mpc$^{-1}$ is used.
Monopoles and dipoles are used except for the first and last bins in the BAO analysis.
Besides, the first ELG bin ($z \in [0.8, 1.1]$) is not used in the analysis.

\begin{table}[htb!]
\centering
\begin{tabular}{|c|c|c|c|c|c|} \hline \hline
$z$ range & $z_\text{eff}$ & $\bar{n}$ $[\times10^-3]$ \\ \hline
$[0.5, 0.7]$ & 0.9 & 3.83 \\
$[0.7, 0.9]$ & 0.8 & 2.08 \\
$[0.9, 1.1]$ & 1.0 & 1.18 \\
$[1.1, 1.3]$ & 1.2 & 0.70 \\
$[1.3, 1.5]$ & 1.4 & 0.39 \\
$[1.5, 1.7]$ & 1.6 & 0.21 \\
$[1.7, 1.9]$ & 1.8 & 0.12 \\
$[1.9, 2.1]$ & 2.0 & 0.07 \\
\hline \hline
\end{tabular}
\caption{Specification for an Eucild-like survey used in our work.}
\label{tab:mockspec_Euclid}
\end{table}

For the Euclid-like survey, it is specified by the parameters in \autoref{tab:mockspec_Euclid}.
The specifications we used are similar to \cite{Chudaykin:2019ock,Farren:2021grl}.
The tracer is ELG with a total of about $5.5\times 10^7$~\cite{EUCLID:2011zbd,Amendola:2016saw}.
All the redshift bins cover the fraction of the sky $f_\text{sky}=0.3636$.
Multipoles with $\ell=0,2,4$ and $k \in [0.01, 1] h$Mpc$^{-1}$ are used in our analysis.

The linear bias on the basis of Eulerian perturbation theory (EPT) for these ELG tracers is assumed to be a function based on~\cite{Orsi:2009mj}:
\begin{equation}
    b_1(z) = 0.9+0.4z \, ,
\end{equation}
and $b_2$ is based on a fitting formula obtained from N-body simulations~\cite{Yankelevich:2018uaz,DiDio:2018unb}:
\begin{equation}
    b_2(z) = -0.704172 - 0.207993 z + 0.183023 z^2 - 0.00771288 z^3
\end{equation}
The bias parameter $b_s$ is based on the relation~\cite{Desjacques:2016bnm}:
\begin{equation}
    b_s = \frac{2}{7} (1-b_1) \, .
\end{equation}
Other higher-order biases are assumed to be negligible or degenerate and absorbed by other parameters.
We then convert them to the basis of Lagrangian perturbation theory (LPT):
\begin{equation}
    b_1^L = b_1 - 1 \,, \quad b_2^L = b_2 - \frac{8}{21} b_1^L \,, \quad b_s^L = b_s \, .
\end{equation}
The values of the counterterms and stochastic terms are the same as those chosen in \cite{Farren:2021grl}.
The specific values of higher-order terms are not expected to significantly change the results.

The covariance is calculated using \texttt{desilike}\footnote{\url{https://github.com/cosmodesi/desilike}}.
We account for statistical errors and treat theoretical uncertainty as negligible.

\section{Validation of BAO results} \label{sec:exp_BAO}

\begin{table}[htb!]
\centering
\begin{tabular}{|c|c|c|c|c|} \hline \hline
Model & Planck2018 & Smith2019 & Hill2021 & Jiang2024 \\ \hline
$r_sh$ & 100.60 & 101.81 & 104.98 & 100.60 \\
$\Omega_m$ & 0.32 & 0.29 & 0.27 & 0.30 \\
$H_0$ & 67.35 & 68.64 & 70.46 & 68.01 \\
\hline \hline
\end{tabular}
\caption{The expected values of parameters obtained from the BAO analysis assuming $\Lambda$CDM.}
\label{tab:exp_BAO}
\end{table}

\begin{figure}[htb!]
    \centering
    \includegraphics[width=\linewidth]{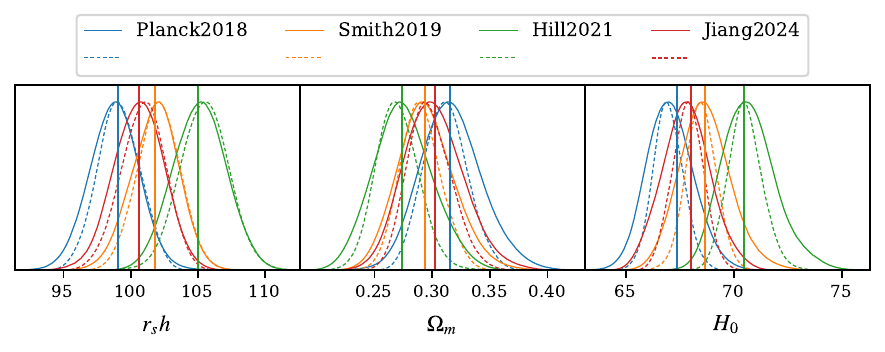}
    \caption{The posterior distribution of the parameters obtained through our BAO analysis pipeline. Solid lines indicate a DESI Y1-like survey and dashed lines indicate an Euclid-like survey. The theoretical values (\autoref{tab:exp_BAO}) of these parameters are indicated by vertical lines in the corresponding colors.}
    \label{fig:exp_BAO}
\end{figure}

In all the models we consider in this paper, the late universe is assumed to be $\Lambda$CDM, so an unbiased BAO analysis pipeline should correctly recover $r_sH_0$ and $\Omega_m$.
Furthermore, considering the mean value of the BBN prior $\omega_b = 0.02218$, we numerically search for the $H_0$ required to recover $r_sH_0$ in the $\Lambda$CDM model.
We summarize the results in \autoref{tab:exp_BAO} and compare them with the posterior distributions of our chains in \autoref{fig:exp_BAO} and find that our BAO analysis pipeline recovers the theoretically expected values well.

\section{Sound horizon-independent constraints for EFTofLSS analysis of a weakened Euclid-like survey} \label{sec:weak_Euclid}

\begin{figure}[htb!]
    \centering
    \includegraphics[width=0.5\linewidth]{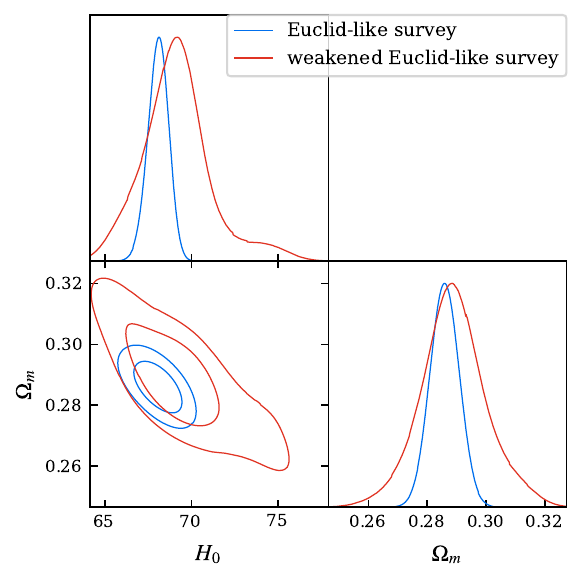}
    \caption{Constraints ($1 \sigma$ and $2 \sigma$ confidence intervals) of an Euclid-like survey and a weakened Euclid-like survey on $H_0$ and $\Omega_m$ with Hill2021 cosmology.}
    \label{fig:weak_Euclid}
\end{figure}

In \autoref{fig:weak_Euclid}, we consider a weakened Euclid-like survey with Hill2021 as the mock cosmology.
The sky area is reduced by half and $\bar{n}$ is reduced to 1/5 of their original value, which reduces the total number of tracers to $\sim 5.5\times 10^6$.
$H_0$ is moved to a higher value in this case.

\bibliography{refs}

\begin{thebibliography}{111}%
\makeatletter
\providecommand \@ifxundefined [1]{%
 \@ifx{#1\undefined}
}%
\providecommand \@ifnum [1]{%
 \ifnum #1\expandafter \@firstoftwo
 \else \expandafter \@secondoftwo
 \fi
}%
\providecommand \@ifx [1]{%
 \ifx #1\expandafter \@firstoftwo
 \else \expandafter \@secondoftwo
 \fi
}%
\providecommand \natexlab [1]{#1}%
\providecommand \enquote  [1]{``#1''}%
\providecommand \bibnamefont  [1]{#1}%
\providecommand \bibfnamefont [1]{#1}%
\providecommand \citenamefont [1]{#1}%
\providecommand \href@noop [0]{\@secondoftwo}%
\providecommand \href [0]{\begingroup \@sanitize@url \@href}%
\providecommand \@href[1]{\@@startlink{#1}\@@href}%
\providecommand \@@href[1]{\endgroup#1\@@endlink}%
\providecommand \@sanitize@url [0]{\catcode `\\12\catcode `\$12\catcode `\&12\catcode `\#12\catcode `\^12\catcode `\_12\catcode `\%12\relax}%
\providecommand \@@startlink[1]{}%
\providecommand \@@endlink[0]{}%
\providecommand \url  [0]{\begingroup\@sanitize@url \@url }%
\providecommand \@url [1]{\endgroup\@href {#1}{\urlprefix }}%
\providecommand \urlprefix  [0]{URL }%
\providecommand \Eprint [0]{\href }%
\providecommand \doibase [0]{https://doi.org/}%
\providecommand \selectlanguage [0]{\@gobble}%
\providecommand \bibinfo  [0]{\@secondoftwo}%
\providecommand \bibfield  [0]{\@secondoftwo}%
\providecommand \translation [1]{[#1]}%
\providecommand \BibitemOpen [0]{}%
\providecommand \bibitemStop [0]{}%
\providecommand \bibitemNoStop [0]{.\EOS\space}%
\providecommand \EOS [0]{\spacefactor3000\relax}%
\providecommand \BibitemShut  [1]{\csname bibitem#1\endcsname}%
\let\auto@bib@innerbib\@empty
\bibitem [{\citenamefont {Verde}\ \emph {et~al.}(2019)\citenamefont {Verde}, \citenamefont {Treu},\ and\ \citenamefont {Riess}}]{Verde:2019ivm}%
  \BibitemOpen
  \bibfield  {author} {\bibinfo {author} {\bibfnamefont {L.}~\bibnamefont {Verde}}, \bibinfo {author} {\bibfnamefont {T.}~\bibnamefont {Treu}},\ and\ \bibinfo {author} {\bibfnamefont {A.~G.}\ \bibnamefont {Riess}},\ }\bibfield  {title} {\bibinfo {title} {{Tensions between the Early and the Late Universe}},\ }\href {https://doi.org/10.1038/s41550-019-0902-0} {\bibfield  {journal} {\bibinfo  {journal} {Nature Astron.}\ }\textbf {\bibinfo {volume} {3}},\ \bibinfo {pages} {891} (\bibinfo {year} {2019})},\ \Eprint {https://arxiv.org/abs/1907.10625} {arXiv:1907.10625 [astro-ph.CO]} \BibitemShut {NoStop}%
\bibitem [{\citenamefont {Di~Valentino}\ \emph {et~al.}(2021)\citenamefont {Di~Valentino}, \citenamefont {Mena}, \citenamefont {Pan}, \citenamefont {Visinelli}, \citenamefont {Yang}, \citenamefont {Melchiorri}, \citenamefont {Mota}, \citenamefont {Riess},\ and\ \citenamefont {Silk}}]{DiValentino:2021izs}%
  \BibitemOpen
  \bibfield  {author} {\bibinfo {author} {\bibfnamefont {E.}~\bibnamefont {Di~Valentino}}, \bibinfo {author} {\bibfnamefont {O.}~\bibnamefont {Mena}}, \bibinfo {author} {\bibfnamefont {S.}~\bibnamefont {Pan}}, \bibinfo {author} {\bibfnamefont {L.}~\bibnamefont {Visinelli}}, \bibinfo {author} {\bibfnamefont {W.}~\bibnamefont {Yang}}, \bibinfo {author} {\bibfnamefont {A.}~\bibnamefont {Melchiorri}}, \bibinfo {author} {\bibfnamefont {D.~F.}\ \bibnamefont {Mota}}, \bibinfo {author} {\bibfnamefont {A.~G.}\ \bibnamefont {Riess}},\ and\ \bibinfo {author} {\bibfnamefont {J.}~\bibnamefont {Silk}},\ }\bibfield  {title} {\bibinfo {title} {{In the realm of the Hubble tension\textemdash{}a review of solutions}},\ }\href {https://doi.org/10.1088/1361-6382/ac086d} {\bibfield  {journal} {\bibinfo  {journal} {Class. Quant. Grav.}\ }\textbf {\bibinfo {volume} {38}},\ \bibinfo {pages} {153001} (\bibinfo {year} {2021})},\ \Eprint {https://arxiv.org/abs/2103.01183} {arXiv:2103.01183 [astro-ph.CO]} \BibitemShut {NoStop}%
\bibitem [{\citenamefont {Perivolaropoulos}\ and\ \citenamefont {Skara}(2022)}]{Perivolaropoulos:2021jda}%
  \BibitemOpen
  \bibfield  {author} {\bibinfo {author} {\bibfnamefont {L.}~\bibnamefont {Perivolaropoulos}}\ and\ \bibinfo {author} {\bibfnamefont {F.}~\bibnamefont {Skara}},\ }\bibfield  {title} {\bibinfo {title} {{Challenges for \ensuremath{\Lambda}CDM: An update}},\ }\href {https://doi.org/10.1016/j.newar.2022.101659} {\bibfield  {journal} {\bibinfo  {journal} {New Astron. Rev.}\ }\textbf {\bibinfo {volume} {95}},\ \bibinfo {pages} {101659} (\bibinfo {year} {2022})},\ \Eprint {https://arxiv.org/abs/2105.05208} {arXiv:2105.05208 [astro-ph.CO]} \BibitemShut {NoStop}%
\bibitem [{\citenamefont {Sch\"oneberg}\ \emph {et~al.}(2022)\citenamefont {Sch\"oneberg}, \citenamefont {Franco~Abell\'an}, \citenamefont {P\'erez~S\'anchez}, \citenamefont {Witte}, \citenamefont {Poulin},\ and\ \citenamefont {Lesgourgues}}]{Schoneberg:2021qvd}%
  \BibitemOpen
  \bibfield  {author} {\bibinfo {author} {\bibfnamefont {N.}~\bibnamefont {Sch\"oneberg}}, \bibinfo {author} {\bibfnamefont {G.}~\bibnamefont {Franco~Abell\'an}}, \bibinfo {author} {\bibfnamefont {A.}~\bibnamefont {P\'erez~S\'anchez}}, \bibinfo {author} {\bibfnamefont {S.~J.}\ \bibnamefont {Witte}}, \bibinfo {author} {\bibfnamefont {V.}~\bibnamefont {Poulin}},\ and\ \bibinfo {author} {\bibfnamefont {J.}~\bibnamefont {Lesgourgues}},\ }\bibfield  {title} {\bibinfo {title} {{The H0 Olympics: A fair ranking of proposed models}},\ }\href {https://doi.org/10.1016/j.physrep.2022.07.001} {\bibfield  {journal} {\bibinfo  {journal} {Phys. Rept.}\ }\textbf {\bibinfo {volume} {984}},\ \bibinfo {pages} {1} (\bibinfo {year} {2022})},\ \Eprint {https://arxiv.org/abs/2107.10291} {arXiv:2107.10291 [astro-ph.CO]} \BibitemShut {NoStop}%
\bibitem [{\citenamefont {Shah}\ \emph {et~al.}(2021)\citenamefont {Shah}, \citenamefont {Lemos},\ and\ \citenamefont {Lahav}}]{Shah:2021onj}%
  \BibitemOpen
  \bibfield  {author} {\bibinfo {author} {\bibfnamefont {P.}~\bibnamefont {Shah}}, \bibinfo {author} {\bibfnamefont {P.}~\bibnamefont {Lemos}},\ and\ \bibinfo {author} {\bibfnamefont {O.}~\bibnamefont {Lahav}},\ }\bibfield  {title} {\bibinfo {title} {{A buyer\textquoteright{}s guide to the Hubble constant}},\ }\href {https://doi.org/10.1007/s00159-021-00137-4} {\bibfield  {journal} {\bibinfo  {journal} {Astron. Astrophys. Rev.}\ }\textbf {\bibinfo {volume} {29}},\ \bibinfo {pages} {9} (\bibinfo {year} {2021})},\ \Eprint {https://arxiv.org/abs/2109.01161} {arXiv:2109.01161 [astro-ph.CO]} \BibitemShut {NoStop}%
\bibitem [{\citenamefont {Abdalla}\ \emph {et~al.}(2022)\citenamefont {Abdalla} \emph {et~al.}}]{Abdalla:2022yfr}%
  \BibitemOpen
  \bibfield  {author} {\bibinfo {author} {\bibfnamefont {E.}~\bibnamefont {Abdalla}} \emph {et~al.},\ }\bibfield  {title} {\bibinfo {title} {{Cosmology intertwined: A review of the particle physics, astrophysics, and cosmology associated with the cosmological tensions and anomalies}},\ }\href {https://doi.org/10.1016/j.jheap.2022.04.002} {\bibfield  {journal} {\bibinfo  {journal} {JHEAp}\ }\textbf {\bibinfo {volume} {34}},\ \bibinfo {pages} {49} (\bibinfo {year} {2022})},\ \Eprint {https://arxiv.org/abs/2203.06142} {arXiv:2203.06142 [astro-ph.CO]} \BibitemShut {NoStop}%
\bibitem [{\citenamefont {Di~Valentino}(2022)}]{DiValentino:2022fjm}%
  \BibitemOpen
  \bibfield  {author} {\bibinfo {author} {\bibfnamefont {E.}~\bibnamefont {Di~Valentino}},\ }\bibfield  {title} {\bibinfo {title} {{Challenges of the Standard Cosmological Model}},\ }\href {https://doi.org/10.3390/universe8080399} {\bibfield  {journal} {\bibinfo  {journal} {Universe}\ }\textbf {\bibinfo {volume} {8}},\ \bibinfo {pages} {399} (\bibinfo {year} {2022})}\BibitemShut {NoStop}%
\bibitem [{\citenamefont {Hu}\ and\ \citenamefont {Wang}(2023)}]{Hu:2023jqc}%
  \BibitemOpen
  \bibfield  {author} {\bibinfo {author} {\bibfnamefont {J.-P.}\ \bibnamefont {Hu}}\ and\ \bibinfo {author} {\bibfnamefont {F.-Y.}\ \bibnamefont {Wang}},\ }\bibfield  {title} {\bibinfo {title} {{Hubble Tension: The Evidence of New Physics}},\ }\href {https://doi.org/10.3390/universe9020094} {\bibfield  {journal} {\bibinfo  {journal} {Universe}\ }\textbf {\bibinfo {volume} {9}},\ \bibinfo {pages} {94} (\bibinfo {year} {2023})},\ \Eprint {https://arxiv.org/abs/2302.05709} {arXiv:2302.05709 [astro-ph.CO]} \BibitemShut {NoStop}%
\bibitem [{\citenamefont {Verde}\ \emph {et~al.}(2024)\citenamefont {Verde}, \citenamefont {Sch\"oneberg},\ and\ \citenamefont {Gil-Mar\'\i{}n}}]{Verde:2023lmm}%
  \BibitemOpen
  \bibfield  {author} {\bibinfo {author} {\bibfnamefont {L.}~\bibnamefont {Verde}}, \bibinfo {author} {\bibfnamefont {N.}~\bibnamefont {Sch\"oneberg}},\ and\ \bibinfo {author} {\bibfnamefont {H.}~\bibnamefont {Gil-Mar\'\i{}n}},\ }\bibfield  {title} {\bibinfo {title} {{A Tale of Many H0}},\ }\href {https://doi.org/10.1146/annurev-astro-052622-033813} {\bibfield  {journal} {\bibinfo  {journal} {Ann. Rev. Astron. Astrophys.}\ }\textbf {\bibinfo {volume} {62}},\ \bibinfo {pages} {287} (\bibinfo {year} {2024})},\ \Eprint {https://arxiv.org/abs/2311.13305} {arXiv:2311.13305 [astro-ph.CO]} \BibitemShut {NoStop}%
\bibitem [{\citenamefont {Aghanim}\ \emph {et~al.}(2020)\citenamefont {Aghanim} \emph {et~al.}}]{Planck:2018vyg}%
  \BibitemOpen
  \bibfield  {author} {\bibinfo {author} {\bibfnamefont {N.}~\bibnamefont {Aghanim}} \emph {et~al.} (\bibinfo {collaboration} {Planck}),\ }\bibfield  {title} {\bibinfo {title} {{Planck 2018 results. VI. Cosmological parameters}},\ }\href {https://doi.org/10.1051/0004-6361/201833910} {\bibfield  {journal} {\bibinfo  {journal} {Astron. Astrophys.}\ }\textbf {\bibinfo {volume} {641}},\ \bibinfo {pages} {A6} (\bibinfo {year} {2020})},\ \bibinfo {note} {[Erratum: Astron.Astrophys. 652, C4 (2021)]},\ \Eprint {https://arxiv.org/abs/1807.06209} {arXiv:1807.06209 [astro-ph.CO]} \BibitemShut {NoStop}%
\bibitem [{\citenamefont {Riess}\ \emph {et~al.}(2022)\citenamefont {Riess} \emph {et~al.}}]{Riess:2021jrx}%
  \BibitemOpen
  \bibfield  {author} {\bibinfo {author} {\bibfnamefont {A.~G.}\ \bibnamefont {Riess}} \emph {et~al.},\ }\bibfield  {title} {\bibinfo {title} {{A Comprehensive Measurement of the Local Value of the Hubble Constant with 1 km s$^{−1}$ Mpc$^{−1}$ Uncertainty from the Hubble Space Telescope and the SH0ES Team}},\ }\href {https://doi.org/10.3847/2041-8213/ac5c5b} {\bibfield  {journal} {\bibinfo  {journal} {Astrophys. J. Lett.}\ }\textbf {\bibinfo {volume} {934}},\ \bibinfo {pages} {L7} (\bibinfo {year} {2022})},\ \Eprint {https://arxiv.org/abs/2112.04510} {arXiv:2112.04510 [astro-ph.CO]} \BibitemShut {NoStop}%
\bibitem [{\citenamefont {Bernal}\ \emph {et~al.}(2016)\citenamefont {Bernal}, \citenamefont {Verde},\ and\ \citenamefont {Riess}}]{Bernal:2016gxb}%
  \BibitemOpen
  \bibfield  {author} {\bibinfo {author} {\bibfnamefont {J.~L.}\ \bibnamefont {Bernal}}, \bibinfo {author} {\bibfnamefont {L.}~\bibnamefont {Verde}},\ and\ \bibinfo {author} {\bibfnamefont {A.~G.}\ \bibnamefont {Riess}},\ }\bibfield  {title} {\bibinfo {title} {{The trouble with $H_0$}},\ }\href {https://doi.org/10.1088/1475-7516/2016/10/019} {\bibfield  {journal} {\bibinfo  {journal} {JCAP}\ }\textbf {\bibinfo {volume} {10}},\ \bibinfo {pages} {019}},\ \Eprint {https://arxiv.org/abs/1607.05617} {arXiv:1607.05617 [astro-ph.CO]} \BibitemShut {NoStop}%
\bibitem [{\citenamefont {Addison}\ \emph {et~al.}(2018)\citenamefont {Addison}, \citenamefont {Watts}, \citenamefont {Bennett}, \citenamefont {Halpern}, \citenamefont {Hinshaw},\ and\ \citenamefont {Weiland}}]{Addison:2017fdm}%
  \BibitemOpen
  \bibfield  {author} {\bibinfo {author} {\bibfnamefont {G.~E.}\ \bibnamefont {Addison}}, \bibinfo {author} {\bibfnamefont {D.~J.}\ \bibnamefont {Watts}}, \bibinfo {author} {\bibfnamefont {C.~L.}\ \bibnamefont {Bennett}}, \bibinfo {author} {\bibfnamefont {M.}~\bibnamefont {Halpern}}, \bibinfo {author} {\bibfnamefont {G.}~\bibnamefont {Hinshaw}},\ and\ \bibinfo {author} {\bibfnamefont {J.~L.}\ \bibnamefont {Weiland}},\ }\bibfield  {title} {\bibinfo {title} {{Elucidating $\Lambda$CDM: Impact of Baryon Acoustic Oscillation Measurements on the Hubble Constant Discrepancy}},\ }\href {https://doi.org/10.3847/1538-4357/aaa1ed} {\bibfield  {journal} {\bibinfo  {journal} {Astrophys. J.}\ }\textbf {\bibinfo {volume} {853}},\ \bibinfo {pages} {119} (\bibinfo {year} {2018})},\ \Eprint {https://arxiv.org/abs/1707.06547} {arXiv:1707.06547 [astro-ph.CO]} \BibitemShut {NoStop}%
\bibitem [{\citenamefont {Lemos}\ \emph {et~al.}(2019)\citenamefont {Lemos}, \citenamefont {Lee}, \citenamefont {Efstathiou},\ and\ \citenamefont {Gratton}}]{Lemos:2018smw}%
  \BibitemOpen
  \bibfield  {author} {\bibinfo {author} {\bibfnamefont {P.}~\bibnamefont {Lemos}}, \bibinfo {author} {\bibfnamefont {E.}~\bibnamefont {Lee}}, \bibinfo {author} {\bibfnamefont {G.}~\bibnamefont {Efstathiou}},\ and\ \bibinfo {author} {\bibfnamefont {S.}~\bibnamefont {Gratton}},\ }\bibfield  {title} {\bibinfo {title} {{Model independent $H(z)$ reconstruction using the cosmic inverse distance ladder}},\ }\href {https://doi.org/10.1093/mnras/sty3082} {\bibfield  {journal} {\bibinfo  {journal} {Mon. Not. Roy. Astron. Soc.}\ }\textbf {\bibinfo {volume} {483}},\ \bibinfo {pages} {4803} (\bibinfo {year} {2019})},\ \Eprint {https://arxiv.org/abs/1806.06781} {arXiv:1806.06781 [astro-ph.CO]} \BibitemShut {NoStop}%
\bibitem [{\citenamefont {Aylor}\ \emph {et~al.}(2019)\citenamefont {Aylor}, \citenamefont {Joy}, \citenamefont {Knox}, \citenamefont {Millea}, \citenamefont {Raghunathan},\ and\ \citenamefont {Wu}}]{Aylor:2018drw}%
  \BibitemOpen
  \bibfield  {author} {\bibinfo {author} {\bibfnamefont {K.}~\bibnamefont {Aylor}}, \bibinfo {author} {\bibfnamefont {M.}~\bibnamefont {Joy}}, \bibinfo {author} {\bibfnamefont {L.}~\bibnamefont {Knox}}, \bibinfo {author} {\bibfnamefont {M.}~\bibnamefont {Millea}}, \bibinfo {author} {\bibfnamefont {S.}~\bibnamefont {Raghunathan}},\ and\ \bibinfo {author} {\bibfnamefont {W.~L.~K.}\ \bibnamefont {Wu}},\ }\bibfield  {title} {\bibinfo {title} {{Sounds Discordant: Classical Distance Ladder \& $\Lambda$CDM -based Determinations of the Cosmological Sound Horizon}},\ }\href {https://doi.org/10.3847/1538-4357/ab0898} {\bibfield  {journal} {\bibinfo  {journal} {Astrophys. J.}\ }\textbf {\bibinfo {volume} {874}},\ \bibinfo {pages} {4} (\bibinfo {year} {2019})},\ \Eprint {https://arxiv.org/abs/1811.00537} {arXiv:1811.00537 [astro-ph.CO]} \BibitemShut {NoStop}%
\bibitem [{\citenamefont {Sch\"oneberg}\ \emph {et~al.}(2019)\citenamefont {Sch\"oneberg}, \citenamefont {Lesgourgues},\ and\ \citenamefont {Hooper}}]{Schoneberg:2019wmt}%
  \BibitemOpen
  \bibfield  {author} {\bibinfo {author} {\bibfnamefont {N.}~\bibnamefont {Sch\"oneberg}}, \bibinfo {author} {\bibfnamefont {J.}~\bibnamefont {Lesgourgues}},\ and\ \bibinfo {author} {\bibfnamefont {D.~C.}\ \bibnamefont {Hooper}},\ }\bibfield  {title} {\bibinfo {title} {{The BAO+BBN take on the Hubble tension}},\ }\href {https://doi.org/10.1088/1475-7516/2019/10/029} {\bibfield  {journal} {\bibinfo  {journal} {JCAP}\ }\textbf {\bibinfo {volume} {10}},\ \bibinfo {pages} {029}},\ \Eprint {https://arxiv.org/abs/1907.11594} {arXiv:1907.11594 [astro-ph.CO]} \BibitemShut {NoStop}%
\bibitem [{\citenamefont {Knox}\ and\ \citenamefont {Millea}(2020)}]{Knox:2019rjx}%
  \BibitemOpen
  \bibfield  {author} {\bibinfo {author} {\bibfnamefont {L.}~\bibnamefont {Knox}}\ and\ \bibinfo {author} {\bibfnamefont {M.}~\bibnamefont {Millea}},\ }\bibfield  {title} {\bibinfo {title} {{Hubble constant hunter\textquoteright{}s guide}},\ }\href {https://doi.org/10.1103/PhysRevD.101.043533} {\bibfield  {journal} {\bibinfo  {journal} {Phys. Rev. D}\ }\textbf {\bibinfo {volume} {101}},\ \bibinfo {pages} {043533} (\bibinfo {year} {2020})},\ \Eprint {https://arxiv.org/abs/1908.03663} {arXiv:1908.03663 [astro-ph.CO]} \BibitemShut {NoStop}%
\bibitem [{\citenamefont {Arendse}\ \emph {et~al.}(2020)\citenamefont {Arendse} \emph {et~al.}}]{Arendse:2019hev}%
  \BibitemOpen
  \bibfield  {author} {\bibinfo {author} {\bibfnamefont {N.}~\bibnamefont {Arendse}} \emph {et~al.},\ }\bibfield  {title} {\bibinfo {title} {{Cosmic dissonance: are new physics or systematics behind a short sound horizon?}},\ }\href {https://doi.org/10.1051/0004-6361/201936720} {\bibfield  {journal} {\bibinfo  {journal} {Astron. Astrophys.}\ }\textbf {\bibinfo {volume} {639}},\ \bibinfo {pages} {A57} (\bibinfo {year} {2020})},\ \Eprint {https://arxiv.org/abs/1909.07986} {arXiv:1909.07986 [astro-ph.CO]} \BibitemShut {NoStop}%
\bibitem [{\citenamefont {Efstathiou}(2021)}]{Efstathiou:2021ocp}%
  \BibitemOpen
  \bibfield  {author} {\bibinfo {author} {\bibfnamefont {G.}~\bibnamefont {Efstathiou}},\ }\bibfield  {title} {\bibinfo {title} {{To H0 or not to H0?}},\ }\href {https://doi.org/10.1093/mnras/stab1588} {\bibfield  {journal} {\bibinfo  {journal} {Mon. Not. Roy. Astron. Soc.}\ }\textbf {\bibinfo {volume} {505}},\ \bibinfo {pages} {3866} (\bibinfo {year} {2021})},\ \Eprint {https://arxiv.org/abs/2103.08723} {arXiv:2103.08723 [astro-ph.CO]} \BibitemShut {NoStop}%
\bibitem [{\citenamefont {Krishnan}\ \emph {et~al.}(2021)\citenamefont {Krishnan}, \citenamefont {Mohayaee}, \citenamefont {Colg\'ain}, \citenamefont {Sheikh-Jabbari},\ and\ \citenamefont {Yin}}]{Krishnan:2021dyb}%
  \BibitemOpen
  \bibfield  {author} {\bibinfo {author} {\bibfnamefont {C.}~\bibnamefont {Krishnan}}, \bibinfo {author} {\bibfnamefont {R.}~\bibnamefont {Mohayaee}}, \bibinfo {author} {\bibfnamefont {E.~O.}\ \bibnamefont {Colg\'ain}}, \bibinfo {author} {\bibfnamefont {M.~M.}\ \bibnamefont {Sheikh-Jabbari}},\ and\ \bibinfo {author} {\bibfnamefont {L.}~\bibnamefont {Yin}},\ }\bibfield  {title} {\bibinfo {title} {{Does Hubble tension signal a breakdown in FLRW cosmology?}},\ }\href {https://doi.org/10.1088/1361-6382/ac1a81} {\bibfield  {journal} {\bibinfo  {journal} {Class. Quant. Grav.}\ }\textbf {\bibinfo {volume} {38}},\ \bibinfo {pages} {184001} (\bibinfo {year} {2021})},\ \Eprint {https://arxiv.org/abs/2105.09790} {arXiv:2105.09790 [astro-ph.CO]} \BibitemShut {NoStop}%
\bibitem [{\citenamefont {Cai}\ \emph {et~al.}(2022)\citenamefont {Cai}, \citenamefont {Guo}, \citenamefont {Wang}, \citenamefont {Yu},\ and\ \citenamefont {Zhou}}]{Cai:2021weh}%
  \BibitemOpen
  \bibfield  {author} {\bibinfo {author} {\bibfnamefont {R.-G.}\ \bibnamefont {Cai}}, \bibinfo {author} {\bibfnamefont {Z.-K.}\ \bibnamefont {Guo}}, \bibinfo {author} {\bibfnamefont {S.-J.}\ \bibnamefont {Wang}}, \bibinfo {author} {\bibfnamefont {W.-W.}\ \bibnamefont {Yu}},\ and\ \bibinfo {author} {\bibfnamefont {Y.}~\bibnamefont {Zhou}},\ }\bibfield  {title} {\bibinfo {title} {{No-go guide for the Hubble tension: Late-time solutions}},\ }\href {https://doi.org/10.1103/PhysRevD.105.L021301} {\bibfield  {journal} {\bibinfo  {journal} {Phys. Rev. D}\ }\textbf {\bibinfo {volume} {105}},\ \bibinfo {pages} {L021301} (\bibinfo {year} {2022})},\ \Eprint {https://arxiv.org/abs/2107.13286} {arXiv:2107.13286 [astro-ph.CO]} \BibitemShut {NoStop}%
\bibitem [{\citenamefont {Keeley}\ and\ \citenamefont {Shafieloo}(2023)}]{Keeley:2022ojz}%
  \BibitemOpen
  \bibfield  {author} {\bibinfo {author} {\bibfnamefont {R.~E.}\ \bibnamefont {Keeley}}\ and\ \bibinfo {author} {\bibfnamefont {A.}~\bibnamefont {Shafieloo}},\ }\bibfield  {title} {\bibinfo {title} {{Ruling Out New Physics at Low Redshift as a Solution to the H0 Tension}},\ }\href {https://doi.org/10.1103/PhysRevLett.131.111002} {\bibfield  {journal} {\bibinfo  {journal} {Phys. Rev. Lett.}\ }\textbf {\bibinfo {volume} {131}},\ \bibinfo {pages} {111002} (\bibinfo {year} {2023})},\ \Eprint {https://arxiv.org/abs/2206.08440} {arXiv:2206.08440 [astro-ph.CO]} \BibitemShut {NoStop}%
\bibitem [{\citenamefont {G\'omez-Valent}\ \emph {et~al.}(2024)\citenamefont {G\'omez-Valent}, \citenamefont {Favale}, \citenamefont {Migliaccio},\ and\ \citenamefont {Sen}}]{Gomez-Valent:2023uof}%
  \BibitemOpen
  \bibfield  {author} {\bibinfo {author} {\bibfnamefont {A.}~\bibnamefont {G\'omez-Valent}}, \bibinfo {author} {\bibfnamefont {A.}~\bibnamefont {Favale}}, \bibinfo {author} {\bibfnamefont {M.}~\bibnamefont {Migliaccio}},\ and\ \bibinfo {author} {\bibfnamefont {A.~A.}\ \bibnamefont {Sen}},\ }\bibfield  {title} {\bibinfo {title} {{Late-time phenomenology required to solve the H0 tension in view of the cosmic ladders and the anisotropic and angular BAO datasets}},\ }\href {https://doi.org/10.1103/PhysRevD.109.023525} {\bibfield  {journal} {\bibinfo  {journal} {Phys. Rev. D}\ }\textbf {\bibinfo {volume} {109}},\ \bibinfo {pages} {023525} (\bibinfo {year} {2024})},\ \Eprint {https://arxiv.org/abs/2309.07795} {arXiv:2309.07795 [astro-ph.CO]} \BibitemShut {NoStop}%
\bibitem [{\citenamefont {Jiang}\ \emph {et~al.}(2024{\natexlab{a}})\citenamefont {Jiang}, \citenamefont {Pedrotti}, \citenamefont {da~Costa},\ and\ \citenamefont {Vagnozzi}}]{Jiang:2024xnu}%
  \BibitemOpen
  \bibfield  {author} {\bibinfo {author} {\bibfnamefont {J.-Q.}\ \bibnamefont {Jiang}}, \bibinfo {author} {\bibfnamefont {D.}~\bibnamefont {Pedrotti}}, \bibinfo {author} {\bibfnamefont {S.~S.}\ \bibnamefont {da~Costa}},\ and\ \bibinfo {author} {\bibfnamefont {S.}~\bibnamefont {Vagnozzi}},\ }\bibfield  {title} {\bibinfo {title} {{Nonparametric late-time expansion history reconstruction and implications for the Hubble tension in light of recent DESI and type Ia supernovae data}},\ }\href {https://doi.org/10.1103/PhysRevD.110.123519} {\bibfield  {journal} {\bibinfo  {journal} {Phys. Rev. D}\ }\textbf {\bibinfo {volume} {110}},\ \bibinfo {pages} {123519} (\bibinfo {year} {2024}{\natexlab{a}})},\ \Eprint {https://arxiv.org/abs/2408.02365} {arXiv:2408.02365 [astro-ph.CO]} \BibitemShut {NoStop}%
\bibitem [{\citenamefont {Roy~Choudhury}\ and\ \citenamefont {Okumura}(2024)}]{RoyChoudhury:2024wri}%
  \BibitemOpen
  \bibfield  {author} {\bibinfo {author} {\bibfnamefont {S.}~\bibnamefont {Roy~Choudhury}}\ and\ \bibinfo {author} {\bibfnamefont {T.}~\bibnamefont {Okumura}},\ }\bibfield  {title} {\bibinfo {title} {{Updated Cosmological Constraints in Extended Parameter Space with Planck PR4, DESI Baryon Acoustic Oscillations, and Supernovae: Dynamical Dark Energy, Neutrino Masses, Lensing Anomaly, and the Hubble Tension}},\ }\href {https://doi.org/10.3847/2041-8213/ad8c26} {\bibfield  {journal} {\bibinfo  {journal} {Astrophys. J. Lett.}\ }\textbf {\bibinfo {volume} {976}},\ \bibinfo {pages} {L11} (\bibinfo {year} {2024})},\ \Eprint {https://arxiv.org/abs/2409.13022} {arXiv:2409.13022 [astro-ph.CO]} \BibitemShut {NoStop}%
\bibitem [{\citenamefont {Roy~Choudhury}\ \emph {et~al.}(2021)\citenamefont {Roy~Choudhury}, \citenamefont {Hannestad},\ and\ \citenamefont {Tram}}]{RoyChoudhury:2020dmd}%
  \BibitemOpen
  \bibfield  {author} {\bibinfo {author} {\bibfnamefont {S.}~\bibnamefont {Roy~Choudhury}}, \bibinfo {author} {\bibfnamefont {S.}~\bibnamefont {Hannestad}},\ and\ \bibinfo {author} {\bibfnamefont {T.}~\bibnamefont {Tram}},\ }\bibfield  {title} {\bibinfo {title} {{Updated constraints on massive neutrino self-interactions from cosmology in light of the $H_0$ tension}},\ }\href {https://doi.org/10.1088/1475-7516/2021/03/084} {\bibfield  {journal} {\bibinfo  {journal} {JCAP}\ }\textbf {\bibinfo {volume} {03}},\ \bibinfo {pages} {084}},\ \Eprint {https://arxiv.org/abs/2012.07519} {arXiv:2012.07519 [astro-ph.CO]} \BibitemShut {NoStop}%
\bibitem [{\citenamefont {Karwal}\ and\ \citenamefont {Kamionkowski}(2016)}]{Karwal:2016vyq}%
  \BibitemOpen
  \bibfield  {author} {\bibinfo {author} {\bibfnamefont {T.}~\bibnamefont {Karwal}}\ and\ \bibinfo {author} {\bibfnamefont {M.}~\bibnamefont {Kamionkowski}},\ }\bibfield  {title} {\bibinfo {title} {{Dark energy at early times, the Hubble parameter, and the string axiverse}},\ }\href {https://doi.org/10.1103/PhysRevD.94.103523} {\bibfield  {journal} {\bibinfo  {journal} {Phys. Rev. D}\ }\textbf {\bibinfo {volume} {94}},\ \bibinfo {pages} {103523} (\bibinfo {year} {2016})},\ \Eprint {https://arxiv.org/abs/1608.01309} {arXiv:1608.01309 [astro-ph.CO]} \BibitemShut {NoStop}%
\bibitem [{\citenamefont {Poulin}\ \emph {et~al.}(2019)\citenamefont {Poulin}, \citenamefont {Smith}, \citenamefont {Karwal},\ and\ \citenamefont {Kamionkowski}}]{Poulin:2018cxd}%
  \BibitemOpen
  \bibfield  {author} {\bibinfo {author} {\bibfnamefont {V.}~\bibnamefont {Poulin}}, \bibinfo {author} {\bibfnamefont {T.~L.}\ \bibnamefont {Smith}}, \bibinfo {author} {\bibfnamefont {T.}~\bibnamefont {Karwal}},\ and\ \bibinfo {author} {\bibfnamefont {M.}~\bibnamefont {Kamionkowski}},\ }\bibfield  {title} {\bibinfo {title} {{Early Dark Energy Can Resolve The Hubble Tension}},\ }\href {https://doi.org/10.1103/PhysRevLett.122.221301} {\bibfield  {journal} {\bibinfo  {journal} {Phys. Rev. Lett.}\ }\textbf {\bibinfo {volume} {122}},\ \bibinfo {pages} {221301} (\bibinfo {year} {2019})},\ \Eprint {https://arxiv.org/abs/1811.04083} {arXiv:1811.04083 [astro-ph.CO]} \BibitemShut {NoStop}%
\bibitem [{\citenamefont {Kaloper}(2019)}]{Kaloper:2019lpl}%
  \BibitemOpen
  \bibfield  {author} {\bibinfo {author} {\bibfnamefont {N.}~\bibnamefont {Kaloper}},\ }\bibfield  {title} {\bibinfo {title} {{Dark energy, $H_0$ and weak gravity conjecture}},\ }\href {https://doi.org/10.1142/S0218271819440176} {\bibfield  {journal} {\bibinfo  {journal} {Int. J. Mod. Phys. D}\ }\textbf {\bibinfo {volume} {28}},\ \bibinfo {pages} {1944017} (\bibinfo {year} {2019})},\ \Eprint {https://arxiv.org/abs/1903.11676} {arXiv:1903.11676 [hep-th]} \BibitemShut {NoStop}%
\bibitem [{\citenamefont {Agrawal}\ \emph {et~al.}(2023)\citenamefont {Agrawal}, \citenamefont {Cyr-Racine}, \citenamefont {Pinner},\ and\ \citenamefont {Randall}}]{Agrawal:2019lmo}%
  \BibitemOpen
  \bibfield  {author} {\bibinfo {author} {\bibfnamefont {P.}~\bibnamefont {Agrawal}}, \bibinfo {author} {\bibfnamefont {F.-Y.}\ \bibnamefont {Cyr-Racine}}, \bibinfo {author} {\bibfnamefont {D.}~\bibnamefont {Pinner}},\ and\ \bibinfo {author} {\bibfnamefont {L.}~\bibnamefont {Randall}},\ }\bibfield  {title} {\bibinfo {title} {{Rock \textquoteleft{}n\textquoteright{} roll solutions to the Hubble tension}},\ }\href {https://doi.org/10.1016/j.dark.2023.101347} {\bibfield  {journal} {\bibinfo  {journal} {Phys. Dark Univ.}\ }\textbf {\bibinfo {volume} {42}},\ \bibinfo {pages} {101347} (\bibinfo {year} {2023})},\ \Eprint {https://arxiv.org/abs/1904.01016} {arXiv:1904.01016 [astro-ph.CO]} \BibitemShut {NoStop}%
\bibitem [{\citenamefont {Lin}\ \emph {et~al.}(2019)\citenamefont {Lin}, \citenamefont {Benevento}, \citenamefont {Hu},\ and\ \citenamefont {Raveri}}]{Lin:2019qug}%
  \BibitemOpen
  \bibfield  {author} {\bibinfo {author} {\bibfnamefont {M.-X.}\ \bibnamefont {Lin}}, \bibinfo {author} {\bibfnamefont {G.}~\bibnamefont {Benevento}}, \bibinfo {author} {\bibfnamefont {W.}~\bibnamefont {Hu}},\ and\ \bibinfo {author} {\bibfnamefont {M.}~\bibnamefont {Raveri}},\ }\bibfield  {title} {\bibinfo {title} {{Acoustic Dark Energy: Potential Conversion of the Hubble Tension}},\ }\href {https://doi.org/10.1103/PhysRevD.100.063542} {\bibfield  {journal} {\bibinfo  {journal} {Phys. Rev. D}\ }\textbf {\bibinfo {volume} {100}},\ \bibinfo {pages} {063542} (\bibinfo {year} {2019})},\ \Eprint {https://arxiv.org/abs/1905.12618} {arXiv:1905.12618 [astro-ph.CO]} \BibitemShut {NoStop}%
\bibitem [{\citenamefont {Smith}\ \emph {et~al.}(2020)\citenamefont {Smith}, \citenamefont {Poulin},\ and\ \citenamefont {Amin}}]{Smith:2019ihp}%
  \BibitemOpen
  \bibfield  {author} {\bibinfo {author} {\bibfnamefont {T.~L.}\ \bibnamefont {Smith}}, \bibinfo {author} {\bibfnamefont {V.}~\bibnamefont {Poulin}},\ and\ \bibinfo {author} {\bibfnamefont {M.~A.}\ \bibnamefont {Amin}},\ }\bibfield  {title} {\bibinfo {title} {{Oscillating scalar fields and the Hubble tension: a resolution with novel signatures}},\ }\href {https://doi.org/10.1103/PhysRevD.101.063523} {\bibfield  {journal} {\bibinfo  {journal} {Phys. Rev. D}\ }\textbf {\bibinfo {volume} {101}},\ \bibinfo {pages} {063523} (\bibinfo {year} {2020})},\ \Eprint {https://arxiv.org/abs/1908.06995} {arXiv:1908.06995 [astro-ph.CO]} \BibitemShut {NoStop}%
\bibitem [{\citenamefont {Niedermann}\ and\ \citenamefont {Sloth}(2021)}]{Niedermann:2019olb}%
  \BibitemOpen
  \bibfield  {author} {\bibinfo {author} {\bibfnamefont {F.}~\bibnamefont {Niedermann}}\ and\ \bibinfo {author} {\bibfnamefont {M.~S.}\ \bibnamefont {Sloth}},\ }\bibfield  {title} {\bibinfo {title} {{New early dark energy}},\ }\href {https://doi.org/10.1103/PhysRevD.103.L041303} {\bibfield  {journal} {\bibinfo  {journal} {Phys. Rev. D}\ }\textbf {\bibinfo {volume} {103}},\ \bibinfo {pages} {L041303} (\bibinfo {year} {2021})},\ \Eprint {https://arxiv.org/abs/1910.10739} {arXiv:1910.10739 [astro-ph.CO]} \BibitemShut {NoStop}%
\bibitem [{\citenamefont {Sakstein}\ and\ \citenamefont {Trodden}(2020)}]{Sakstein:2019fmf}%
  \BibitemOpen
  \bibfield  {author} {\bibinfo {author} {\bibfnamefont {J.}~\bibnamefont {Sakstein}}\ and\ \bibinfo {author} {\bibfnamefont {M.}~\bibnamefont {Trodden}},\ }\bibfield  {title} {\bibinfo {title} {{Early Dark Energy from Massive Neutrinos as a Natural Resolution of the Hubble Tension}},\ }\href {https://doi.org/10.1103/PhysRevLett.124.161301} {\bibfield  {journal} {\bibinfo  {journal} {Phys. Rev. Lett.}\ }\textbf {\bibinfo {volume} {124}},\ \bibinfo {pages} {161301} (\bibinfo {year} {2020})},\ \Eprint {https://arxiv.org/abs/1911.11760} {arXiv:1911.11760 [astro-ph.CO]} \BibitemShut {NoStop}%
\bibitem [{\citenamefont {Ye}\ and\ \citenamefont {Piao}(2020{\natexlab{a}})}]{Ye:2020btb}%
  \BibitemOpen
  \bibfield  {author} {\bibinfo {author} {\bibfnamefont {G.}~\bibnamefont {Ye}}\ and\ \bibinfo {author} {\bibfnamefont {Y.-S.}\ \bibnamefont {Piao}},\ }\bibfield  {title} {\bibinfo {title} {{Is the Hubble tension a hint of AdS phase around recombination?}},\ }\href {https://doi.org/10.1103/PhysRevD.101.083507} {\bibfield  {journal} {\bibinfo  {journal} {Phys. Rev. D}\ }\textbf {\bibinfo {volume} {101}},\ \bibinfo {pages} {083507} (\bibinfo {year} {2020}{\natexlab{a}})},\ \Eprint {https://arxiv.org/abs/2001.02451} {arXiv:2001.02451 [astro-ph.CO]} \BibitemShut {NoStop}%
\bibitem [{\citenamefont {Gogoi}\ \emph {et~al.}(2021)\citenamefont {Gogoi}, \citenamefont {Sharma}, \citenamefont {Chanda},\ and\ \citenamefont {Das}}]{Gogoi:2020qif}%
  \BibitemOpen
  \bibfield  {author} {\bibinfo {author} {\bibfnamefont {A.}~\bibnamefont {Gogoi}}, \bibinfo {author} {\bibfnamefont {R.~K.}\ \bibnamefont {Sharma}}, \bibinfo {author} {\bibfnamefont {P.}~\bibnamefont {Chanda}},\ and\ \bibinfo {author} {\bibfnamefont {S.}~\bibnamefont {Das}},\ }\bibfield  {title} {\bibinfo {title} {{Early Mass-varying Neutrino Dark Energy: Nugget Formation and Hubble Anomaly}},\ }\href {https://doi.org/10.3847/1538-4357/abfe5b} {\bibfield  {journal} {\bibinfo  {journal} {Astrophys. J.}\ }\textbf {\bibinfo {volume} {915}},\ \bibinfo {pages} {132} (\bibinfo {year} {2021})},\ \Eprint {https://arxiv.org/abs/2005.11889} {arXiv:2005.11889 [astro-ph.CO]} \BibitemShut {NoStop}%
\bibitem [{\citenamefont {Braglia}\ \emph {et~al.}(2020)\citenamefont {Braglia}, \citenamefont {Emond}, \citenamefont {Finelli}, \citenamefont {Gumrukcuoglu},\ and\ \citenamefont {Koyama}}]{Braglia:2020bym}%
  \BibitemOpen
  \bibfield  {author} {\bibinfo {author} {\bibfnamefont {M.}~\bibnamefont {Braglia}}, \bibinfo {author} {\bibfnamefont {W.~T.}\ \bibnamefont {Emond}}, \bibinfo {author} {\bibfnamefont {F.}~\bibnamefont {Finelli}}, \bibinfo {author} {\bibfnamefont {A.~E.}\ \bibnamefont {Gumrukcuoglu}},\ and\ \bibinfo {author} {\bibfnamefont {K.}~\bibnamefont {Koyama}},\ }\bibfield  {title} {\bibinfo {title} {{Unified framework for early dark energy from $\alpha$-attractors}},\ }\href {https://doi.org/10.1103/PhysRevD.102.083513} {\bibfield  {journal} {\bibinfo  {journal} {Phys. Rev. D}\ }\textbf {\bibinfo {volume} {102}},\ \bibinfo {pages} {083513} (\bibinfo {year} {2020})},\ \Eprint {https://arxiv.org/abs/2005.14053} {arXiv:2005.14053 [astro-ph.CO]} \BibitemShut {NoStop}%
\bibitem [{\citenamefont {Ye}\ and\ \citenamefont {Piao}(2020{\natexlab{b}})}]{Ye:2020oix}%
  \BibitemOpen
  \bibfield  {author} {\bibinfo {author} {\bibfnamefont {G.}~\bibnamefont {Ye}}\ and\ \bibinfo {author} {\bibfnamefont {Y.-S.}\ \bibnamefont {Piao}},\ }\bibfield  {title} {\bibinfo {title} {{$T_0$ censorship of early dark energy and AdS vacua}},\ }\href {https://doi.org/10.1103/PhysRevD.102.083523} {\bibfield  {journal} {\bibinfo  {journal} {Phys. Rev. D}\ }\textbf {\bibinfo {volume} {102}},\ \bibinfo {pages} {083523} (\bibinfo {year} {2020}{\natexlab{b}})},\ \Eprint {https://arxiv.org/abs/2008.10832} {arXiv:2008.10832 [astro-ph.CO]} \BibitemShut {NoStop}%
\bibitem [{\citenamefont {Lin}\ \emph {et~al.}(2020)\citenamefont {Lin}, \citenamefont {Hu},\ and\ \citenamefont {Raveri}}]{Lin:2020jcb}%
  \BibitemOpen
  \bibfield  {author} {\bibinfo {author} {\bibfnamefont {M.-X.}\ \bibnamefont {Lin}}, \bibinfo {author} {\bibfnamefont {W.}~\bibnamefont {Hu}},\ and\ \bibinfo {author} {\bibfnamefont {M.}~\bibnamefont {Raveri}},\ }\bibfield  {title} {\bibinfo {title} {{Testing $H_0$ in Acoustic Dark Energy with Planck and ACT Polarization}},\ }\href {https://doi.org/10.1103/PhysRevD.102.123523} {\bibfield  {journal} {\bibinfo  {journal} {Phys. Rev. D}\ }\textbf {\bibinfo {volume} {102}},\ \bibinfo {pages} {123523} (\bibinfo {year} {2020})},\ \Eprint {https://arxiv.org/abs/2009.08974} {arXiv:2009.08974 [astro-ph.CO]} \BibitemShut {NoStop}%
\bibitem [{\citenamefont {Odintsov}\ \emph {et~al.}(2021)\citenamefont {Odintsov}, \citenamefont {S\'aez-Chill\'on~G\'omez},\ and\ \citenamefont {Sharov}}]{Odintsov:2020qzd}%
  \BibitemOpen
  \bibfield  {author} {\bibinfo {author} {\bibfnamefont {S.~D.}\ \bibnamefont {Odintsov}}, \bibinfo {author} {\bibfnamefont {D.}~\bibnamefont {S\'aez-Chill\'on~G\'omez}},\ and\ \bibinfo {author} {\bibfnamefont {G.~S.}\ \bibnamefont {Sharov}},\ }\bibfield  {title} {\bibinfo {title} {{Analyzing the $H_0$ tension in $F(R)$ gravity models}},\ }\href {https://doi.org/10.1016/j.nuclphysb.2021.115377} {\bibfield  {journal} {\bibinfo  {journal} {Nucl. Phys. B}\ }\textbf {\bibinfo {volume} {966}},\ \bibinfo {pages} {115377} (\bibinfo {year} {2021})},\ \Eprint {https://arxiv.org/abs/2011.03957} {arXiv:2011.03957 [gr-qc]} \BibitemShut {NoStop}%
\bibitem [{\citenamefont {Seto}\ and\ \citenamefont {Toda}(2021)}]{Seto:2021xua}%
  \BibitemOpen
  \bibfield  {author} {\bibinfo {author} {\bibfnamefont {O.}~\bibnamefont {Seto}}\ and\ \bibinfo {author} {\bibfnamefont {Y.}~\bibnamefont {Toda}},\ }\bibfield  {title} {\bibinfo {title} {{Comparing early dark energy and extra radiation solutions to the Hubble tension with BBN}},\ }\href {https://doi.org/10.1103/PhysRevD.103.123501} {\bibfield  {journal} {\bibinfo  {journal} {Phys. Rev. D}\ }\textbf {\bibinfo {volume} {103}},\ \bibinfo {pages} {123501} (\bibinfo {year} {2021})},\ \Eprint {https://arxiv.org/abs/2101.03740} {arXiv:2101.03740 [astro-ph.CO]} \BibitemShut {NoStop}%
\bibitem [{\citenamefont {Nojiri}\ \emph {et~al.}(2021)\citenamefont {Nojiri}, \citenamefont {Odintsov}, \citenamefont {Saez-Chillon~Gomez},\ and\ \citenamefont {Sharov}}]{Nojiri:2021dze}%
  \BibitemOpen
  \bibfield  {author} {\bibinfo {author} {\bibfnamefont {S.}~\bibnamefont {Nojiri}}, \bibinfo {author} {\bibfnamefont {S.~D.}\ \bibnamefont {Odintsov}}, \bibinfo {author} {\bibfnamefont {D.}~\bibnamefont {Saez-Chillon~Gomez}},\ and\ \bibinfo {author} {\bibfnamefont {G.~S.}\ \bibnamefont {Sharov}},\ }\bibfield  {title} {\bibinfo {title} {{Modeling and testing the equation of state for (Early) dark energy}},\ }\href {https://doi.org/10.1016/j.dark.2021.100837} {\bibfield  {journal} {\bibinfo  {journal} {Phys. Dark Univ.}\ }\textbf {\bibinfo {volume} {32}},\ \bibinfo {pages} {100837} (\bibinfo {year} {2021})},\ \Eprint {https://arxiv.org/abs/2103.05304} {arXiv:2103.05304 [gr-qc]} \BibitemShut {NoStop}%
\bibitem [{\citenamefont {Karwal}\ \emph {et~al.}(2022)\citenamefont {Karwal}, \citenamefont {Raveri}, \citenamefont {Jain}, \citenamefont {Khoury},\ and\ \citenamefont {Trodden}}]{Karwal:2021vpk}%
  \BibitemOpen
  \bibfield  {author} {\bibinfo {author} {\bibfnamefont {T.}~\bibnamefont {Karwal}}, \bibinfo {author} {\bibfnamefont {M.}~\bibnamefont {Raveri}}, \bibinfo {author} {\bibfnamefont {B.}~\bibnamefont {Jain}}, \bibinfo {author} {\bibfnamefont {J.}~\bibnamefont {Khoury}},\ and\ \bibinfo {author} {\bibfnamefont {M.}~\bibnamefont {Trodden}},\ }\bibfield  {title} {\bibinfo {title} {{Chameleon early dark energy and the Hubble tension}},\ }\href {https://doi.org/10.1103/PhysRevD.105.063535} {\bibfield  {journal} {\bibinfo  {journal} {Phys. Rev. D}\ }\textbf {\bibinfo {volume} {105}},\ \bibinfo {pages} {063535} (\bibinfo {year} {2022})},\ \Eprint {https://arxiv.org/abs/2106.13290} {arXiv:2106.13290 [astro-ph.CO]} \BibitemShut {NoStop}%
\bibitem [{\citenamefont {Jiang}\ and\ \citenamefont {Piao}(2021)}]{Jiang:2021bab}%
  \BibitemOpen
  \bibfield  {author} {\bibinfo {author} {\bibfnamefont {J.-Q.}\ \bibnamefont {Jiang}}\ and\ \bibinfo {author} {\bibfnamefont {Y.-S.}\ \bibnamefont {Piao}},\ }\bibfield  {title} {\bibinfo {title} {{Testing AdS early dark energy with Planck, SPTpol, and LSS data}},\ }\href {https://doi.org/10.1103/PhysRevD.104.103524} {\bibfield  {journal} {\bibinfo  {journal} {Phys. Rev. D}\ }\textbf {\bibinfo {volume} {104}},\ \bibinfo {pages} {103524} (\bibinfo {year} {2021})},\ \Eprint {https://arxiv.org/abs/2107.07128} {arXiv:2107.07128 [astro-ph.CO]} \BibitemShut {NoStop}%
\bibitem [{\citenamefont {Ye}\ \emph {et~al.}(2023)\citenamefont {Ye}, \citenamefont {Zhang},\ and\ \citenamefont {Piao}}]{Ye:2021iwa}%
  \BibitemOpen
  \bibfield  {author} {\bibinfo {author} {\bibfnamefont {G.}~\bibnamefont {Ye}}, \bibinfo {author} {\bibfnamefont {J.}~\bibnamefont {Zhang}},\ and\ \bibinfo {author} {\bibfnamefont {Y.-S.}\ \bibnamefont {Piao}},\ }\bibfield  {title} {\bibinfo {title} {{Alleviating both H0 and S8 tensions: Early dark energy lifts the CMB-lockdown on ultralight axion}},\ }\href {https://doi.org/10.1016/j.physletb.2023.137770} {\bibfield  {journal} {\bibinfo  {journal} {Phys. Lett. B}\ }\textbf {\bibinfo {volume} {839}},\ \bibinfo {pages} {137770} (\bibinfo {year} {2023})},\ \Eprint {https://arxiv.org/abs/2107.13391} {arXiv:2107.13391 [astro-ph.CO]} \BibitemShut {NoStop}%
\bibitem [{\citenamefont {Wang}\ and\ \citenamefont {Piao}(2022)}]{Wang:2022jpo}%
  \BibitemOpen
  \bibfield  {author} {\bibinfo {author} {\bibfnamefont {H.}~\bibnamefont {Wang}}\ and\ \bibinfo {author} {\bibfnamefont {Y.-S.}\ \bibnamefont {Piao}},\ }\bibfield  {title} {\bibinfo {title} {{Testing dark energy after pre-recombination early dark energy}},\ }\href {https://doi.org/10.1016/j.physletb.2022.137244} {\bibfield  {journal} {\bibinfo  {journal} {Phys. Lett. B}\ }\textbf {\bibinfo {volume} {832}},\ \bibinfo {pages} {137244} (\bibinfo {year} {2022})},\ \Eprint {https://arxiv.org/abs/2201.07079} {arXiv:2201.07079 [astro-ph.CO]} \BibitemShut {NoStop}%
\bibitem [{\citenamefont {Rezazadeh}\ \emph {et~al.}(2024)\citenamefont {Rezazadeh}, \citenamefont {Ashoorioon},\ and\ \citenamefont {Grin}}]{Rezazadeh:2022lsf}%
  \BibitemOpen
  \bibfield  {author} {\bibinfo {author} {\bibfnamefont {K.}~\bibnamefont {Rezazadeh}}, \bibinfo {author} {\bibfnamefont {A.}~\bibnamefont {Ashoorioon}},\ and\ \bibinfo {author} {\bibfnamefont {D.}~\bibnamefont {Grin}},\ }\bibfield  {title} {\bibinfo {title} {{Cascading Dark Energy}},\ }\href {https://doi.org/10.3847/1538-4357/ad7b16} {\bibfield  {journal} {\bibinfo  {journal} {Astrophys. J.}\ }\textbf {\bibinfo {volume} {975}},\ \bibinfo {pages} {137} (\bibinfo {year} {2024})},\ \Eprint {https://arxiv.org/abs/2208.07631} {arXiv:2208.07631 [astro-ph.CO]} \BibitemShut {NoStop}%
\bibitem [{\citenamefont {Poulin}\ \emph {et~al.}(2023)\citenamefont {Poulin}, \citenamefont {Smith},\ and\ \citenamefont {Karwal}}]{Poulin:2023lkg}%
  \BibitemOpen
  \bibfield  {author} {\bibinfo {author} {\bibfnamefont {V.}~\bibnamefont {Poulin}}, \bibinfo {author} {\bibfnamefont {T.~L.}\ \bibnamefont {Smith}},\ and\ \bibinfo {author} {\bibfnamefont {T.}~\bibnamefont {Karwal}},\ }\bibfield  {title} {\bibinfo {title} {{The Ups and Downs of Early Dark Energy solutions to the Hubble tension: A review of models, hints and constraints circa 2023}},\ }\href {https://doi.org/10.1016/j.dark.2023.101348} {\bibfield  {journal} {\bibinfo  {journal} {Phys. Dark Univ.}\ }\textbf {\bibinfo {volume} {42}},\ \bibinfo {pages} {101348} (\bibinfo {year} {2023})},\ \Eprint {https://arxiv.org/abs/2302.09032} {arXiv:2302.09032 [astro-ph.CO]} \BibitemShut {NoStop}%
\bibitem [{\citenamefont {Odintsov}\ \emph {et~al.}(2023)\citenamefont {Odintsov}, \citenamefont {Oikonomou},\ and\ \citenamefont {Sharov}}]{Odintsov:2023cli}%
  \BibitemOpen
  \bibfield  {author} {\bibinfo {author} {\bibfnamefont {S.~D.}\ \bibnamefont {Odintsov}}, \bibinfo {author} {\bibfnamefont {V.~K.}\ \bibnamefont {Oikonomou}},\ and\ \bibinfo {author} {\bibfnamefont {G.~S.}\ \bibnamefont {Sharov}},\ }\bibfield  {title} {\bibinfo {title} {{Early dark energy with power-law F(R) gravity}},\ }\href {https://doi.org/10.1016/j.physletb.2023.137988} {\bibfield  {journal} {\bibinfo  {journal} {Phys. Lett. B}\ }\textbf {\bibinfo {volume} {843}},\ \bibinfo {pages} {137988} (\bibinfo {year} {2023})},\ \Eprint {https://arxiv.org/abs/2305.17513} {arXiv:2305.17513 [gr-qc]} \BibitemShut {NoStop}%
\bibitem [{\citenamefont {Baxter}\ and\ \citenamefont {Sherwin}(2021)}]{Baxter:2020qlr}%
  \BibitemOpen
  \bibfield  {author} {\bibinfo {author} {\bibfnamefont {E.~J.}\ \bibnamefont {Baxter}}\ and\ \bibinfo {author} {\bibfnamefont {B.~D.}\ \bibnamefont {Sherwin}},\ }\bibfield  {title} {\bibinfo {title} {{Determining the Hubble Constant without the Sound Horizon Scale: Measurements from CMB Lensing}},\ }\href {https://doi.org/10.1093/mnras/staa3706} {\bibfield  {journal} {\bibinfo  {journal} {Mon. Not. Roy. Astron. Soc.}\ }\textbf {\bibinfo {volume} {501}},\ \bibinfo {pages} {1823} (\bibinfo {year} {2021})},\ \Eprint {https://arxiv.org/abs/2007.04007} {arXiv:2007.04007 [astro-ph.CO]} \BibitemShut {NoStop}%
\bibitem [{\citenamefont {Philcox}\ \emph {et~al.}(2021)\citenamefont {Philcox}, \citenamefont {Sherwin}, \citenamefont {Farren},\ and\ \citenamefont {Baxter}}]{Philcox:2020xbv}%
  \BibitemOpen
  \bibfield  {author} {\bibinfo {author} {\bibfnamefont {O.~H.~E.}\ \bibnamefont {Philcox}}, \bibinfo {author} {\bibfnamefont {B.~D.}\ \bibnamefont {Sherwin}}, \bibinfo {author} {\bibfnamefont {G.~S.}\ \bibnamefont {Farren}},\ and\ \bibinfo {author} {\bibfnamefont {E.~J.}\ \bibnamefont {Baxter}},\ }\bibfield  {title} {\bibinfo {title} {{Determining the Hubble Constant without the Sound Horizon: Measurements from Galaxy Surveys}},\ }\href {https://doi.org/10.1103/PhysRevD.103.023538} {\bibfield  {journal} {\bibinfo  {journal} {Phys. Rev. D}\ }\textbf {\bibinfo {volume} {103}},\ \bibinfo {pages} {023538} (\bibinfo {year} {2021})},\ \Eprint {https://arxiv.org/abs/2008.08084} {arXiv:2008.08084 [astro-ph.CO]} \BibitemShut {NoStop}%
\bibitem [{\citenamefont {Farren}\ \emph {et~al.}(2022)\citenamefont {Farren}, \citenamefont {Philcox},\ and\ \citenamefont {Sherwin}}]{Farren:2021grl}%
  \BibitemOpen
  \bibfield  {author} {\bibinfo {author} {\bibfnamefont {G.~S.}\ \bibnamefont {Farren}}, \bibinfo {author} {\bibfnamefont {O.~H.~E.}\ \bibnamefont {Philcox}},\ and\ \bibinfo {author} {\bibfnamefont {B.~D.}\ \bibnamefont {Sherwin}},\ }\bibfield  {title} {\bibinfo {title} {{Determining the Hubble constant without the sound horizon: Perspectives with future galaxy surveys}},\ }\href {https://doi.org/10.1103/PhysRevD.105.063503} {\bibfield  {journal} {\bibinfo  {journal} {Phys. Rev. D}\ }\textbf {\bibinfo {volume} {105}},\ \bibinfo {pages} {063503} (\bibinfo {year} {2022})},\ \Eprint {https://arxiv.org/abs/2112.10749} {arXiv:2112.10749 [astro-ph.CO]} \BibitemShut {NoStop}%
\bibitem [{\citenamefont {Brieden}\ \emph {et~al.}(2021)\citenamefont {Brieden}, \citenamefont {Gil-Mar\'\i{}n},\ and\ \citenamefont {Verde}}]{Brieden:2021edu}%
  \BibitemOpen
  \bibfield  {author} {\bibinfo {author} {\bibfnamefont {S.}~\bibnamefont {Brieden}}, \bibinfo {author} {\bibfnamefont {H.}~\bibnamefont {Gil-Mar\'\i{}n}},\ and\ \bibinfo {author} {\bibfnamefont {L.}~\bibnamefont {Verde}},\ }\bibfield  {title} {\bibinfo {title} {{ShapeFit: extracting the power spectrum shape information in galaxy surveys beyond BAO and RSD}},\ }\href {https://doi.org/10.1088/1475-7516/2021/12/054} {\bibfield  {journal} {\bibinfo  {journal} {JCAP}\ }\textbf {\bibinfo {volume} {12}}\bibfield  {number} {\bibinfo  {number} { (12)},\ \bibinfo {pages} {054}},\ }\Eprint {https://arxiv.org/abs/2106.07641} {arXiv:2106.07641 [astro-ph.CO]} \BibitemShut {NoStop}%
\bibitem [{\citenamefont {Brieden}\ \emph {et~al.}(2023)\citenamefont {Brieden}, \citenamefont {Gil-Mar\'\i{}n},\ and\ \citenamefont {Verde}}]{Brieden:2022heh}%
  \BibitemOpen
  \bibfield  {author} {\bibinfo {author} {\bibfnamefont {S.}~\bibnamefont {Brieden}}, \bibinfo {author} {\bibfnamefont {H.}~\bibnamefont {Gil-Mar\'\i{}n}},\ and\ \bibinfo {author} {\bibfnamefont {L.}~\bibnamefont {Verde}},\ }\bibfield  {title} {\bibinfo {title} {{A tale of two (or more) h's}},\ }\href {https://doi.org/10.1088/1475-7516/2023/04/023} {\bibfield  {journal} {\bibinfo  {journal} {JCAP}\ }\textbf {\bibinfo {volume} {04}},\ \bibinfo {pages} {023}},\ \Eprint {https://arxiv.org/abs/2212.04522} {arXiv:2212.04522 [astro-ph.CO]} \BibitemShut {NoStop}%
\bibitem [{\citenamefont {Bahr-Kalus}\ \emph {et~al.}(2023)\citenamefont {Bahr-Kalus}, \citenamefont {Parkinson},\ and\ \citenamefont {Mueller}}]{Bahr-Kalus:2023ebd}%
  \BibitemOpen
  \bibfield  {author} {\bibinfo {author} {\bibfnamefont {B.}~\bibnamefont {Bahr-Kalus}}, \bibinfo {author} {\bibfnamefont {D.}~\bibnamefont {Parkinson}},\ and\ \bibinfo {author} {\bibfnamefont {E.-M.}\ \bibnamefont {Mueller}},\ }\bibfield  {title} {\bibinfo {title} {{Measurement of the matter-radiation equality scale using the extended baryon oscillation spectroscopic survey quasar sample}},\ }\href {https://doi.org/10.1093/mnras/stad1867} {\bibfield  {journal} {\bibinfo  {journal} {Mon. Not. Roy. Astron. Soc.}\ }\textbf {\bibinfo {volume} {524}},\ \bibinfo {pages} {2463} (\bibinfo {year} {2023})},\ \bibinfo {note} {[Erratum: Mon.Not.Roy.Astron.Soc. 526, 3248--3249 (2023)]},\ \Eprint {https://arxiv.org/abs/2302.07484} {arXiv:2302.07484 [astro-ph.CO]} \BibitemShut {NoStop}%
\bibitem [{\citenamefont {Krolewski}\ \emph {et~al.}(2024)\citenamefont {Krolewski}, \citenamefont {Percival},\ and\ \citenamefont {Woodfinden}}]{Krolewski:2024jwj}%
  \BibitemOpen
  \bibfield  {author} {\bibinfo {author} {\bibfnamefont {A.}~\bibnamefont {Krolewski}}, \bibinfo {author} {\bibfnamefont {W.~J.}\ \bibnamefont {Percival}},\ and\ \bibinfo {author} {\bibfnamefont {A.}~\bibnamefont {Woodfinden}},\ }\bibfield  {title} {\bibinfo {title} {{A new method to determine $H_0$ from cosmological energy-density measurements}},\ }\href@noop {} {\  (\bibinfo {year} {2024})},\ \Eprint {https://arxiv.org/abs/2403.19227} {arXiv:2403.19227 [astro-ph.CO]} \BibitemShut {NoStop}%
\bibitem [{\citenamefont {Smith}\ \emph {et~al.}(2023)\citenamefont {Smith}, \citenamefont {Poulin},\ and\ \citenamefont {Simon}}]{Smith:2022iax}%
  \BibitemOpen
  \bibfield  {author} {\bibinfo {author} {\bibfnamefont {T.~L.}\ \bibnamefont {Smith}}, \bibinfo {author} {\bibfnamefont {V.}~\bibnamefont {Poulin}},\ and\ \bibinfo {author} {\bibfnamefont {T.}~\bibnamefont {Simon}},\ }\bibfield  {title} {\bibinfo {title} {{Assessing the robustness of sound horizon-free determinations of the Hubble constant}},\ }\href {https://doi.org/10.1103/PhysRevD.108.103525} {\bibfield  {journal} {\bibinfo  {journal} {Phys. Rev. D}\ }\textbf {\bibinfo {volume} {108}},\ \bibinfo {pages} {103525} (\bibinfo {year} {2023})},\ \Eprint {https://arxiv.org/abs/2208.12992} {arXiv:2208.12992 [astro-ph.CO]} \BibitemShut {NoStop}%
\bibitem [{\citenamefont {Zaborowski}\ \emph {et~al.}(2024)\citenamefont {Zaborowski} \emph {et~al.}}]{Zaborowski:2024wpo}%
  \BibitemOpen
  \bibfield  {author} {\bibinfo {author} {\bibfnamefont {E.~A.}\ \bibnamefont {Zaborowski}} \emph {et~al.},\ }\bibfield  {title} {\bibinfo {title} {{A Sound Horizon-Free Measurement of $H_0$ in DESI 2024}},\ }\href@noop {} {\  (\bibinfo {year} {2024})},\ \Eprint {https://arxiv.org/abs/2411.16677} {arXiv:2411.16677 [astro-ph.CO]} \BibitemShut {NoStop}%
\bibitem [{\citenamefont {Philcox}\ \emph {et~al.}(2022)\citenamefont {Philcox}, \citenamefont {Farren}, \citenamefont {Sherwin}, \citenamefont {Baxter},\ and\ \citenamefont {Brout}}]{Philcox:2022sgj}%
  \BibitemOpen
  \bibfield  {author} {\bibinfo {author} {\bibfnamefont {O.~H.~E.}\ \bibnamefont {Philcox}}, \bibinfo {author} {\bibfnamefont {G.~S.}\ \bibnamefont {Farren}}, \bibinfo {author} {\bibfnamefont {B.~D.}\ \bibnamefont {Sherwin}}, \bibinfo {author} {\bibfnamefont {E.~J.}\ \bibnamefont {Baxter}},\ and\ \bibinfo {author} {\bibfnamefont {D.~J.}\ \bibnamefont {Brout}},\ }\bibfield  {title} {\bibinfo {title} {{Determining the Hubble constant without the sound horizon: A 3.6\% constraint on H0 from galaxy surveys, CMB lensing, and supernovae}},\ }\href {https://doi.org/10.1103/PhysRevD.106.063530} {\bibfield  {journal} {\bibinfo  {journal} {Phys. Rev. D}\ }\textbf {\bibinfo {volume} {106}},\ \bibinfo {pages} {063530} (\bibinfo {year} {2022})},\ \Eprint {https://arxiv.org/abs/2204.02984} {arXiv:2204.02984 [astro-ph.CO]} \BibitemShut {NoStop}%
\bibitem [{\citenamefont {Adame}\ \emph {et~al.}(2024{\natexlab{a}})\citenamefont {Adame} \emph {et~al.}}]{DESI:2024uvr}%
  \BibitemOpen
  \bibfield  {author} {\bibinfo {author} {\bibfnamefont {A.~G.}\ \bibnamefont {Adame}} \emph {et~al.} (\bibinfo {collaboration} {DESI}),\ }\bibfield  {title} {\bibinfo {title} {{DESI 2024 III: Baryon Acoustic Oscillations from Galaxies and Quasars}},\ }\href@noop {} {\  (\bibinfo {year} {2024}{\natexlab{a}})},\ \Eprint {https://arxiv.org/abs/2404.03000} {arXiv:2404.03000 [astro-ph.CO]} \BibitemShut {NoStop}%
\bibitem [{\citenamefont {Brieden}\ \emph {et~al.}(2022)\citenamefont {Brieden}, \citenamefont {Gil-Mar\'\i{}n},\ and\ \citenamefont {Verde}}]{Brieden:2022lsd}%
  \BibitemOpen
  \bibfield  {author} {\bibinfo {author} {\bibfnamefont {S.}~\bibnamefont {Brieden}}, \bibinfo {author} {\bibfnamefont {H.}~\bibnamefont {Gil-Mar\'\i{}n}},\ and\ \bibinfo {author} {\bibfnamefont {L.}~\bibnamefont {Verde}},\ }\bibfield  {title} {\bibinfo {title} {{Model-agnostic interpretation of 10 billion years of cosmic evolution traced by BOSS and eBOSS data}},\ }\href {https://doi.org/10.1088/1475-7516/2022/08/024} {\bibfield  {journal} {\bibinfo  {journal} {JCAP}\ }\textbf {\bibinfo {volume} {08}}\bibfield  {number} {\bibinfo  {number} { (08)},\ \bibinfo {pages} {024}},\ }\Eprint {https://arxiv.org/abs/2204.11868} {arXiv:2204.11868 [astro-ph.CO]} \BibitemShut {NoStop}%
\bibitem [{\citenamefont {Vagnozzi}\ \emph {et~al.}(2017)\citenamefont {Vagnozzi}, \citenamefont {Giusarma}, \citenamefont {Mena}, \citenamefont {Freese}, \citenamefont {Gerbino}, \citenamefont {Ho},\ and\ \citenamefont {Lattanzi}}]{Vagnozzi:2017ovm}%
  \BibitemOpen
  \bibfield  {author} {\bibinfo {author} {\bibfnamefont {S.}~\bibnamefont {Vagnozzi}}, \bibinfo {author} {\bibfnamefont {E.}~\bibnamefont {Giusarma}}, \bibinfo {author} {\bibfnamefont {O.}~\bibnamefont {Mena}}, \bibinfo {author} {\bibfnamefont {K.}~\bibnamefont {Freese}}, \bibinfo {author} {\bibfnamefont {M.}~\bibnamefont {Gerbino}}, \bibinfo {author} {\bibfnamefont {S.}~\bibnamefont {Ho}},\ and\ \bibinfo {author} {\bibfnamefont {M.}~\bibnamefont {Lattanzi}},\ }\bibfield  {title} {\bibinfo {title} {{Unveiling $\nu$ secrets with cosmological data: neutrino masses and mass hierarchy}},\ }\href {https://doi.org/10.1103/PhysRevD.96.123503} {\bibfield  {journal} {\bibinfo  {journal} {Phys. Rev. D}\ }\textbf {\bibinfo {volume} {96}},\ \bibinfo {pages} {123503} (\bibinfo {year} {2017})},\ \Eprint {https://arxiv.org/abs/1701.08172} {arXiv:1701.08172 [astro-ph.CO]} \BibitemShut {NoStop}%
\bibitem [{\citenamefont {Wang}\ \emph {et~al.}(2024{\natexlab{a}})\citenamefont {Wang}, \citenamefont {Mena}, \citenamefont {Di~Valentino},\ and\ \citenamefont {Gariazzo}}]{Wang:2024hen}%
  \BibitemOpen
  \bibfield  {author} {\bibinfo {author} {\bibfnamefont {D.}~\bibnamefont {Wang}}, \bibinfo {author} {\bibfnamefont {O.}~\bibnamefont {Mena}}, \bibinfo {author} {\bibfnamefont {E.}~\bibnamefont {Di~Valentino}},\ and\ \bibinfo {author} {\bibfnamefont {S.}~\bibnamefont {Gariazzo}},\ }\bibfield  {title} {\bibinfo {title} {{Updating neutrino mass constraints with background measurements}},\ }\href {https://doi.org/10.1103/PhysRevD.110.103536} {\bibfield  {journal} {\bibinfo  {journal} {Phys. Rev. D}\ }\textbf {\bibinfo {volume} {110}},\ \bibinfo {pages} {103536} (\bibinfo {year} {2024}{\natexlab{a}})},\ \Eprint {https://arxiv.org/abs/2405.03368} {arXiv:2405.03368 [astro-ph.CO]} \BibitemShut {NoStop}%
\bibitem [{\citenamefont {Naredo-Tuero}\ \emph {et~al.}(2024)\citenamefont {Naredo-Tuero}, \citenamefont {Escudero}, \citenamefont {Fern\'andez-Mart\'\i{}nez}, \citenamefont {Marcano},\ and\ \citenamefont {Poulin}}]{Naredo-Tuero:2024sgf}%
  \BibitemOpen
  \bibfield  {author} {\bibinfo {author} {\bibfnamefont {D.}~\bibnamefont {Naredo-Tuero}}, \bibinfo {author} {\bibfnamefont {M.}~\bibnamefont {Escudero}}, \bibinfo {author} {\bibfnamefont {E.}~\bibnamefont {Fern\'andez-Mart\'\i{}nez}}, \bibinfo {author} {\bibfnamefont {X.}~\bibnamefont {Marcano}},\ and\ \bibinfo {author} {\bibfnamefont {V.}~\bibnamefont {Poulin}},\ }\bibfield  {title} {\bibinfo {title} {{Critical look at the cosmological neutrino mass bound}},\ }\href {https://doi.org/10.1103/PhysRevD.110.123537} {\bibfield  {journal} {\bibinfo  {journal} {Phys. Rev. D}\ }\textbf {\bibinfo {volume} {110}},\ \bibinfo {pages} {123537} (\bibinfo {year} {2024})},\ \Eprint {https://arxiv.org/abs/2407.13831} {arXiv:2407.13831 [astro-ph.CO]} \BibitemShut {NoStop}%
\bibitem [{\citenamefont {Du}\ \emph {et~al.}(2024)\citenamefont {Du}, \citenamefont {Wu}, \citenamefont {Li},\ and\ \citenamefont {Zhang}}]{Du:2024pai}%
  \BibitemOpen
  \bibfield  {author} {\bibinfo {author} {\bibfnamefont {G.-H.}\ \bibnamefont {Du}}, \bibinfo {author} {\bibfnamefont {P.-J.}\ \bibnamefont {Wu}}, \bibinfo {author} {\bibfnamefont {T.-N.}\ \bibnamefont {Li}},\ and\ \bibinfo {author} {\bibfnamefont {X.}~\bibnamefont {Zhang}},\ }\bibfield  {title} {\bibinfo {title} {{Impacts of dark energy on weighing neutrinos after DESI BAO}},\ }\href@noop {} {\  (\bibinfo {year} {2024})},\ \Eprint {https://arxiv.org/abs/2407.15640} {arXiv:2407.15640 [astro-ph.CO]} \BibitemShut {NoStop}%
\bibitem [{\citenamefont {Jiang}\ \emph {et~al.}(2024{\natexlab{b}})\citenamefont {Jiang}, \citenamefont {Giar\`e}, \citenamefont {Gariazzo}, \citenamefont {Dainotti}, \citenamefont {Di~Valentino}, \citenamefont {Mena}, \citenamefont {Pedrotti}, \citenamefont {da~Costa},\ and\ \citenamefont {Vagnozzi}}]{Jiang:2024viw}%
  \BibitemOpen
  \bibfield  {author} {\bibinfo {author} {\bibfnamefont {J.-Q.}\ \bibnamefont {Jiang}}, \bibinfo {author} {\bibfnamefont {W.}~\bibnamefont {Giar\`e}}, \bibinfo {author} {\bibfnamefont {S.}~\bibnamefont {Gariazzo}}, \bibinfo {author} {\bibfnamefont {M.~G.}\ \bibnamefont {Dainotti}}, \bibinfo {author} {\bibfnamefont {E.}~\bibnamefont {Di~Valentino}}, \bibinfo {author} {\bibfnamefont {O.}~\bibnamefont {Mena}}, \bibinfo {author} {\bibfnamefont {D.}~\bibnamefont {Pedrotti}}, \bibinfo {author} {\bibfnamefont {S.~S.}\ \bibnamefont {da~Costa}},\ and\ \bibinfo {author} {\bibfnamefont {S.}~\bibnamefont {Vagnozzi}},\ }\bibfield  {title} {\bibinfo {title} {{Neutrino cosmology after DESI: tightest mass upper limits, preference for the normal ordering, and tension with terrestrial observations}},\ }\href@noop {} {\  (\bibinfo {year} {2024}{\natexlab{b}})},\ \Eprint {https://arxiv.org/abs/2407.18047} {arXiv:2407.18047 [astro-ph.CO]} \BibitemShut {NoStop}%
\bibitem [{\citenamefont {Roy~Choudhury}\ and\ \citenamefont {Choubey}(2018)}]{RoyChoudhury:2018gay}%
  \BibitemOpen
  \bibfield  {author} {\bibinfo {author} {\bibfnamefont {S.}~\bibnamefont {Roy~Choudhury}}\ and\ \bibinfo {author} {\bibfnamefont {S.}~\bibnamefont {Choubey}},\ }\bibfield  {title} {\bibinfo {title} {{Updated Bounds on Sum of Neutrino Masses in Various Cosmological Scenarios}},\ }\href {https://doi.org/10.1088/1475-7516/2018/09/017} {\bibfield  {journal} {\bibinfo  {journal} {JCAP}\ }\textbf {\bibinfo {volume} {09}},\ \bibinfo {pages} {017}},\ \Eprint {https://arxiv.org/abs/1806.10832} {arXiv:1806.10832 [astro-ph.CO]} \BibitemShut {NoStop}%
\bibitem [{\citenamefont {Roy~Choudhury}\ and\ \citenamefont {Hannestad}(2020)}]{RoyChoudhury:2019hls}%
  \BibitemOpen
  \bibfield  {author} {\bibinfo {author} {\bibfnamefont {S.}~\bibnamefont {Roy~Choudhury}}\ and\ \bibinfo {author} {\bibfnamefont {S.}~\bibnamefont {Hannestad}},\ }\bibfield  {title} {\bibinfo {title} {{Updated results on neutrino mass and mass hierarchy from cosmology with Planck 2018 likelihoods}},\ }\href {https://doi.org/10.1088/1475-7516/2020/07/037} {\bibfield  {journal} {\bibinfo  {journal} {JCAP}\ }\textbf {\bibinfo {volume} {07}},\ \bibinfo {pages} {037}},\ \Eprint {https://arxiv.org/abs/1907.12598} {arXiv:1907.12598 [astro-ph.CO]} \BibitemShut {NoStop}%
\bibitem [{\citenamefont {Adame}\ \emph {et~al.}(2024{\natexlab{b}})\citenamefont {Adame} \emph {et~al.}}]{DESI:2024jis}%
  \BibitemOpen
  \bibfield  {author} {\bibinfo {author} {\bibfnamefont {A.~G.}\ \bibnamefont {Adame}} \emph {et~al.} (\bibinfo {collaboration} {DESI}),\ }\bibfield  {title} {\bibinfo {title} {{DESI 2024 V: Full-Shape Galaxy Clustering from Galaxies and Quasars}},\ }\href@noop {} {\  (\bibinfo {year} {2024}{\natexlab{b}})},\ \Eprint {https://arxiv.org/abs/2411.12021} {arXiv:2411.12021 [astro-ph.CO]} \BibitemShut {NoStop}%
\bibitem [{\citenamefont {Adame}\ \emph {et~al.}(2024{\natexlab{c}})\citenamefont {Adame} \emph {et~al.}}]{DESI:2024hhd}%
  \BibitemOpen
  \bibfield  {author} {\bibinfo {author} {\bibfnamefont {A.~G.}\ \bibnamefont {Adame}} \emph {et~al.} (\bibinfo {collaboration} {DESI}),\ }\bibfield  {title} {\bibinfo {title} {{DESI 2024 VII: Cosmological Constraints from the Full-Shape Modeling of Clustering Measurements}},\ }\href@noop {} {\  (\bibinfo {year} {2024}{\natexlab{c}})},\ \Eprint {https://arxiv.org/abs/2411.12022} {arXiv:2411.12022 [astro-ph.CO]} \BibitemShut {NoStop}%
\bibitem [{\citenamefont {Chen}\ \emph {et~al.}(2020)\citenamefont {Chen}, \citenamefont {Vlah},\ and\ \citenamefont {White}}]{Chen:2020fxs}%
  \BibitemOpen
  \bibfield  {author} {\bibinfo {author} {\bibfnamefont {S.-F.}\ \bibnamefont {Chen}}, \bibinfo {author} {\bibfnamefont {Z.}~\bibnamefont {Vlah}},\ and\ \bibinfo {author} {\bibfnamefont {M.}~\bibnamefont {White}},\ }\bibfield  {title} {\bibinfo {title} {{Consistent Modeling of Velocity Statistics and Redshift-Space Distortions in One-Loop Perturbation Theory}},\ }\href {https://doi.org/10.1088/1475-7516/2020/07/062} {\bibfield  {journal} {\bibinfo  {journal} {JCAP}\ }\textbf {\bibinfo {volume} {07}},\ \bibinfo {pages} {062}},\ \Eprint {https://arxiv.org/abs/2005.00523} {arXiv:2005.00523 [astro-ph.CO]} \BibitemShut {NoStop}%
\bibitem [{\citenamefont {Chen}\ \emph {et~al.}(2021)\citenamefont {Chen}, \citenamefont {Vlah}, \citenamefont {Castorina},\ and\ \citenamefont {White}}]{Chen:2020zjt}%
  \BibitemOpen
  \bibfield  {author} {\bibinfo {author} {\bibfnamefont {S.-F.}\ \bibnamefont {Chen}}, \bibinfo {author} {\bibfnamefont {Z.}~\bibnamefont {Vlah}}, \bibinfo {author} {\bibfnamefont {E.}~\bibnamefont {Castorina}},\ and\ \bibinfo {author} {\bibfnamefont {M.}~\bibnamefont {White}},\ }\bibfield  {title} {\bibinfo {title} {{Redshift-Space Distortions in Lagrangian Perturbation Theory}},\ }\href {https://doi.org/10.1088/1475-7516/2021/03/100} {\bibfield  {journal} {\bibinfo  {journal} {JCAP}\ }\textbf {\bibinfo {volume} {03}},\ \bibinfo {pages} {100}},\ \Eprint {https://arxiv.org/abs/2012.04636} {arXiv:2012.04636 [astro-ph.CO]} \BibitemShut {NoStop}%
\bibitem [{\citenamefont {Lesgourgues}\ and\ \citenamefont {Pastor}(2006)}]{Lesgourgues:2006nd}%
  \BibitemOpen
  \bibfield  {author} {\bibinfo {author} {\bibfnamefont {J.}~\bibnamefont {Lesgourgues}}\ and\ \bibinfo {author} {\bibfnamefont {S.}~\bibnamefont {Pastor}},\ }\bibfield  {title} {\bibinfo {title} {{Massive neutrinos and cosmology}},\ }\href {https://doi.org/10.1016/j.physrep.2006.04.001} {\bibfield  {journal} {\bibinfo  {journal} {Phys. Rept.}\ }\textbf {\bibinfo {volume} {429}},\ \bibinfo {pages} {307} (\bibinfo {year} {2006})},\ \Eprint {https://arxiv.org/abs/astro-ph/0603494} {arXiv:astro-ph/0603494} \BibitemShut {NoStop}%
\bibitem [{\citenamefont {Simon}\ \emph {et~al.}(2023)\citenamefont {Simon}, \citenamefont {Zhang}, \citenamefont {Poulin},\ and\ \citenamefont {Smith}}]{Simon:2022lde}%
  \BibitemOpen
  \bibfield  {author} {\bibinfo {author} {\bibfnamefont {T.}~\bibnamefont {Simon}}, \bibinfo {author} {\bibfnamefont {P.}~\bibnamefont {Zhang}}, \bibinfo {author} {\bibfnamefont {V.}~\bibnamefont {Poulin}},\ and\ \bibinfo {author} {\bibfnamefont {T.~L.}\ \bibnamefont {Smith}},\ }\bibfield  {title} {\bibinfo {title} {{Consistency of effective field theory analyses of the BOSS power spectrum}},\ }\href {https://doi.org/10.1103/PhysRevD.107.123530} {\bibfield  {journal} {\bibinfo  {journal} {Phys. Rev. D}\ }\textbf {\bibinfo {volume} {107}},\ \bibinfo {pages} {123530} (\bibinfo {year} {2023})},\ \Eprint {https://arxiv.org/abs/2208.05929} {arXiv:2208.05929 [astro-ph.CO]} \BibitemShut {NoStop}%
\bibitem [{\citenamefont {Holm}\ \emph {et~al.}(2023)\citenamefont {Holm}, \citenamefont {Herold}, \citenamefont {Simon}, \citenamefont {Ferreira}, \citenamefont {Hannestad}, \citenamefont {Poulin},\ and\ \citenamefont {Tram}}]{Holm:2023laa}%
  \BibitemOpen
  \bibfield  {author} {\bibinfo {author} {\bibfnamefont {E.~B.}\ \bibnamefont {Holm}}, \bibinfo {author} {\bibfnamefont {L.}~\bibnamefont {Herold}}, \bibinfo {author} {\bibfnamefont {T.}~\bibnamefont {Simon}}, \bibinfo {author} {\bibfnamefont {E.~G.~M.}\ \bibnamefont {Ferreira}}, \bibinfo {author} {\bibfnamefont {S.}~\bibnamefont {Hannestad}}, \bibinfo {author} {\bibfnamefont {V.}~\bibnamefont {Poulin}},\ and\ \bibinfo {author} {\bibfnamefont {T.}~\bibnamefont {Tram}},\ }\bibfield  {title} {\bibinfo {title} {{Bayesian and frequentist investigation of prior effects in EFT of LSS analyses of full-shape BOSS and eBOSS data}},\ }\href {https://doi.org/10.1103/PhysRevD.108.123514} {\bibfield  {journal} {\bibinfo  {journal} {Phys. Rev. D}\ }\textbf {\bibinfo {volume} {108}},\ \bibinfo {pages} {123514} (\bibinfo {year} {2023})},\ \Eprint {https://arxiv.org/abs/2309.04468} {arXiv:2309.04468 [astro-ph.CO]} \BibitemShut {NoStop}%
\bibitem [{\citenamefont {{Jeffreys}}(1946)}]{1946RSPSA.186..453J}%
  \BibitemOpen
  \bibfield  {author} {\bibinfo {author} {\bibfnamefont {H.}~\bibnamefont {{Jeffreys}}},\ }\bibfield  {title} {\bibinfo {title} {{An Invariant Form for the Prior Probability in Estimation Problems}},\ }\href {https://doi.org/10.1098/rspa.1946.0056} {\bibfield  {journal} {\bibinfo  {journal} {Proceedings of the Royal Society of London Series A}\ }\textbf {\bibinfo {volume} {186}},\ \bibinfo {pages} {453} (\bibinfo {year} {1946})}\BibitemShut {NoStop}%
\bibitem [{\citenamefont {Maus}\ \emph {et~al.}(2024)\citenamefont {Maus} \emph {et~al.}}]{Maus:2024dzi}%
  \BibitemOpen
  \bibfield  {author} {\bibinfo {author} {\bibfnamefont {M.}~\bibnamefont {Maus}} \emph {et~al.},\ }\bibfield  {title} {\bibinfo {title} {{An analysis of parameter compression and full-modeling techniques with Velocileptors for DESI 2024 and beyond}},\ }\href@noop {} {\  (\bibinfo {year} {2024})},\ \Eprint {https://arxiv.org/abs/2404.07312} {arXiv:2404.07312 [astro-ph.CO]} \BibitemShut {NoStop}%
\bibitem [{\citenamefont {Hadzhiyska}\ \emph {et~al.}(2023)\citenamefont {Hadzhiyska}, \citenamefont {Wolz}, \citenamefont {Azzoni}, \citenamefont {Alonso}, \citenamefont {Garc\'\i{}a-Garc\'\i{}a}, \citenamefont {Ruiz-Zapatero},\ and\ \citenamefont {Slosar}}]{Hadzhiyska:2023wae}%
  \BibitemOpen
  \bibfield  {author} {\bibinfo {author} {\bibfnamefont {B.}~\bibnamefont {Hadzhiyska}}, \bibinfo {author} {\bibfnamefont {K.}~\bibnamefont {Wolz}}, \bibinfo {author} {\bibfnamefont {S.}~\bibnamefont {Azzoni}}, \bibinfo {author} {\bibfnamefont {D.}~\bibnamefont {Alonso}}, \bibinfo {author} {\bibfnamefont {C.}~\bibnamefont {Garc\'\i{}a-Garc\'\i{}a}}, \bibinfo {author} {\bibfnamefont {J.}~\bibnamefont {Ruiz-Zapatero}},\ and\ \bibinfo {author} {\bibfnamefont {A.}~\bibnamefont {Slosar}},\ }\bibfield  {title} {\bibinfo {title} {{Cosmology with 6 parameters in the Stage-IV era: efficient marginalisation over nuisance parameters}}\ }\href {https://doi.org/10.21105/astro.2301.11895} {10.21105/astro.2301.11895} (\bibinfo {year} {2023}),\ \Eprint {https://arxiv.org/abs/2301.11895} {arXiv:2301.11895 [astro-ph.CO]} \BibitemShut {NoStop}%
\bibitem [{\citenamefont {Sch\"oneberg}(2024)}]{Schoneberg:2024ifp}%
  \BibitemOpen
  \bibfield  {author} {\bibinfo {author} {\bibfnamefont {N.}~\bibnamefont {Sch\"oneberg}},\ }\bibfield  {title} {\bibinfo {title} {{The 2024 BBN baryon abundance update}},\ }\href {https://doi.org/10.1088/1475-7516/2024/06/006} {\bibfield  {journal} {\bibinfo  {journal} {JCAP}\ }\textbf {\bibinfo {volume} {06}},\ \bibinfo {pages} {006}},\ \Eprint {https://arxiv.org/abs/2401.15054} {arXiv:2401.15054 [astro-ph.CO]} \BibitemShut {NoStop}%
\bibitem [{\citenamefont {Hoffman}\ and\ \citenamefont {Gelman}(2011)}]{Hoffman:2011ukg}%
  \BibitemOpen
  \bibfield  {author} {\bibinfo {author} {\bibfnamefont {M.~D.}\ \bibnamefont {Hoffman}}\ and\ \bibinfo {author} {\bibfnamefont {A.}~\bibnamefont {Gelman}},\ }\bibfield  {title} {\bibinfo {title} {{The No-U-Turn Sampler: Adaptively Setting Path Lengths in Hamiltonian Monte Carlo}},\ }\href@noop {} {\  (\bibinfo {year} {2011})},\ \Eprint {https://arxiv.org/abs/1111.4246} {arXiv:1111.4246 [stat.CO]} \BibitemShut {NoStop}%
\bibitem [{\citenamefont {Torrado}\ and\ \citenamefont {Lewis}(2021)}]{Torrado:2020dgo}%
  \BibitemOpen
  \bibfield  {author} {\bibinfo {author} {\bibfnamefont {J.}~\bibnamefont {Torrado}}\ and\ \bibinfo {author} {\bibfnamefont {A.}~\bibnamefont {Lewis}},\ }\bibfield  {title} {\bibinfo {title} {{Cobaya: Code for Bayesian Analysis of hierarchical physical models}},\ }\href {https://doi.org/10.1088/1475-7516/2021/05/057} {\bibfield  {journal} {\bibinfo  {journal} {JCAP}\ }\textbf {\bibinfo {volume} {05}},\ \bibinfo {pages} {057}},\ \Eprint {https://arxiv.org/abs/2005.05290} {arXiv:2005.05290 [astro-ph.IM]} \BibitemShut {NoStop}%
\bibitem [{\citenamefont {Carron}\ \emph {et~al.}(2022)\citenamefont {Carron}, \citenamefont {Mirmelstein},\ and\ \citenamefont {Lewis}}]{Carron:2022eyg}%
  \BibitemOpen
  \bibfield  {author} {\bibinfo {author} {\bibfnamefont {J.}~\bibnamefont {Carron}}, \bibinfo {author} {\bibfnamefont {M.}~\bibnamefont {Mirmelstein}},\ and\ \bibinfo {author} {\bibfnamefont {A.}~\bibnamefont {Lewis}},\ }\bibfield  {title} {\bibinfo {title} {{CMB lensing from Planck PR4~maps}},\ }\href {https://doi.org/10.1088/1475-7516/2022/09/039} {\bibfield  {journal} {\bibinfo  {journal} {JCAP}\ }\textbf {\bibinfo {volume} {09}},\ \bibinfo {pages} {039}},\ \Eprint {https://arxiv.org/abs/2206.07773} {arXiv:2206.07773 [astro-ph.CO]} \BibitemShut {NoStop}%
\bibitem [{\citenamefont {Qu}\ \emph {et~al.}(2024)\citenamefont {Qu} \emph {et~al.}}]{ACT:2023dou}%
  \BibitemOpen
  \bibfield  {author} {\bibinfo {author} {\bibfnamefont {F.~J.}\ \bibnamefont {Qu}} \emph {et~al.} (\bibinfo {collaboration} {ACT}),\ }\bibfield  {title} {\bibinfo {title} {{The Atacama Cosmology Telescope: A Measurement of the DR6 CMB Lensing Power Spectrum and Its Implications for Structure Growth}},\ }\href {https://doi.org/10.3847/1538-4357/acfe06} {\bibfield  {journal} {\bibinfo  {journal} {Astrophys. J.}\ }\textbf {\bibinfo {volume} {962}},\ \bibinfo {pages} {112} (\bibinfo {year} {2024})},\ \Eprint {https://arxiv.org/abs/2304.05202} {arXiv:2304.05202 [astro-ph.CO]} \BibitemShut {NoStop}%
\bibitem [{\citenamefont {Madhavacheril}\ \emph {et~al.}(2024)\citenamefont {Madhavacheril} \emph {et~al.}}]{ACT:2023kun}%
  \BibitemOpen
  \bibfield  {author} {\bibinfo {author} {\bibfnamefont {M.~S.}\ \bibnamefont {Madhavacheril}} \emph {et~al.} (\bibinfo {collaboration} {ACT}),\ }\bibfield  {title} {\bibinfo {title} {{The Atacama Cosmology Telescope: DR6 Gravitational Lensing Map and Cosmological Parameters}},\ }\href {https://doi.org/10.3847/1538-4357/acff5f} {\bibfield  {journal} {\bibinfo  {journal} {Astrophys. J.}\ }\textbf {\bibinfo {volume} {962}},\ \bibinfo {pages} {113} (\bibinfo {year} {2024})},\ \Eprint {https://arxiv.org/abs/2304.05203} {arXiv:2304.05203 [astro-ph.CO]} \BibitemShut {NoStop}%
\bibitem [{\citenamefont {Scolnic}\ \emph {et~al.}(2022)\citenamefont {Scolnic} \emph {et~al.}}]{Scolnic:2021amr}%
  \BibitemOpen
  \bibfield  {author} {\bibinfo {author} {\bibfnamefont {D.}~\bibnamefont {Scolnic}} \emph {et~al.},\ }\bibfield  {title} {\bibinfo {title} {{The Pantheon+ Analysis: The Full Data Set and Light-curve Release}},\ }\href {https://doi.org/10.3847/1538-4357/ac8b7a} {\bibfield  {journal} {\bibinfo  {journal} {Astrophys. J.}\ }\textbf {\bibinfo {volume} {938}},\ \bibinfo {pages} {113} (\bibinfo {year} {2022})},\ \Eprint {https://arxiv.org/abs/2112.03863} {arXiv:2112.03863 [astro-ph.CO]} \BibitemShut {NoStop}%
\bibitem [{\citenamefont {Hill}\ \emph {et~al.}(2022)\citenamefont {Hill} \emph {et~al.}}]{Hill:2021yec}%
  \BibitemOpen
  \bibfield  {author} {\bibinfo {author} {\bibfnamefont {J.~C.}\ \bibnamefont {Hill}} \emph {et~al.},\ }\bibfield  {title} {\bibinfo {title} {{Atacama Cosmology Telescope: Constraints on prerecombination early dark energy}},\ }\href {https://doi.org/10.1103/PhysRevD.105.123536} {\bibfield  {journal} {\bibinfo  {journal} {Phys. Rev. D}\ }\textbf {\bibinfo {volume} {105}},\ \bibinfo {pages} {123536} (\bibinfo {year} {2022})},\ \Eprint {https://arxiv.org/abs/2109.04451} {arXiv:2109.04451 [astro-ph.CO]} \BibitemShut {NoStop}%
\bibitem [{\citenamefont {Jiang}(2024)}]{Jiang:2024nha}%
  \BibitemOpen
  \bibfield  {author} {\bibinfo {author} {\bibfnamefont {J.-Q.}\ \bibnamefont {Jiang}},\ }\bibfield  {title} {\bibinfo {title} {{Scale-dependence in $\Lambda$CDM parameters inferred from the CMB: a possible sign of Early Dark Energy}},\ }\href@noop {} {\  (\bibinfo {year} {2024})},\ \Eprint {https://arxiv.org/abs/2410.10559} {arXiv:2410.10559 [astro-ph.CO]} \BibitemShut {NoStop}%
\bibitem [{\citenamefont {Chudaykin}\ and\ \citenamefont {Ivanov}(2019)}]{Chudaykin:2019ock}%
  \BibitemOpen
  \bibfield  {author} {\bibinfo {author} {\bibfnamefont {A.}~\bibnamefont {Chudaykin}}\ and\ \bibinfo {author} {\bibfnamefont {M.~M.}\ \bibnamefont {Ivanov}},\ }\bibfield  {title} {\bibinfo {title} {{Measuring neutrino masses with large-scale structure: Euclid forecast with controlled theoretical error}},\ }\href {https://doi.org/10.1088/1475-7516/2019/11/034} {\bibfield  {journal} {\bibinfo  {journal} {JCAP}\ }\textbf {\bibinfo {volume} {11}},\ \bibinfo {pages} {034}},\ \Eprint {https://arxiv.org/abs/1907.06666} {arXiv:1907.06666 [astro-ph.CO]} \BibitemShut {NoStop}%
\bibitem [{\citenamefont {Ye}\ \emph {et~al.}(2021)\citenamefont {Ye}, \citenamefont {Hu},\ and\ \citenamefont {Piao}}]{Ye:2021nej}%
  \BibitemOpen
  \bibfield  {author} {\bibinfo {author} {\bibfnamefont {G.}~\bibnamefont {Ye}}, \bibinfo {author} {\bibfnamefont {B.}~\bibnamefont {Hu}},\ and\ \bibinfo {author} {\bibfnamefont {Y.-S.}\ \bibnamefont {Piao}},\ }\bibfield  {title} {\bibinfo {title} {{Implication of the Hubble tension for the primordial Universe in light of recent cosmological data}},\ }\href {https://doi.org/10.1103/PhysRevD.104.063510} {\bibfield  {journal} {\bibinfo  {journal} {Phys. Rev. D}\ }\textbf {\bibinfo {volume} {104}},\ \bibinfo {pages} {063510} (\bibinfo {year} {2021})},\ \Eprint {https://arxiv.org/abs/2103.09729} {arXiv:2103.09729 [astro-ph.CO]} \BibitemShut {NoStop}%
\bibitem [{\citenamefont {Jiang}\ and\ \citenamefont {Piao}(2022)}]{Jiang:2022uyg}%
  \BibitemOpen
  \bibfield  {author} {\bibinfo {author} {\bibfnamefont {J.-Q.}\ \bibnamefont {Jiang}}\ and\ \bibinfo {author} {\bibfnamefont {Y.-S.}\ \bibnamefont {Piao}},\ }\bibfield  {title} {\bibinfo {title} {{Toward early dark energy and ns=1 with Planck, ACT, and SPT observations}},\ }\href {https://doi.org/10.1103/PhysRevD.105.103514} {\bibfield  {journal} {\bibinfo  {journal} {Phys. Rev. D}\ }\textbf {\bibinfo {volume} {105}},\ \bibinfo {pages} {103514} (\bibinfo {year} {2022})},\ \Eprint {https://arxiv.org/abs/2202.13379} {arXiv:2202.13379 [astro-ph.CO]} \BibitemShut {NoStop}%
\bibitem [{\citenamefont {Cruz}\ \emph {et~al.}(2023)\citenamefont {Cruz}, \citenamefont {Niedermann},\ and\ \citenamefont {Sloth}}]{Cruz:2022oqk}%
  \BibitemOpen
  \bibfield  {author} {\bibinfo {author} {\bibfnamefont {J.~S.}\ \bibnamefont {Cruz}}, \bibinfo {author} {\bibfnamefont {F.}~\bibnamefont {Niedermann}},\ and\ \bibinfo {author} {\bibfnamefont {M.~S.}\ \bibnamefont {Sloth}},\ }\bibfield  {title} {\bibinfo {title} {{A grounded perspective on new early dark energy using ACT, SPT, and BICEP/Keck}},\ }\href {https://doi.org/10.1088/1475-7516/2023/02/041} {\bibfield  {journal} {\bibinfo  {journal} {JCAP}\ }\textbf {\bibinfo {volume} {02}},\ \bibinfo {pages} {041}},\ \Eprint {https://arxiv.org/abs/2209.02708} {arXiv:2209.02708 [astro-ph.CO]} \BibitemShut {NoStop}%
\bibitem [{\citenamefont {Jiang}\ \emph {et~al.}(2023)\citenamefont {Jiang}, \citenamefont {Ye},\ and\ \citenamefont {Piao}}]{Jiang:2022qlj}%
  \BibitemOpen
  \bibfield  {author} {\bibinfo {author} {\bibfnamefont {J.-Q.}\ \bibnamefont {Jiang}}, \bibinfo {author} {\bibfnamefont {G.}~\bibnamefont {Ye}},\ and\ \bibinfo {author} {\bibfnamefont {Y.-S.}\ \bibnamefont {Piao}},\ }\bibfield  {title} {\bibinfo {title} {{Return of Harrison\textendash{}Zeldovich spectrum in light of recent cosmological tensions}},\ }\href {https://doi.org/10.1093/mnrasl/slad137} {\bibfield  {journal} {\bibinfo  {journal} {Mon. Not. Roy. Astron. Soc.}\ }\textbf {\bibinfo {volume} {527}},\ \bibinfo {pages} {L54} (\bibinfo {year} {2023})},\ \Eprint {https://arxiv.org/abs/2210.06125} {arXiv:2210.06125 [astro-ph.CO]} \BibitemShut {NoStop}%
\bibitem [{\citenamefont {Jiang}\ \emph {et~al.}(2024{\natexlab{c}})\citenamefont {Jiang}, \citenamefont {Ye},\ and\ \citenamefont {Piao}}]{Jiang:2023bsz}%
  \BibitemOpen
  \bibfield  {author} {\bibinfo {author} {\bibfnamefont {J.-Q.}\ \bibnamefont {Jiang}}, \bibinfo {author} {\bibfnamefont {G.}~\bibnamefont {Ye}},\ and\ \bibinfo {author} {\bibfnamefont {Y.-S.}\ \bibnamefont {Piao}},\ }\bibfield  {title} {\bibinfo {title} {{Impact of the Hubble tension on the r \ensuremath{-} ns contour}},\ }\href {https://doi.org/10.1016/j.physletb.2024.138588} {\bibfield  {journal} {\bibinfo  {journal} {Phys. Lett. B}\ }\textbf {\bibinfo {volume} {851}},\ \bibinfo {pages} {138588} (\bibinfo {year} {2024}{\natexlab{c}})},\ \Eprint {https://arxiv.org/abs/2303.12345} {arXiv:2303.12345 [astro-ph.CO]} \BibitemShut {NoStop}%
\bibitem [{\citenamefont {Peng}\ and\ \citenamefont {Piao}(2024)}]{Peng:2023bik}%
  \BibitemOpen
  \bibfield  {author} {\bibinfo {author} {\bibfnamefont {Z.-Y.}\ \bibnamefont {Peng}}\ and\ \bibinfo {author} {\bibfnamefont {Y.-S.}\ \bibnamefont {Piao}},\ }\bibfield  {title} {\bibinfo {title} {{Testing the ns-H0 scaling relation with Planck-independent CMB data}},\ }\href {https://doi.org/10.1103/PhysRevD.109.023519} {\bibfield  {journal} {\bibinfo  {journal} {Phys. Rev. D}\ }\textbf {\bibinfo {volume} {109}},\ \bibinfo {pages} {023519} (\bibinfo {year} {2024})},\ \Eprint {https://arxiv.org/abs/2308.01012} {arXiv:2308.01012 [astro-ph.CO]} \BibitemShut {NoStop}%
\bibitem [{\citenamefont {Wang}\ \emph {et~al.}(2024{\natexlab{b}})\citenamefont {Wang}, \citenamefont {Ye}, \citenamefont {Jiang},\ and\ \citenamefont {Piao}}]{Wang:2024tjd}%
  \BibitemOpen
  \bibfield  {author} {\bibinfo {author} {\bibfnamefont {H.}~\bibnamefont {Wang}}, \bibinfo {author} {\bibfnamefont {G.}~\bibnamefont {Ye}}, \bibinfo {author} {\bibfnamefont {J.-Q.}\ \bibnamefont {Jiang}},\ and\ \bibinfo {author} {\bibfnamefont {Y.-S.}\ \bibnamefont {Piao}},\ }\bibfield  {title} {\bibinfo {title} {{Towards primordial gravitational waves and $n_s=1$ in light of BICEP/Keck, DESI BAO and Hubble tension}},\ }\href@noop {} {\  (\bibinfo {year} {2024}{\natexlab{b}})},\ \Eprint {https://arxiv.org/abs/2409.17879} {arXiv:2409.17879 [astro-ph.CO]} \BibitemShut {NoStop}%
\bibitem [{\citenamefont {Wang}\ and\ \citenamefont {Piao}(2024)}]{Wang:2024dka}%
  \BibitemOpen
  \bibfield  {author} {\bibinfo {author} {\bibfnamefont {H.}~\bibnamefont {Wang}}\ and\ \bibinfo {author} {\bibfnamefont {Y.-S.}\ \bibnamefont {Piao}},\ }\bibfield  {title} {\bibinfo {title} {{Dark energy in light of recent DESI BAO and Hubble tension}},\ }\href@noop {} {\  (\bibinfo {year} {2024})},\ \Eprint {https://arxiv.org/abs/2404.18579} {arXiv:2404.18579 [astro-ph.CO]} \BibitemShut {NoStop}%
\bibitem [{\citenamefont {Kallosh}\ and\ \citenamefont {Linde}(2022)}]{Kallosh:2022ggf}%
  \BibitemOpen
  \bibfield  {author} {\bibinfo {author} {\bibfnamefont {R.}~\bibnamefont {Kallosh}}\ and\ \bibinfo {author} {\bibfnamefont {A.}~\bibnamefont {Linde}},\ }\bibfield  {title} {\bibinfo {title} {{Hybrid cosmological attractors}},\ }\href {https://doi.org/10.1103/PhysRevD.106.023522} {\bibfield  {journal} {\bibinfo  {journal} {Phys. Rev. D}\ }\textbf {\bibinfo {volume} {106}},\ \bibinfo {pages} {023522} (\bibinfo {year} {2022})},\ \Eprint {https://arxiv.org/abs/2204.02425} {arXiv:2204.02425 [hep-th]} \BibitemShut {NoStop}%
\bibitem [{\citenamefont {Ye}\ \emph {et~al.}(2022)\citenamefont {Ye}, \citenamefont {Jiang},\ and\ \citenamefont {Piao}}]{Ye:2022efx}%
  \BibitemOpen
  \bibfield  {author} {\bibinfo {author} {\bibfnamefont {G.}~\bibnamefont {Ye}}, \bibinfo {author} {\bibfnamefont {J.-Q.}\ \bibnamefont {Jiang}},\ and\ \bibinfo {author} {\bibfnamefont {Y.-S.}\ \bibnamefont {Piao}},\ }\bibfield  {title} {\bibinfo {title} {{Toward inflation with ns=1 in light of the Hubble tension and implications for primordial gravitational waves}},\ }\href {https://doi.org/10.1103/PhysRevD.106.103528} {\bibfield  {journal} {\bibinfo  {journal} {Phys. Rev. D}\ }\textbf {\bibinfo {volume} {106}},\ \bibinfo {pages} {103528} (\bibinfo {year} {2022})},\ \Eprint {https://arxiv.org/abs/2205.02478} {arXiv:2205.02478 [astro-ph.CO]} \BibitemShut {NoStop}%
\bibitem [{\citenamefont {Vagnozzi}(2023)}]{Vagnozzi:2023nrq}%
  \BibitemOpen
  \bibfield  {author} {\bibinfo {author} {\bibfnamefont {S.}~\bibnamefont {Vagnozzi}},\ }\bibfield  {title} {\bibinfo {title} {{Seven Hints That Early-Time New Physics Alone Is Not Sufficient to Solve the Hubble Tension}},\ }\href {https://doi.org/10.3390/universe9090393} {\bibfield  {journal} {\bibinfo  {journal} {Universe}\ }\textbf {\bibinfo {volume} {9}},\ \bibinfo {pages} {393} (\bibinfo {year} {2023})},\ \Eprint {https://arxiv.org/abs/2308.16628} {arXiv:2308.16628 [astro-ph.CO]} \BibitemShut {NoStop}%
\bibitem [{\citenamefont {Bargiacchi}\ \emph {et~al.}(2022)\citenamefont {Bargiacchi}, \citenamefont {Benetti}, \citenamefont {Capozziello}, \citenamefont {Lusso}, \citenamefont {Risaliti},\ and\ \citenamefont {Signorini}}]{Bargiacchi:2021hdp}%
  \BibitemOpen
  \bibfield  {author} {\bibinfo {author} {\bibfnamefont {G.}~\bibnamefont {Bargiacchi}}, \bibinfo {author} {\bibfnamefont {M.}~\bibnamefont {Benetti}}, \bibinfo {author} {\bibfnamefont {S.}~\bibnamefont {Capozziello}}, \bibinfo {author} {\bibfnamefont {E.}~\bibnamefont {Lusso}}, \bibinfo {author} {\bibfnamefont {G.}~\bibnamefont {Risaliti}},\ and\ \bibinfo {author} {\bibfnamefont {M.}~\bibnamefont {Signorini}},\ }\bibfield  {title} {\bibinfo {title} {{Quasar cosmology: dark energy evolution and spatial curvature}},\ }\href {https://doi.org/10.1093/mnras/stac1941} {\bibfield  {journal} {\bibinfo  {journal} {Mon. Not. Roy. Astron. Soc.}\ }\textbf {\bibinfo {volume} {515}},\ \bibinfo {pages} {1795} (\bibinfo {year} {2022})},\ \Eprint {https://arxiv.org/abs/2111.02420} {arXiv:2111.02420 [astro-ph.CO]} \BibitemShut {NoStop}%
\bibitem [{\citenamefont {Bargiacchi}\ \emph {et~al.}(2023)\citenamefont {Bargiacchi}, \citenamefont {Dainotti},\ and\ \citenamefont {Capozziello}}]{Bargiacchi:2023rfd}%
  \BibitemOpen
  \bibfield  {author} {\bibinfo {author} {\bibfnamefont {G.}~\bibnamefont {Bargiacchi}}, \bibinfo {author} {\bibfnamefont {M.~G.}\ \bibnamefont {Dainotti}},\ and\ \bibinfo {author} {\bibfnamefont {S.}~\bibnamefont {Capozziello}},\ }\bibfield  {title} {\bibinfo {title} {{Tensions with the flat $\boldsymbol{\Lambda}$CDM model from high-redshift cosmography}},\ }\href {https://doi.org/10.1093/mnras/stad2326} {\bibfield  {journal} {\bibinfo  {journal} {Mon. Not. Roy. Astron. Soc.}\ }\textbf {\bibinfo {volume} {525}},\ \bibinfo {pages} {3104} (\bibinfo {year} {2023})},\ \Eprint {https://arxiv.org/abs/2307.15359} {arXiv:2307.15359 [astro-ph.CO]} \BibitemShut {NoStop}%
\bibitem [{\citenamefont {Adame}\ \emph {et~al.}(2024{\natexlab{d}})\citenamefont {Adame} \emph {et~al.}}]{DESI:2024mwx}%
  \BibitemOpen
  \bibfield  {author} {\bibinfo {author} {\bibfnamefont {A.~G.}\ \bibnamefont {Adame}} \emph {et~al.} (\bibinfo {collaboration} {DESI}),\ }\bibfield  {title} {\bibinfo {title} {{DESI 2024 VI: Cosmological Constraints from the Measurements of Baryon Acoustic Oscillations}},\ }\href@noop {} {\  (\bibinfo {year} {2024}{\natexlab{d}})},\ \Eprint {https://arxiv.org/abs/2404.03002} {arXiv:2404.03002 [astro-ph.CO]} \BibitemShut {NoStop}%
\bibitem [{\citenamefont {Sakr}(2025{\natexlab{a}})}]{Sakr:2025daj}%
  \BibitemOpen
  \bibfield  {author} {\bibinfo {author} {\bibfnamefont {Z.}~\bibnamefont {Sakr}},\ }\bibfield  {title} {\bibinfo {title} {{Uncovering the bias in the evidence for dynamical dark energy through minimal and generalized modeling approaches}},\ }\href@noop {} {\  (\bibinfo {year} {2025}{\natexlab{a}})},\ \Eprint {https://arxiv.org/abs/2501.14366} {arXiv:2501.14366 [astro-ph.CO]} \BibitemShut {NoStop}%
\bibitem [{\citenamefont {Sakr}(2025{\natexlab{b}})}]{Sakr:2025fay}%
  \BibitemOpen
  \bibfield  {author} {\bibinfo {author} {\bibfnamefont {Z.}~\bibnamefont {Sakr}},\ }\bibfield  {title} {\bibinfo {title} {{The case for a low dark matter density in dynamical dark energy model from local probes}},\ }in\ \href@noop {} {\emph {\bibinfo {booktitle} {{17th Marcel Grossmann Meeting}: {On Recent Developments in Theoretical and Experimental General Relativity, Gravitation, and Relativistic Field Theories}}}}\ (\bibinfo {year} {2025})\ \Eprint {https://arxiv.org/abs/2501.08915} {arXiv:2501.08915 [astro-ph.CO]} \BibitemShut {NoStop}%
\bibitem [{\citenamefont {Adame}\ \emph {et~al.}(2024{\natexlab{e}})\citenamefont {Adame} \emph {et~al.}}]{DESI:2024aax}%
  \BibitemOpen
  \bibfield  {author} {\bibinfo {author} {\bibfnamefont {A.~G.}\ \bibnamefont {Adame}} \emph {et~al.} (\bibinfo {collaboration} {DESI}),\ }\bibfield  {title} {\bibinfo {title} {{DESI 2024 II: Sample Definitions, Characteristics, and Two-point Clustering Statistics}},\ }\href@noop {} {\  (\bibinfo {year} {2024}{\natexlab{e}})},\ \Eprint {https://arxiv.org/abs/2411.12020} {arXiv:2411.12020 [astro-ph.CO]} \BibitemShut {NoStop}%
\bibitem [{\citenamefont {Laureijs}\ \emph {et~al.}(2011)\citenamefont {Laureijs} \emph {et~al.}}]{EUCLID:2011zbd}%
  \BibitemOpen
  \bibfield  {author} {\bibinfo {author} {\bibfnamefont {R.}~\bibnamefont {Laureijs}} \emph {et~al.} (\bibinfo {collaboration} {EUCLID}),\ }\bibfield  {title} {\bibinfo {title} {{Euclid Definition Study Report}},\ }\href@noop {} {\  (\bibinfo {year} {2011})},\ \Eprint {https://arxiv.org/abs/1110.3193} {arXiv:1110.3193 [astro-ph.CO]} \BibitemShut {NoStop}%
\bibitem [{\citenamefont {Amendola}\ \emph {et~al.}(2018)\citenamefont {Amendola} \emph {et~al.}}]{Amendola:2016saw}%
  \BibitemOpen
  \bibfield  {author} {\bibinfo {author} {\bibfnamefont {L.}~\bibnamefont {Amendola}} \emph {et~al.},\ }\bibfield  {title} {\bibinfo {title} {{Cosmology and fundamental physics with the Euclid satellite}},\ }\href {https://doi.org/10.1007/s41114-017-0010-3} {\bibfield  {journal} {\bibinfo  {journal} {Living Rev. Rel.}\ }\textbf {\bibinfo {volume} {21}},\ \bibinfo {pages} {2} (\bibinfo {year} {2018})},\ \Eprint {https://arxiv.org/abs/1606.00180} {arXiv:1606.00180 [astro-ph.CO]} \BibitemShut {NoStop}%
\bibitem [{\citenamefont {Orsi}\ \emph {et~al.}(2010)\citenamefont {Orsi}, \citenamefont {Baugh}, \citenamefont {Lacey}, \citenamefont {Cimatti}, \citenamefont {Wang},\ and\ \citenamefont {Zamorani}}]{Orsi:2009mj}%
  \BibitemOpen
  \bibfield  {author} {\bibinfo {author} {\bibfnamefont {A.}~\bibnamefont {Orsi}}, \bibinfo {author} {\bibfnamefont {C.~M.}\ \bibnamefont {Baugh}}, \bibinfo {author} {\bibfnamefont {C.~G.}\ \bibnamefont {Lacey}}, \bibinfo {author} {\bibfnamefont {A.}~\bibnamefont {Cimatti}}, \bibinfo {author} {\bibfnamefont {Y.}~\bibnamefont {Wang}},\ and\ \bibinfo {author} {\bibfnamefont {G.}~\bibnamefont {Zamorani}},\ }\bibfield  {title} {\bibinfo {title} {{Probing dark energy with future redshift surveys: A comparison of emission line and broad band selection in the near infrared}},\ }\href {https://doi.org/10.1111/j.1365-2966.2010.16585.x} {\bibfield  {journal} {\bibinfo  {journal} {Mon. Not. Roy. Astron. Soc.}\ }\textbf {\bibinfo {volume} {405}},\ \bibinfo {pages} {1006} (\bibinfo {year} {2010})},\ \Eprint {https://arxiv.org/abs/0911.0669} {arXiv:0911.0669 [astro-ph.CO]} \BibitemShut {NoStop}%
\bibitem [{\citenamefont {Yankelevich}\ and\ \citenamefont {Porciani}(2019)}]{Yankelevich:2018uaz}%
  \BibitemOpen
  \bibfield  {author} {\bibinfo {author} {\bibfnamefont {V.}~\bibnamefont {Yankelevich}}\ and\ \bibinfo {author} {\bibfnamefont {C.}~\bibnamefont {Porciani}},\ }\bibfield  {title} {\bibinfo {title} {{Cosmological information in the redshift-space bispectrum}},\ }\href {https://doi.org/10.1093/mnras/sty3143} {\bibfield  {journal} {\bibinfo  {journal} {Mon. Not. Roy. Astron. Soc.}\ }\textbf {\bibinfo {volume} {483}},\ \bibinfo {pages} {2078} (\bibinfo {year} {2019})},\ \Eprint {https://arxiv.org/abs/1807.07076} {arXiv:1807.07076 [astro-ph.CO]} \BibitemShut {NoStop}%
\bibitem [{\citenamefont {Di~Dio}\ \emph {et~al.}(2019)\citenamefont {Di~Dio}, \citenamefont {Durrer}, \citenamefont {Maartens}, \citenamefont {Montanari},\ and\ \citenamefont {Umeh}}]{DiDio:2018unb}%
  \BibitemOpen
  \bibfield  {author} {\bibinfo {author} {\bibfnamefont {E.}~\bibnamefont {Di~Dio}}, \bibinfo {author} {\bibfnamefont {R.}~\bibnamefont {Durrer}}, \bibinfo {author} {\bibfnamefont {R.}~\bibnamefont {Maartens}}, \bibinfo {author} {\bibfnamefont {F.}~\bibnamefont {Montanari}},\ and\ \bibinfo {author} {\bibfnamefont {O.}~\bibnamefont {Umeh}},\ }\bibfield  {title} {\bibinfo {title} {{The Full-Sky Angular Bispectrum in Redshift Space}},\ }\href {https://doi.org/10.1088/1475-7516/2019/04/053} {\bibfield  {journal} {\bibinfo  {journal} {JCAP}\ }\textbf {\bibinfo {volume} {04}},\ \bibinfo {pages} {053}},\ \Eprint {https://arxiv.org/abs/1812.09297} {arXiv:1812.09297 [astro-ph.CO]} \BibitemShut {NoStop}%
\bibitem [{\citenamefont {Desjacques}\ \emph {et~al.}(2018)\citenamefont {Desjacques}, \citenamefont {Jeong},\ and\ \citenamefont {Schmidt}}]{Desjacques:2016bnm}%
  \BibitemOpen
  \bibfield  {author} {\bibinfo {author} {\bibfnamefont {V.}~\bibnamefont {Desjacques}}, \bibinfo {author} {\bibfnamefont {D.}~\bibnamefont {Jeong}},\ and\ \bibinfo {author} {\bibfnamefont {F.}~\bibnamefont {Schmidt}},\ }\bibfield  {title} {\bibinfo {title} {{Large-Scale Galaxy Bias}},\ }\href {https://doi.org/10.1016/j.physrep.2017.12.002} {\bibfield  {journal} {\bibinfo  {journal} {Phys. Rept.}\ }\textbf {\bibinfo {volume} {733}},\ \bibinfo {pages} {1} (\bibinfo {year} {2018})},\ \Eprint {https://arxiv.org/abs/1611.09787} {arXiv:1611.09787 [astro-ph.CO]} \BibitemShut {NoStop}%
\end{thebibliography}%
\end{document}